\documentclass[11pt]{article}
\usepackage{jheppub}
\usepackage{amsmath,amssymb,amsfonts,graphicx,slashed,amsthm,mathtools,upgreek, enumerate, tensor}
\usepackage[dvipsnames]{xcolor}

\usepackage{longtable}
\usepackage{booktabs}
\usepackage{arydshln}

\usepackage{amsthm}

\newtheorem{lemma}{Lemma}

\usepackage{float}

\usepackage{tikz}
\usetikzlibrary{calc,arrows,decorations.pathreplacing}
\usepackage{caption}

\usepackage{comment}
\usepackage{hyperref}
\usepackage[utf8]{inputenc}
\usepackage[titletoc]{appendix}

\usepackage{cleveref}

\usepackage{placeins}

\usepackage[normalem]{ulem}

\usepackage{xparse}

\NewDocumentCommand{\GM}{o}{%
  \mathrm{GM}\IfValueT{#1}{^{(#1)}}%
}

\usepackage{braket}
\usepackage{bm}
\usepackage{cancel}
\usepackage{multirow}

\usepackage{subcaption}

\numberwithin{equation}{section}

\allowdisplaybreaks

\title{Genuine Multi-Entropy in Abelian Chern--Simons Theory: Exact Key-Ring Collapse and Its Breakdown for Generic Link States}

\author[\spadesuit]{Yen-Cheng Chang}

\author[\spadesuit]{Kaberi Goswami}

\author[\spadesuit,\diamondsuit]{Norihiro Iizuka}
 
\author[\diamondsuit]{Akihiro Miyata}

\affiliation[\spadesuit]{\it Department of Physics, National Tsing Hua University, Hsinchu 300044, Taiwan}

\affiliation[\diamondsuit]{\it Yukawa Institute for Theoretical Physics, Kyoto University, Kyoto 606-8502, Japan}

\emailAdd{yenchengchang23@gmail.com}
\emailAdd{kaberi@gapp.nthu.edu.tw}
\emailAdd{iizuka@phys.nthu.edu.tw}
\emailAdd{akihiro.miyata@yukawa.kyoto-u.ac.jp}

\abstract{
We study genuine multi-entropy in Abelian $U(1)_k$ Chern-Simons theory. For key-ring link states, where only the linking numbers between one distinguished component $K$ and the remaining $\mathtt{q}-1$ components are nonzero, we derive an exact closed-form expression for the $\mathtt{q}$-partite R\'enyi multi-entropy for general $\mathtt{q}$, level $k$, and R\'enyi index $n$. For $\mathtt{q}=4$, this shows that the genuine multi-entropy $\mathrm{GM}^{(4)}_n$ collapses exactly onto the tripartite information $I_{3,n}$ for all $n$, while for $\mathtt{q}=5$ it is likewise completely determined, for all $n$, by a linear combination of tripartite and bipartite R\'enyi multi-entropies. We then go beyond the key-ring class and study general four-component link states with arbitrary pairwise linking numbers. A numerical scan over Chern--Simons levels $2\leq k\leq24$ shows that the all-$n$ collapse found analytically for key-ring states does not survive for generic link states. Remarkably, the collapse remains exact at $n=2$ for every level examined. At $n=3$, violations occur, within the scanned range, only when $3\mid k$, with a further dependence on the $3$-adic valuation of $k$. At $n=4$ and $n=5$, violations occur for every level examined, with rates that vary strongly with $k$. These results show that the breakdown is not controlled simply by the zero-divisor structure of composite $\mathbb Z_k$, but instead exhibits a nontrivial joint dependence on the R\'enyi index and the arithmetic structure of the Chern--Simons level.
}


\preprint{}

\begin{document}

\maketitle

\parskip=10pt

\section{Introduction}

Quantum entanglement is a useful tool for probing nonlocal structures in quantum field theory. Three-dimensional Chern--Simons theory provides a particularly natural setting in which topology and entanglement meet. In link states, topological information of a link is encoded in the quantum state \cite{Witten:1988hf} and can therefore be probed through its entanglement structure. Bipartite entanglement measures of such states have been extensively studied \cite{Balasubramanian:2016sro, Dwivedi:2017rnj, Balasubramanian:2018por, Dwivedi:2020jyx}, whereas their multipartite entanglement structure remains much less explored \cite{Salton:2016qpp, Balasubramanian:2025kaf}. Link states thus provide a concrete setting for asking which topological data are detected by multipartite entanglement quantities.

To address this question, we focus on \emph{genuine} multi-entropy. Multi-entropy \cite{Gadde:2022cqi} is a symmetric generalization of the R\'enyi entropy to multipartite systems, characterized by the number of parties $\mathtt q$ and a R\'enyi index $n$. It is constructed from a canonical family of replica permutations acting on a $\mathtt q$-partite state. Although it is intrinsically multipartite, multi-entropy is generally sensitive to entanglement involving any subset of the parties, so a nonzero value does not by itself imply genuinely $\mathtt q$-partite correlations. Genuine multi-entropy \cite{Iizuka:2025ioc,Iizuka:2025caq} was introduced to remove these lower-partite contributions and isolate the irreducible $\mathtt q$-partite component. Applied to link states, it therefore provides a natural probe of whether there exists topological linking information that is intrinsically multipartite, rather than reducible to lower-partite data. Note that this is in sharp contrast to local gapped systems, where genuine multipartite entanglement has recently been shown to be effectively localized near junctions of multipartite partitions within a region set by the correlation length~\cite{Iizuka:2026qqg, Iizuka:2026ahd}.

For Abelian Chern–Simons theories, this question becomes particularly tractable. The corresponding link states are determined by pairwise linking numbers modulo the Chern–Simons level, and their entanglement quantities often reduce to arithmetic expressions involving greatest common divisors of these linking numbers. Recent work directly studying multipartite information and multi-entropy in Abelian Chern–Simons link states \cite{Yuan:2025dgx}, primarily for the tripartite ($\mathtt q=3$) case\footnote{Ref.~\cite{Yuan:2025dgx} also considered the $\mathtt q=4$ case at R\'enyi index $n=2$. Their result agrees exactly with the collapse of genuine multi-entropy onto the tripartite information $I_3$ discussed in this paper.}, has shown that these quantities are likewise governed by such gcd data. Related results for $\mathtt q=3$ were also obtained from the stabilizer-state perspective \cite{Akella:2025owv}, exploiting the graph-state description of Abelian Chern–Simons link states. These results leave open how this structure extends beyond the tripartite setting. Abelian link states therefore provide a natural and exactly solvable setting for asking whether, for higher $\mathtt q$, genuine multi-entropy reduces to lower-partite topological information or instead detects intrinsically multipartite linking structures.

A closely related question has recently been addressed for qubit stabilizer states. An earlier graph-state analysis \cite{Iizuka:2025pqq} showed that, for $\mathtt q=4$ and $n=2$, genuine multi-entropy collapses to a quantity proportional to the tripartite information $I_3$. At first sight, this reduction might be attributed to the special nature of the $n=2$ case: the corresponding multi-entropy belongs to the Coxeter class of multi-invariants and, for stabilizer states, reduces to a combination of bipartite entanglement entropies \cite{Akella:2026xza}. Remarkably, however, the collapse was subsequently shown to persist at $n=3$ \cite{Akella:2026rbe}. More generally, Ref.~\cite{Akella:2026rbe} established that, for $\mathtt q=4$ qubit stabilizer states, genuine multi-entropy reduces to lower-partite data for all $n<\mathtt q$.

The Abelian Chern–Simons link states studied here are stabilizer states at arbitrary Chern–Simons level $k$, as we show explicitly in Sec.\ref{sec:graphstates}. It is therefore natural to ask whether the same threshold at $n=\mathtt q$ persists beyond the qubit case. To our knowledge, this has not been established analytically for general qudit dimension $k$. Testing this threshold numerically, including at composite $k$, where the underlying arithmetic differs qualitatively from that of the qubit and prime-$k$ cases, is one of the central goals of Sec.\ref{sec:numerics}.

In this paper, we study genuine multi-entropy in Abelian $U(1)_k$ Chern–Simons link states along two complementary lines. We first analyze \emph{key-ring} link states, in which one distinguished component plays the role of the key ring and only the linking numbers between this component and the remaining components can be nonzero. For this class, the replica partition functions can be evaluated analytically for arbitrary numbers of parties, R\'enyi index $n$, and Chern–Simons level $k$. For $\mathtt q=4$, we show that genuine multi-entropy is proportional to the tripartite information $I_3$ for \emph{arbitrary $n$ and $k$}. Note that this is stronger than the general qubit-stabilizer result of Ref.~\cite{Akella:2026rbe}, where the proportionality holds for $n=2$ and $n=3$ but breaks down at $n=4$: within the special key-ring family, the collapse onto $I_3$ persists for all R\'enyi indices. We find an analogous simplification for $\mathtt q=5$, where, again for arbitrary $n$, genuine multi-entropy reduces to a linear combination of lower-partite quantities, namely tripartite and bipartite multi-entropies. Thus, within the key-ring class, genuine multi-entropy provides no independent $\mathtt q$-partite diagnostic beyond lower-partite quantities.

The key-ring family is analytically tractable but highly special, and this analytic control does not extend to generic linking configurations. We therefore turn to general four-component link states and investigate numerically when the collapse of genuine multi-entropy to $I_3$ ceases to hold. In particular, we identify the linking configurations for which genuine multi-entropy is no longer proportional to the tripartite information $I_3$, thereby isolating genuinely four-partite linking structures that are absent in the key-ring family.

Having identified genuinely four-partite linking structures in generic
link states, we then address the qudit-dimension question raised above.
While the threshold at $n=\mathtt q$ is known for qubit stabilizer
states ($k=2$) \cite{Akella:2026rbe}, it is not known whether this
pattern persists for general qudit dimension. We therefore investigate
generic four-component link states at higher Chern--Simons levels,
beginning with the representative cases $k=3$ and $k=4$ and then
extending the analysis to $2\leq k\leq24$. This allows us to test both
the role of composite-level arithmetic, including the zero-divisor
structure of $\mathbb Z_4$, and the broader dependence of the collapse
on the R\'enyi index and the arithmetic structure of $k$. As we will
see, the resulting pattern is not governed simply by whether $k$ is
prime or composite, but exhibits a nontrivial dependence on both the
R\'enyi index and the arithmetic structure of $k$.

The remainder of this paper is organized as follows.
In Sec.~\ref{sec:Review}, we review Abelian $U(1)_k$ Chern--Simons
link states, multi-entropy, and genuine multi-entropy, together with
the multipartite partition-function formulas used throughout the paper.
In Sec.~\ref{sec:keyring}, we analyze key-ring link states and derive
their multi-entropies and genuine multi-entropies. In particular, for
$\mathtt q=4$, we show that the genuine multi-entropy collapses onto
the tripartite information $I_{3,n}$ for arbitrary $n$ and $k$, while
for $\mathtt q=5$ it is completely determined by a linear combination
of tripartite and bipartite R\'enyi multi-entropies, again for arbitrary
$n$ and $k$. In Sec.~\ref{sec:numerics}, we move beyond the key-ring
class and study generic four-component link states numerically. After
using the analytic key-ring results as benchmarks, we analyze the
representative levels $k=2,3,4$ and then extend the scan to
$2\leq k\leq24$, identifying when the collapse to lower-partite
quantities persists and how its breakdown depends on the R\'enyi index
and the arithmetic structure of $k$. We conclude with a discussion in
Sec.~\ref{sec:discussion}.
Additional derivations, details of the equivalence relations and replica
contractions, and supplementary numerical data are collected in the
Appendices.

\section{Abelian Chern–Simons link states and genuine multi-entropy}
\label{sec:Review}

\subsection{Notation}

As we will see below, we consider three-dimensional Chern--Simons theory on a manifold with $N$ torus boundaries associated with an $N$-component link in $S^3$. We divide these $N$ boundary tori into $\mathtt{q}$ parties and study the multipartite entanglement among them. We first fix the notation used throughout the paper.

We denote the $\mathtt{q}$ parties by
\begin{equation}
\mathcal{A}_1\,,\mathcal{A}_2\,,\ldots\,,\mathcal{A}_{\mathtt{q}}.
\end{equation}

Let $N_i$ denote the number of component tori contained in the party $\mathcal{A}_i$. Then 
\begin{equation}
N=N_1+N_2+\cdots+N_{\mathtt{q}}.
\end{equation}
Thus, we always use the following notation throughout the paper:
\begin{align}
\begin{split}
N &: \textnormal{the total number of component tori}, \\
\mathtt{q} &: \textnormal{the number of parties}, \\
n &: \textnormal{the R\'enyi index}, \\
\mathcal{A}_i &: \textnormal{the $i$-th party}, \\
N_i &: \textnormal{the number of component tori contained in $\mathcal{A}_i$}.
\end{split}
\end{align}
In general,
\begin{equation}
\mathtt{q}\leq N.
\end{equation}
The equality
\begin{equation}
\mathtt{q}=N
\end{equation}
holds only in the highest-partite case, in which each party consists of a single component torus.

\subsection{Genuine multi-entropy}

Given a pure state defined on $\mathtt q$ parties
$\mathcal A_1,\ldots,\mathcal A_{\mathtt q}$,
the $n$-th R\'enyi multi-entropy is defined by \cite{Gadde:2022cqi}
\begin{align}
\begin{split}
S_n^{(\mathtt q)}
(\mathcal A_1:\cdots:\mathcal A_{\mathtt q})
& =
\frac{1}{1-n}\frac{1}{n^{\mathtt q-2}}
\log
\frac{\mathcal Z_n^{(\mathtt q)}}
{\left(\mathcal Z_1^{(\mathtt q)}\right)^{n^{\mathtt q-1}}}, \\
\mathcal Z_n^{(\mathtt q)} &= \bra{\psi}^{\otimes n^{\mathtt{q}-1}} \Sigma_1(g_1)\Sigma_2(g_2)\dots\Sigma_\mathtt{q}(g_\mathtt{q})\ket{\psi}^{\otimes n^{\mathtt{q}-1}},
\end{split}
\end{align}
where $\Sigma_\mathtt{k}(g_\mathtt{k})$ are twist operators for the permutation action of \(g_\mathtt{k}\) on indices of density matrices for $A_\mathtt{k}$. The action of \(g_\mathtt{k}\) can be expressed as
\begin{align}
\label{gkdefinition}
g_{\mathtt{k}} & \cdot (x_1,\dots,x_\mathtt{k},\dots,x_{\mathtt{q}-1}) = (x_1,\dots,x_{\mathtt{k}}+1,\dots,x_{\mathtt{q}-1}), \quad 1\le \mathtt{k} \le \mathtt{q}-1, \\
g_\mathtt{q} &= e ,
\end{align}
where \((x_1,x_2,\dots,x_{\mathtt{q}-1})\) represents an integer lattice point on a $(\mathtt{q}-1)$-dimensional hypercube of length \(n\) with identification of \(x_\mathtt{k}= n + 1 \) and \(x_\mathtt{k}=1\).

Although multi-entropy $S_n^{(\mathtt{q})}$ is a new multi-partite quantity, this by itself is not a great measure of $\texttt{q}$-partite entanglement since it is also sensitive to all $\mathtt{\tilde{q}}$-partite entanglements for $\mathtt{\tilde{q}}<\mathtt{q}$. 
To study genuine $\mathtt{q}$-partite entanglements, we define the genuine $\mathtt{q}$-partite R\'enyi multi-entropy $\GM[{\mathtt{q}}]_n(A_1:A_2:\dots:A_\mathtt{q})$  with the following properties:

\begin{itemize}
    \item $\GM[{\mathtt{q}}]_n(A_1:A_2:\dots:A_\mathtt{q})$ includes the $\mathtt{q}$-partite R\'enyi multi-entropy $S^{(\mathtt{q})}_n(A_1:A_2:\dots:A_\mathtt{q})$.
    \item $\GM[{\mathtt{q}}]_n(A_1:A_2:\dots:A_\mathtt{q})$ vanishes for all states that can be factorized as 
    \begin{align}
    \label{genuinecondition}
    \ket{\psi_\mathtt{q}}_{A_1 \cdots A_\mathtt{q}}=\ket{\psi_\mathtt{\tilde{q}}}_{A_1 \cdots A_\mathtt{\tilde{q}}} \otimes \ket{\psi_{\mathtt{q}-\mathtt{\tilde{q}}}}_{A_\mathtt{\tilde{q}+1} \cdots A_\mathtt{q}} 
    \end{align}
    where $\mathtt{\tilde{q}} = 1, 2, \cdots \mathtt{q}-1$. 
\end{itemize}
\noindent
In this way, $\GM[\mathtt q]_n$ extracts the irreducible $\mathtt q$-partite contribution from the multi-entropy. Explicit formulas for $\GM[\mathtt q]_n$ were derived in Refs.~\cite{Iizuka:2025ioc,Iizuka:2025caq}. We briefly review the $\mathtt q=3,4,$ and $5$ cases below.

\subsubsection{\texorpdfstring{$\mathtt q=3$}{q=3}}

For three parties, the genuine R\'enyi multi-entropy is uniquely given by \cite{Harper:2024ker, Iizuka:2025ioc}
\begin{equation}
\begin{split}
\GM^{(3)}_n(A\!:\!B\!:\!C)
&= S^{(3)}_n(A\!:\!B\!:\!C) \\
&\quad -\frac{1}{2}\Big(S_n^{(2)}(AB\!:\!C) 
+ S_n^{(2)}(AC\!:\!B)
+S_n^{(2)}(BC\!:\!A)\Big), \\
\end{split}
\label{eq:GM3-definition}
\end{equation}

\subsubsection{\texorpdfstring{$\mathtt q=4$}{q=4}}

For four parties, the genuine R\'enyi multi-entropy contains one free
parameter $a$ \cite{Iizuka:2025ioc}:
\begin{equation}
\begin{split}
\GM^{(4)}_n(A\!:\!B\!:\!C\!:\!D)
&=S^{(4)}_n[1:1:1:1]
-\frac{1}{3}S^{(3)}_n[2:1:1]+\frac{1}{3}S^{(2)}_n[3:1] -a\, I_{3,n}(A\!:\!B\!:\!C\!:\!D) \\
&= S_n^{(4)}(A\!:\!B\!:\!C\!:\!D)  \\
 & \quad  -\frac{1}{3}\Big(
S_n^{(3)}(AB\!:\!C\!:\!D)
+S_n^{(3)}(AC\!:\!B\!:\!D) 
+S_n^{(3)}(AD\!:\!B\!:\!C)  \\
& \quad  \qquad
+S_n^{(3)}(BC\!:\!A\!:\!D) 
+S_n^{(3)}(BD\!:\!A\!:\!C)
+S_n^{(3)}(CD\!:\!A\!:\!B)\Big)
 \\[-2pt]
&\quad  +\frac{1}{3}\Big(
S_n^{(2)}(ABC\!:\!D)
+S_n^{(2)}(ABD\!:\!C) 
+S_n^{(2)}(ACD\!:\!B)
+S_n^{(2)}(BCD\!:\!A)\Big)   \\
&\quad  -a\, I_{3,n}(A\!:\!B\!:\!C\!:\!D)\,,
\end{split}
\label{eq:GM4-definition}
\end{equation}
where $I_{3,n}$ denotes the R\'enyi tripartite information defined as 
\begin{align}
\begin{split}
 I_{3,n}(A\!:\!B\!:\!C\!:\!D) &=S^{(2)}_n[3:1]-S^{(2)}_n[2:2] \\
&= S_n^{(2)}(BCD\!:\!A) +S_n^{(2)}(ACD\!:\!B) +S_n^{(2)}(ABD\!:\!C)   + S_n^{(2)}(ABC\!:\!D) 
 \\
& \quad - S_n^{(2)}(AB\!:\!CD)-S_n^{(2)}(AC\!:\!BD)-S_n^{(2)}(AD\!:\!BC). 
\end{split}
\label{eq:I3n_def}
\end{align}
The fact that $I_{3,n}$ is a diagnostic of genuine quadripartite entanglement is consistent with the freedom to add an arbitrary multiple of $I_{3,n}$ to $\GM^{(4)}_n$ without spoiling its genuine character. Indeed, $I_{3,n}$ itself provides a diagnostic of genuine quadripartite entanglement \cite{Balasubramanian:2014hda}. Consequently, the quadripartite genuine multi-entropy yields two independent diagnostics: $\GM^{(4)}_n$ evaluated at a fixed value of $a$, and $I_{3,n}$ \cite{Iizuka:2025caq}.

\subsubsection{\texorpdfstring{$\mathtt q=5$}{q=5}}

For five parties, the genuine R\'enyi multi-entropy likewise contains one
free parameter, denoted by $b$ \cite{Iizuka:2025caq}:
\begin{equation}
\begin{split}
\GM[5]_n
={}&S_n^{(5)}[1:1:1:1:1]
-\frac{1}{4}S_n^{(4)}[2:1:1:1] +\frac{1+4b}{10}S_n^{(3)}[2:2:1] \\
&+\frac{1-16b}{20}S_n^{(3)}[3:1:1]
-\frac{1+4b}{20}S_n^{(2)}[3:2]
+b\,S_n^{(2)}[4:1]\,, 
\end{split}
\label{eq:GM5-definition}
\end{equation}
where explicit expressions for $S_n^{(5)}[1:1:1:1:1]$, $S_n^{(4)}[2:1:1:1]$, and the remaining terms are given in Appendix~\ref{detailGM5}.

It is useful to decompose this quantity as
\begin{equation}
 \GM[5]_n 
=
\left. \GM[5]_n \right|_{b=0}
+b\,\partial_b \GM[5]_n,
\end{equation}
where
\begin{equation}
\partial_b\GM[5]_n
=
\frac{2}{5}S_n^{(3)}[2:2:1]
-\frac{4}{5}S_n^{(3)}[3:1:1]
-\frac{1}{5}S_n^{(2)}[3:2]
+S_n^{(2)}[4:1].
\label{eq:partial-GM5}
\end{equation}

Both $\left.\GM[5]_n\right|_{b=0}$ and $\partial_b\GM[5]_n$ vanish on states containing only lower-partite entanglement and therefore provide independent diagnostics of genuine pentapartite entanglement, just as in the $\mathtt q=4$ case, where both $\left.\GM[4]_n\right|_{a=0}$ and $\partial_a\GM[4]_n=-I_{3,n}$ provide independent diagnostics of genuine quadripartite entanglement. We note that, while $\partial_a\GM[4]_n=-I_{3,n}$ is a linear combination of bipartite entropies, $\partial_b\GM[5]_n$ is constructed entirely from tripartite multi-entropies and bipartite entropies.

\subsection{Abelian Chern--Simons theory}

We consider Abelian Chern--Simons theory with gauge group $U(1)$ at level $k$, whose Euclidean action on a three-manifold $\mathcal{M}$ is
\begin{equation}
S_{\rm CS}[A]
=
\frac{k}{4\pi}
\int_{\mathcal{M}}
A\wedge dA.
\end{equation}
When $\mathcal{M}$ is a closed manifold such as $S^3$, the Euclidean path integral computes a topological invariant. If $\mathcal{M}$ has a boundary, on the other hand, the Euclidean path integral defines a quantum state in the Hilbert space associated with the boundary.

In this paper, we restrict our attention to link states prepared by the Euclidean path integral on the complement of a link in $S^3$. Let
$
\mathcal L=\mathcal C_1\cup\mathcal C_2\cup\cdots\cup\mathcal C_N
$
be an $N$-component link embedded in $S^3$. The bulk manifold is
\begin{equation}
\mathcal M=S^3\setminus \mathcal N(\mathcal L),
\end{equation}
where $\mathcal N(\mathcal L)$ denotes a tubular neighborhood of the link. Its boundary consists of $N$ disjoint tori,
\begin{equation}
\partial\mathcal M=\bigsqcup_{i=1}^{N}T_i^2.
\end{equation}

The Euclidean path integral on $\mathcal M$ prepares a quantum state in
\begin{equation}
\mathcal H_{\partial\mathcal M}
=
\mathcal H_{T_1^2}
\otimes
\mathcal H_{T_2^2}
\otimes
\cdots
\otimes
\mathcal H_{T_N^2}.
\end{equation}
This state is referred to as the \emph{link state} associated with $\mathcal L$.

For $U(1)_k$ Chern--Simons theory, the link state is given by
\begin{equation}
\ket{\mathcal L}
=
\frac{1}{k^{N/2}}
\sum_{\alpha_1,\ldots,\alpha_N=0}^{k-1}
\exp\left[
\frac{2\pi i}{k}
\sum_{i<j}
L_{ij}\alpha_i\alpha_j
\right]
\ket{\alpha_1,\ldots,\alpha_N},
\label{eq:U1-link-state}
\end{equation}
where $\alpha_i\in\mathbb Z_k$ labels the basis state on the $i$-th boundary torus and $L_{ij}$ denotes the Gauss linking number between the $i$-th and $j$-th components of the link. Therefore, the wavefunction is completely determined by the pairwise linking numbers modulo $k$.

\subsection{Previous results for Abelian link states}

Given the definition of genuine multi-entropy and the explicit form of the link-state wavefunction reviewed above, the corresponding multi-entropies and genuine multi-entropies can, in principle, be computed directly. In practice, however, the replica contractions and the associated counting of solutions to the resulting congruence constraints are somewhat involved.

In this subsection, we collect the results of Ref.~\cite{Yuan:2025dgx} that will be used in the subsequent sections. We quote only the final formulas and refer to Ref.~\cite{Yuan:2025dgx} for their derivations.
From now on, we denote the three boundary tori by
\begin{equation}
T_1=A,\qquad
T_2=B,\qquad
T_3=C \,.
\end{equation}

For a three-component link state in Abelian $U(1)_k$ Chern--Simons theory, with pairwise linking numbers
$
L_{AB}$, $L_{BC}$, $L_{CA}$, 
the following expression for the tripartite R\'enyi multi-entropy was conjectured in \cite{Yuan:2025dgx}:
\begin{equation}
\begin{aligned}
S_n^{(3)}(A:B:C)
={}&
\frac{1}{n}
\log
\frac{k^3}{
(k,L_{CA},L_{AB})
(k,L_{AB},L_{BC})
(k,L_{BC},L_{CA})
}
\\
&+
\left(1-\frac{2}{n}\right)
\log
\frac{k}{
(k,L_{AB},L_{BC},L_{CA})
}.
\end{aligned}
\label{eq:previous-tripartite-multi-entropy}
\end{equation}
Here and below,
\begin{equation}
(k,L_{AB},L_{BC},\ldots) = \mbox{gcd}(k,L_{AB},L_{BC},\ldots) \,,
\end{equation}
namely $(k,L_{AB},L_{BC},\ldots)$ denotes the greatest common divisor of its arguments. Equation~\eqref{eq:previous-tripartite-multi-entropy} reproduces the exact result at $n=2$ and the analytically tractable key-ring limit, and was further supported by numerical calculations.

The corresponding bipartite R\'enyi entropies are independent of the R\'enyi index $n$. For example,
\begin{equation}
S_n^{(2)}(A:BC)
=
\log
\frac{k}{(k,L_{AB},L_{AC})},
\label{eq:previous-bipartite-entropy}
\end{equation}
with analogous formulas obtained by permutations of $A,B,C$.

These results determine the tripartite genuine R\'enyi multi-entropy. The corresponding tripartite genuine R\'enyi multi-entropy is 
\begin{equation}
\begin{aligned}
 \GM[3]_n(A:B:C) 
=
\left(
\frac{1}{n}-\frac{1}{2}
\right)
\log
\frac{
k\,(k,L_{AB},L_{BC},L_{CA})^2
}{
(k,L_{AB},L_{AC})
(k,L_{AB},L_{BC})
(k,L_{AC},L_{BC})
}.
\end{aligned}
\label{eq:previous-tripartite-GM}
\end{equation}

Ref.~\cite{Yuan:2025dgx} also obtained an exact expression for the second R\'enyi multi-entropy of a general multipartite division of an Abelian link state. In the tripartite case, where the component tori are divided into three parties
$\mathcal A_1$,
$\mathcal A_2$, and 
$\mathcal A_3$, 
the result can be written in terms of the kernels of the linking matrices as
\begin{equation}
S_2^{(3)}
(\mathcal A_1:\mathcal A_2:\mathcal A_3)
=
\frac{1}{2}
\log
\frac{
k^N
}{
\left|\ker G_{\mathcal A_2\mathcal A_3,\mathcal A_1}\right|
\left|\ker G_{\mathcal A_3\mathcal A_1,\mathcal A_2}\right|
\left|\ker G_{\mathcal A_1\mathcal A_2,\mathcal A_3}\right|
}.
\label{eq:previous-general-tripartite-n2}
\end{equation}
Here $N$ is the total number of component tori, and each linking matrix contains the linking numbers connecting one party to the other two parties.

For a four-component link state, we identify
\begin{equation}
T_1=A,\qquad
T_2=B,\qquad
T_3=C,\qquad
T_4=D.
\end{equation}
the second R\'enyi quadripartite multi-entropy is given by
\begin{equation}
\begin{aligned}
S_2^{(4)}(A:B:C:D)
={}&
\frac{1}{4}
\log
\frac{
k^4
}{
|\ker G_{BCD,A}|
|\ker G_{CDA,B}|
|\ker G_{DAB,C}|
|\ker G_{ABC,D}|
}
\\
&+
\frac{1}{4}
\log
\frac{
k^6
}{
|\ker G_{AB,CD}|
|\ker G_{AC,BD}|
|\ker G_{AD,BC}|
}.
\end{aligned}
\label{eq:previous-quadripartite-n2}
\end{equation}
Equivalently, this can be written as
\begin{equation}
\begin{aligned}
S_2^{(4)}(A:B:C:D)
=
\frac{1}{4}
\Big[
&S_2(A:BCD)
+S_2(B:CDA)
+S_2(C:DAB)
+S_2(D:ABC)
\\
&+S_2(AB:CD)
+S_2(AC:BD)
+S_2(AD:BC)
\Big].
\end{aligned}
\label{eq:previous-quadripartite-entropy-relation}
\end{equation}

Thus, with the choice $a=1/12$ used in Ref.~\cite{Yuan:2025dgx}, the
quadripartite genuine multi-entropy vanishes at $n=2$:
\begin{equation}
\left. \GM[4]_{n=2}(A:B:C:D) \right|_{a=1/12} = 0 \,.
\label{YUanetalsq4result}
\end{equation}
As we will see in the next subsection, this vanishing is in fact
equivalent to the $n=2$ collapse relation found in
Ref.~\cite{Iizuka:2025pqq}.

\subsection{Connection to graph states}
\label{sec:graphstates}

The link state \eqref{eq:U1-link-state} contains only bilinear cross terms $L_{ij}\alpha_i\alpha_j$, and no self-interaction term of the form $\alpha_i^2$. This structure identifies $\ket{\mathcal L}$ as a \emph{qudit graph state}, for arbitrary Chern--Simons level $k$ and not only for $k=2$.

To see this explicitly, introduce the qudit Weyl--Heisenberg (generalized Pauli) operators acting on $\mathcal H_{T_i^2}\cong\mathbb C^k$,
\begin{equation}
X\ket{a}=\ket{a+1 \bmod k}\,,\qquad
Z\ket{a}=\omega^{a}\ket{a}\,,\qquad \omega\equiv e^{2\pi i/k}\,,
\label{eq:weyl-heisenberg}
\end{equation}
satisfying $XZ=\omega^{-1}ZX$. For each pair $(i,j)$, define the generalized controlled-phase gate
\begin{equation}
\mathrm{CZ}_{ij}(m)\ket{a_i,a_j}
=\omega^{m\,a_i a_j}\ket{a_i,a_j}\,,
\qquad a_i, a_j\in\mathbb Z_k,\ m\in\mathbb Z_k\,.
\label{eq:CZ-def}
\end{equation}
A direct computation gives the conjugation relations
\begin{equation}
\mathrm{CZ}_{ij}(m)\,X_i\,\mathrm{CZ}_{ij}(m)^\dagger=X_i\,Z_j^{\,m}\,,\qquad
\mathrm{CZ}_{ij}(m)\,Z_i\,\mathrm{CZ}_{ij}(m)^\dagger=Z_i\,,
\label{eq:CZ-conjugation}
\end{equation}
so that $\mathrm{CZ}_{ij}(m)$ is a Clifford operation for every $k$ and every $m\in\mathbb Z_k$; no restriction on the parity of $k$ is needed, precisely because \eqref{eq:U1-link-state} contains no diagonal $\alpha_i^2$ phase (a term whose Cliffordness would instead require special care for even $k$).

With this notation, the link state \eqref{eq:U1-link-state} can be written as
\begin{equation}
\ket{\mathcal L}
=\prod_{i<j} \mathrm{CZ}_{ij}(L_{ij})\;
\ket{+}^{\otimes N}\,,
\qquad
\ket{+}\equiv\frac{1}{\sqrt k}\sum_{a=0}^{k-1}\ket{a}\,,
\label{eq:L-as-graphstate}
\end{equation}
which is precisely the standard construction of a graph state on $N$ qudits \cite{Bahramgiri:2006yab,Schlingemann:2001zyo}, with adjacency matrix given by the linking numbers $L_{ij} \bmod k$. Since $\ket{+}$ is stabilized by $X$, conjugating by the product of $\mathrm{CZ}_{ij}$ gates in \eqref{eq:L-as-graphstate} and using \eqref{eq:CZ-conjugation} shows that $\ket{\mathcal L}$ is stabilized by the $N$ commuting operators
\begin{equation}
g_i \equiv X_i \prod_{j\neq i} Z_j^{\,L_{ij}}\,,
\qquad
g_i\ket{\mathcal L}=\ket{\mathcal L}\,,\qquad i=1,\dots,N\,.
\label{eq:stabilizer-generators}
\end{equation}
The group generated by $\{g_i\}$ has order $k^N$ and is abelian, so $\ket{\mathcal L}$ is a genuine stabilizer state for every Chern--Simons level $k$ and for every choice of the linking numbers $L_{ij}$. This identification holds independently of whether $k$ is prime, and independently of the choice of parties $\mathcal A_1,\dots,\mathcal A_{\mathtt q}$ into which the $N$ components are grouped.

For $k=2$, the above identification reduces the $U(1)_2$ link state
to an ordinary qubit graph state. In fact, for qubit graph and
stabilizer states, the quadripartite genuine multi-entropy was found
to collapse to the tripartite information $I_{3,n}$ at $n=2$
\cite{Iizuka:2025pqq} and $n=3$ \cite{Akella:2026rbe}.
For $\mathtt q=4$, these results can be written as
\begin{align}
\left. \GM^{(4)}_n \right|_{a}
&=
\left. \GM^{(4)}_n \right|_{a=0}
-a I_{3,n}
=
\left(
\frac{n^2-3n+3}{3n^2}-a
\right) I_{3,n}\,,
\qquad \mbox{for $n=2,3$} .
\label{eq:qubit-stabilizer-collapse}
\end{align}

In particular, for $n=2$,
\begin{equation}
\frac{n^2-3n+3}{3n^2}\bigg|_{n=2}
=\frac{1}{12}.
\end{equation}
Therefore, the vanishing of the quadripartite genuine multi-entropy
at $a=1/12$ in Eq.~\eqref{YUanetalsq4result} is precisely equivalent
to the $n=2$ collapse relation of Ref.~\cite{Iizuka:2025pqq}.

\section{Genuine multi-entropy for key-ring link states at generic $\mathtt{q}$ and $n$}
\label{sec:keyring}

In this section, we focus on key-ring link states and study their multi-entropy and genuine multi-entropy for a general number of components $N$, a general number of parties $\mathtt q$, and a general R\'enyi index $n$.
After introducing the general key-ring configuration and evaluating the corresponding multipartite partition function, we derive the multi-entropy in terms of the key-ring linking data. We then apply the general result to the cases $\mathtt{q}=4$ and $\mathtt{q}=5$ and show that, for these states, the genuine multi-entropy satisfies the conjectured relations expressing its collapse to lower-partite diagnostics. As we will see in Sec.~4, however, this collapse is a special property of the key-ring configuration: for generic four-component link states, the corresponding relation fails at $n=\mathtt{q}=4$, revealing genuinely four-partite linking information.

\subsection{Key-ring link state}

Let us first define the key-ring configuration. Let the $N$ component tori be denoted by
\begin{equation}
K,\ T_1,\ T_2,\ldots,T_{N-1},
\end{equation}
where $K$ is the key-ring torus and $T_1,T_2,\ldots,T_{N-1}$ are the key tori. A key-ring link is defined by the condition
\begin{equation}
L_{T_aT_b}=0,
\qquad
1\leq a<b\leq N-1.
\end{equation}
Thus, the only linking numbers that may be non-vanishing are
\begin{equation}
L_{KT_1},L_{KT_2},\ldots,L_{KT_{N-1}}.
\end{equation}

For a general $\mathtt{q}$-partite partition, these $N$ component tori are grouped into the parties
\begin{equation}
\mathcal{A}_1,\mathcal{A}_2,\ldots,\mathcal{A}_{\mathtt{q}}.
\end{equation}
The party containing the key-ring torus $K$ need not consist of $K$ alone. In general, it may also contain one or more key tori.

Consider an $N$-component Abelian link state associated with
$K,T_1,\ldots,T_{N-1}$:
\begin{equation}
\label{Thewavefn}
|\mathcal{L}\rangle=\frac{1}{k^{N/2}}\sum_{
\scalebox{0.5}{$
\begin{matrix}
\alpha=0\\\beta_1,\ldots,\beta_{N-1}=0
\end{matrix}
$}
}^{k-1}
\exp\left[
\frac{2\pi i}{k}
\left(\sum_a^{N-1} L_{KT_a}\alpha\beta_a+\sum_{a<b}^{N-1} L_{T_aT_b} \beta_a\beta_b\right)
\right]
|\alpha,\beta_1,\ldots,\beta_{N-1}\rangle .
\end{equation}
The corresponding reduced density operator of the subsystem $T_1, T_2 , \cdots , T_{N-1}$ is
\begin{align}
\hat{\rho}_{T_1T_2\cdots T_{N-1}}=\operatorname{Tr}_K\left(|\mathcal{L} \rangle\langle\mathcal{L}|\right)
\end{align}
\begin{align}
(\rho_{T_1\cdots T_{N-1}})^{\beta'_1\beta'_2\cdots\beta'_{N-1}}_{\beta_1\beta_2\cdots\beta_{N-1}}&=\sum_{\alpha=0}^{k-1}\frac{1}{k^N}\exp\left(
\frac{2\pi i}{k}
\left(\sum_{a<b}^{N-1} L_{T_aT_b} (\beta'_a \beta'_b-\beta_a \beta_b)
+\sum_a^{N-1}L_{KT_a}\alpha(\beta'_a-\beta_a)\right)\right)\notag\\
&=\frac{1}{k^{N-1}}\exp\left(
\frac{2\pi i}{k}
\sum_{a<b}^{N-1} L_{T_aT_b} (\beta'_a \beta'_b-\beta_a \beta_b)
\right)\eta\left[\sum_a^{N-1}(\beta'_a-\beta_a)L_{KT_a}\right]
\end{align}
where the $\eta$-constraint function,
\begin{equation}
\label{eta constraint definition}
\eta[x]
=
\frac{1}{k}
\sum_{\alpha=0}^{k-1}
\exp\left(\frac{2\pi i}{k}\alpha x\right)
=
\begin{cases}
1, & x\equiv 0 \pmod{k},\\
0, & x\not\equiv 0 \pmod{k}.
\end{cases}
\end{equation}
The $\eta$-constraint function~\eqref{eta constraint definition} satisfies the following properties:
\begin{align}\label{eta constraint properties}
    \eta[x]=\eta[-x], \hspace{4mm} \eta[x]^{2}=\eta[x], \hspace{4mm}\eta[x]\eta[y]=\eta[x+y] \eta[x]
\end{align}
To compute the Rényi multi-entropy, we need to perform a contraction for each index by different permutations. However, performing the contraction with a phase factor will be very difficult even for the case $\mathtt{q}=3$. Therefore, it would be very convenient if we could first get rid of the phase factor.

From the above derivation, we observe that the phase factor depends only on those linking numbers $(L_{T_aT_b})$ that are independent of the components traced out ($K$) in constructing the reduced density operator. By restricting to the key-ring configuration, for which
$L_{T_aT_b}=0$ for all $a<b$, the phase factor disappears and the
reduced density operator simplifies to 
\begin{align}
(\rho_{T_1\cdots T_{N-1}})^{\beta'_1\beta'_2\cdots\beta'_{N-1}}_{\beta_1\beta_2\cdots\beta_{N-1}}=\frac{1}{k^{N-1}} \eta\left[\sum_{a=1}^{N-1}(\beta'_a-\beta_a)L_{KT_a}\right]
\label{eq:rho of key-ring}
\end{align}
Then any $\mathtt{q}$-partite partition function for this $N$-component key-ring state is constructed from a product of $\eta$-constraint functions together with the appropriate contractions of their indices. Consequently, the problem is reduced to counting the independent index assignments that satisfy these $\eta$-constraints.

\subsection{\texorpdfstring{$\mathtt q=N$}{q=N} partition function for an \texorpdfstring{$N$}{N}-component key-ring state}
\label{sub:q partite partition fun for keyring}

We first consider the highest-partite division, in which every component torus forms a separate party, so that $\mathtt q=N$. We illustrate the counting in the tripartite case and then generalize the result to arbitrary $\mathtt q$ and $n$.

\subsubsection{Tripartite example}
\label{subsec:tripartite partition function for key-ring}

Consider the three-component key-ring state associated with $K,T_1,$ and $T_2$. From \eqref{eq:rho of key-ring}, the reduced density matrix obtained by tracing out $K$ is
\begin{equation}
\left(\rho_{T_1T_2}\right)^{\beta_1'\beta_2'}_{\beta_1\beta_2}
=
\frac{1}{k^2}
\eta\left[
(\beta_1'-\beta_1)L_{KT_1}
+
(\beta_2'-\beta_2)L_{KT_2}
\right].
\label{eq:rho of K3}
\end{equation}

The replica contraction defining $\mathcal Z_n^{(3)}$ contains $n^2$ copies of this reduced density matrix, arranged on an $n\times n$ periodic lattice. The two lattice directions correspond to the replica permutations acting on the $T_1$ and $T_2$ indices, respectively. Before simplifying the constraints, each replica contributes an $\eta$-function of the form
\begin{equation}
\eta\left[
\Delta_1\beta_1\,L_{KT_1}
+
\Delta_2\beta_2\,L_{KT_2}
\right],
\end{equation}
where $\Delta_1\beta_1$ and $\Delta_2\beta_2$ denote the differences of the $T_1$ and $T_2$ indices along their respective replica-lattice directions.

\begin{figure}[H]
\centering
\includegraphics[width=0.5\textwidth]{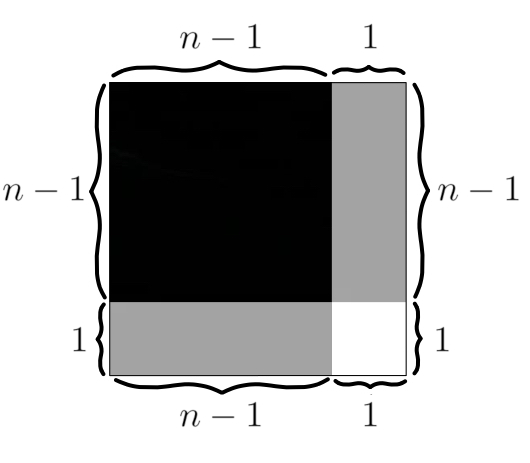}
\caption{Replica-lattice organization of the constraints for the tripartite partition function of a three-component key-ring state. The bulk and the two boundary directions generate the different greatest-common-divisor factors in \eqref{eq:Z^3_n our key}.}
\label{fig:Tripartite Keyring}
\end{figure}

Applying the identities in \eqref{eta constraint properties} successively along each row eliminates one $\beta_1$-difference per row. Repeating the same procedure along each column eliminates one $\beta_2$-difference per column. The resulting constraints are therefore organized into the following classes, as illustrated in Fig.~\ref{fig:Tripartite Keyring}:
\begin{itemize}
\item an $(n-1)\times(n-1)$ bulk block containing $(n-1)^2$ constraints, each involving both $L_{KT_1}$ and $L_{KT_2}$;
\item a boundary row containing $n-1$ constraints involving only $L_{KT_1}$;
\item a boundary column containing $n-1$ constraints involving only $L_{KT_2}$;
\item one trivial corner constraint $\eta[0]=1$.
\end{itemize}
We give an explicit illustration of the replica contraction and the reduction of the constraints in Appendix~\ref{Appendix_keyring_tripartite}.

Each bulk constraint contains two independent difference variables, one associated with $T_1$ and one associated with $T_2$. Hence, the bulk block contains
\begin{equation}
2(n-1)^2
\end{equation}
independent difference variables. Each boundary constraint contains only one difference variable, so the two boundaries contribute
\begin{equation}
(n-1)+(n-1)=2(n-1)
\end{equation}
additional difference variables.

Since each of the $n^2$ replicas carries one index associated with $T_1$ and one associated with $T_2$, the original contraction contains $2n^2$ indices. The total number of independent difference variables appearing in the constraints is therefore
\begin{equation}
2(n-1)^2+2(n-1)=2n(n-1).
\end{equation}
The remaining
\begin{equation}
2n^2-2n(n-1)=2n
\end{equation}
reference indices are unconstrained and contribute a factor of $k^{2n}$.

Using the counting formula proven in Appendix~\ref{app:one_eta_counting},
\begin{equation}
\sum_{\alpha_1,\ldots,\alpha_D=0}^{k-1}
\eta\left[
\sum_{i=1}^{D}\alpha_iL_i
\right]
=
k^{D-1}(k,L_1,\ldots,L_D),
\label{eq:eta-counting}
\end{equation}
where $(k,L_1,\ldots,L_D)$ denotes the greatest common divisor of its arguments, the bulk constraints contribute
\begin{equation}
\left[
k(k,L_{KT_1},L_{KT_2})
\right]^{(n-1)^2},
\end{equation}
while the two sets of boundary constraints contribute
\begin{equation}
(k,L_{KT_1})^{n-1}
(k,L_{KT_2})^{n-1}.
\end{equation}
Combining these factors with the normalization of the $n^2$ reduced density matrices, we obtain
\begin{equation}
\begin{split}
\mathcal Z_n^{(3)}(K:T_1:T_2)
={}&
\frac{k^{2n}}{k^{2n^2}}
\left[
k(k,L_{KT_1},L_{KT_2})
\right]^{(n-1)^2}
(k,L_{KT_1})^{n-1}
(k,L_{KT_2})^{n-1}
\\
={}&
\frac{1}{k^{n^2-1}}
(k,L_{KT_1})^{n-1}
(k,L_{KT_2})^{n-1}
(k,L_{KT_1},L_{KT_2})^{(n-1)^2}.
\end{split}
\label{eq:Z^3_n our key}
\end{equation}

\subsubsection{Generalization to arbitrary \texorpdfstring{$\mathtt q=N$}{q=N}}
\label{subsec:q=N partition function for key-ring}

For an $N$-component key-ring state with $\mathtt q=N$, the reduced
density matrix carries one pair of indices, one bra index and one ket
index, for each of the $\mathtt q-1$ key tori. The replica contraction
defining $\mathcal Z_n^{(\mathtt q)}$ contains
$n^{\mathtt q-1}$ copies of this reduced density matrix. These copies
can be arranged on an $(\mathtt q-1)$-dimensional periodic hypercubic
lattice, with one lattice direction associated with each key torus
$T_i$. The tripartite example corresponds to a two-dimensional lattice. In
that case, after reducing the constraints along the two periodic
directions, the lattice separates into a two-dimensional bulk region,
two one-dimensional boundary regions, and one trivial corner. The same
structure extends recursively to higher dimensions.

\begin{figure}[H]
\centering
\includegraphics[width=\textwidth]{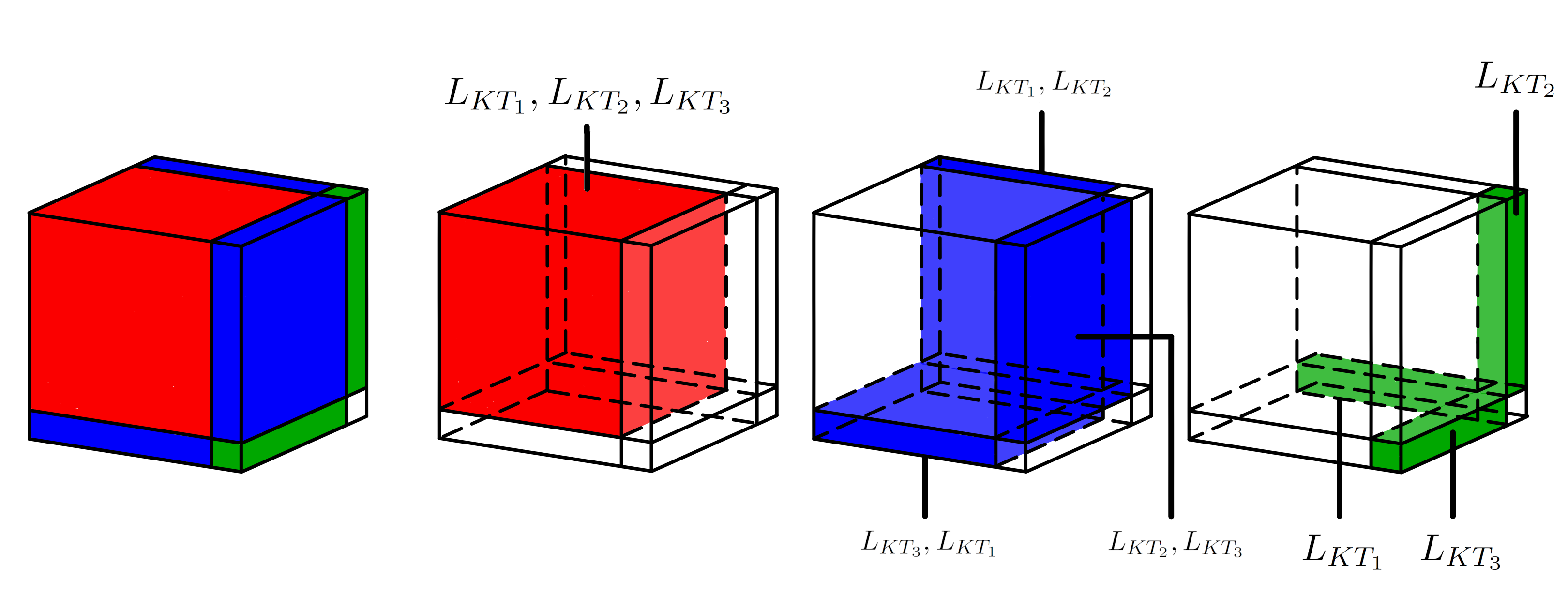}
\caption{Pictorial representation of the computation of the quadripartite partition function for the four-component key-ring state, organized in a hypercubic structure. The three-dimensional bulk, two-dimensional faces, and one-dimensional edges correspond respectively to greatest common divisors involving three, two, and one linking numbers.}
\label{fig:Quadripartite Keyring}
\end{figure}

For example, for $\mathtt q=4$, the replicas form a three-dimensional
periodic lattice, as illustrated in
Fig.~\ref{fig:Quadripartite Keyring}. After applying the
$\eta$-constraint identities along all three directions, the resulting
constraints are organized into a three-dimensional bulk, three
two-dimensional faces, three one-dimensional edges, and one trivial
corner. The bulk constraints involve all three linking numbers, each
face involves a particular pair of linking numbers, and each edge
involves a single linking number. See Appendix~\ref{Appendix_keyring_quadripartite} for an explicit analysis of the $\mathtt q=4$ case.

More generally, after the constraint reduction, the lattice decomposes
into regions of dimensions
\begin{equation}
\mathtt D=1,2,\ldots,\mathtt q-1.
\end{equation}
A $\mathtt D$-dimensional region is obtained by choosing
$\mathtt D$ directions from the original $\mathtt q-1$ lattice
directions. Equivalently, it is associated with choosing
$\mathtt D$ linking numbers from
\begin{equation}
L_{KT_1},L_{KT_2},\ldots,L_{KT_{\mathtt q-1}}.
\end{equation}
There are therefore
\begin{equation}
\binom{\mathtt q-1}{\mathtt D}
\end{equation}
distinct $\mathtt D$-dimensional regions.

Each such region has side length $n-1$ in each of its
$\mathtt D$ directions and hence contains
\begin{equation}
(n-1)^{\mathtt D}
\end{equation}
independent $\eta$-constraints. Every constraint in this region
contains $\mathtt D$ independent difference variables and involves the
corresponding $\mathtt D$ linking numbers
\begin{equation}
L_{KT_i},L_{KT_j},\ldots .
\end{equation}

It follows that the total number of independent difference variables
appearing in the constraints is
\begin{equation}
\begin{split}
\sum_{\mathtt D=1}^{\mathtt q-1}
\binom{\mathtt q-1}{\mathtt D}
\mathtt D(n-1)^{\mathtt D}
=
(\mathtt q-1)(n-1)n^{\mathtt q-2}.
\end{split}
\label{eq:number-constrained-indices}
\end{equation}
Here, the factor $\mathtt D$ arises because each constraint in a
$\mathtt D$-dimensional region contains $\mathtt D$ independent
difference variables.

On the other hand, each of the $n^{\mathtt q-1}$ replicas carries one
index for each of the $\mathtt q-1$ key tori. Therefore, the original
replica contraction contains
\begin{equation}
(\mathtt q-1)n^{\mathtt q-1}
\end{equation}
indices. The number of unconstrained reference indices is consequently
\begin{equation}
\begin{split}
(\mathtt q-1)n^{\mathtt q-1}
-(\mathtt q-1)(n-1)n^{\mathtt q-2}
=
(\mathtt q-1)n^{\mathtt q-2}.
\end{split}
\end{equation}
Since each of these indices can be summed freely over $\mathbb Z_k$,
they contribute the factor
\begin{equation}
k^{(\mathtt q-1)n^{\mathtt q-2}}.
\end{equation}

Using the counting formula in \eqref{eq:eta-counting}, the number
of assignments satisfying a single $\eta$-constraint involving
$\mathtt D$ independent variables and $\mathtt D$ linking numbers is
\begin{equation}
k^{\mathtt D-1}
\left(
k,L_{KT_i},L_{KT_j},\ldots
\right).
\end{equation}
Since each $\mathtt D$-dimensional region contains
$(n-1)^{\mathtt D}$ such constraints, and there are
$\binom{\mathtt q-1}{\mathtt D}$ possible choices of the corresponding
linking numbers, the full partition function takes the form
\begin{equation}
\begin{split}
\mathcal Z_n^{(\mathtt q)}
(K:T_1:\cdots:T_{\mathtt q-1})
={}&
\frac{1}{k^{(\mathtt q-1)n^{\mathtt q-1}}}
k^{(\mathtt q-1)n^{\mathtt q-2}}
\\
&\times
\prod_{\mathtt D=1}^{\mathtt q-1}
\prod_{\substack{
1\leq i_1<\cdots<i_{\mathtt D}\leq\mathtt q-1
}}
\left[
k^{\mathtt D-1}
\left(
k,L_{KT_{i_1}},\ldots,L_{KT_{i_{\mathtt D}}}
\right)
\right]^{(n-1)^{\mathtt D}}.
\end{split}
\label{eq:key-ring-partition-before-k-simplification}
\end{equation}

The total power of $k$ in \eqref{eq:key-ring-partition-before-k-simplification} is
\begin{equation}
\begin{split}
-(\mathtt q-1)n^{\mathtt q-1}
+(\mathtt q-1)n^{\mathtt q-2}
+\sum_{\mathtt D=1}^{\mathtt q-1}
(\mathtt D-1)
\binom{\mathtt q-1}{\mathtt D}
(n-1)^{\mathtt D}
=
-\left(n^{\mathtt q-1}-1\right).
\end{split}
\label{eq:key-ring-total-k-power}
\end{equation}

Therefore, the $\mathtt q$-partite partition function of the
$\mathtt q$-component key-ring state is
\begin{equation}
\mathcal Z_n^{(\mathtt q)}
(K:T_1:\cdots:T_{\mathtt q-1})
=
\frac{1}{k^{n^{\mathtt q-1}-1}}
\prod_{\mathtt D=1}^{\mathtt q-1}
\prod_{\substack{
1\leq i_1<\cdots<i_{\mathtt D}\leq\mathtt q-1
}}
\left(
k,L_{KT_{i_1}},\ldots,L_{KT_{i_{\mathtt D}}}
\right)^{(n-1)^{\mathtt D}}.
\label{eq:q=N key partition}
\end{equation}

\subsection{Lower partite partition function for key-ring state}
\label{sub:lower partite for key ring}
In Sec.~\ref{sub:q partite partition fun for keyring}, we derived the
$n$-replica partition function for the highest-partite division of an
$N$-component key-ring state, in which each component torus forms a
separate party and hence $\mathtt q=N$. To compute the genuine
R\'enyi multi-entropy, however, we also need lower-partite partition
functions with $\mathtt q<N$.

We therefore consider
\begin{equation}
\mathcal Z_n^{(\mathtt q)}
(\mathcal A_1:\mathcal A_2:\cdots:\mathcal A_{\mathtt q}),
\end{equation}
where the component tori
\begin{equation}
K,T_1,\ldots,T_{N-1}
\end{equation}
are grouped into $\mathtt q$ parties. Without loss of generality, we
denote the party containing the key-ring torus $K$ by
$\mathcal A_1$.

There are two independent features that characterize a general
lower-partite division.

First, the party $\mathcal A_1$ containing the key-ring torus may
consist of $K$ alone, or it may contain $K$ together with one or more key tori. As we will see below in Sec.~\ref{subsubsec:Adding key tori to the key-ring party}, the linking numbers associated with the key tori contained in $\mathcal A_1$ do not appear in the resulting partition function.

Second, each party $\mathcal A_i$, with $i=2,\ldots,\mathtt q$, may contain either a single key torus or several key tori. In the latter case, as we will see below in Sec.~\ref{subsubsec:Adding key tori to a non-key-ring party} and \ref{effectivelinkingn},  the corresponding linking numbers are replaced by an effective linking number defined by their greatest common divisor.

These two features are independent and may occur simultaneously.

Equivalently, we may start from an $N=\mathtt q$ component key-ring state in which each component torus forms a separate party. We then add $M$ additional key tori while keeping the number of parties fixed at $\mathtt q$, so that the resulting state has $N=\mathtt q+M$ component tori. Additional key tori assigned to the party $\mathcal A_1$, which contains the key-ring torus $K$, do not introduce new linking-number dependence in the partition function, whereas additional key tori assigned to a party $\mathcal A_i$, $i\geq2$, are incorporated through the corresponding effective linking number as we will see.

\subsubsection{Adding key tori to the key-ring party $\mathcal A_1$}
\label{subsubsec:Adding key tori to the key-ring party}
The first  feature is that adding key tori to the key-ring party $\mathcal A_1$ changes only the overall normalization of the full density operator. This change is exactly canceled by tracing over the added key tori, leaving the reduced density operator unchanged. Consequently, the $\mathtt{q}$-partite partition function remains the same.

Initially, the $N=\mathtt{q}$ parties are
\begin{align}
\mathcal{A}_1=K\,,\qquad\mathcal{A}_2=T_1\,,\qquad\cdots\,,\qquad\mathcal{A}_\mathtt{q}=T_{N-1}=T_{\mathtt{q}-1}
\end{align}
Its density operator is
\begin{align}
\rho^{\alpha'\beta'_1\beta'_2\cdots\beta'_{\mathtt{q}-1}}_{\alpha\beta_1\beta_2\cdots\beta_{\mathtt{q}-1}}=\frac{1}{k^\mathtt{q}}\exp{\left(\frac{2\pi i}{k}\sum_{a=1}^{\mathtt{q}-1}L_{KT_a}\left(\alpha'\beta'_a-\alpha\beta_a\right)\right)}
\end{align}
Tracing out the first party, which is the key-ring torus, we obtain the reduced density operator in \eqref{eq:rho of key-ring}
\begin{align}
(\rho_{\mathcal{A}_2\mathcal{A}_3\cdots\mathcal{A}_\mathtt{q}})^{\beta'_1\beta'_2\cdots\beta'_{\mathtt{q}-1}}_{\beta_1\beta_2\cdots\beta_{\mathtt{q}-1}}=\rho^{\alpha\beta'_1\beta'_2\cdots\beta'_{\mathtt{q}-1}}_{\alpha\beta_1\beta_2\cdots\beta_{\mathtt{q}-1}}=\frac{1}{k^{\mathtt{q}-1}}\eta\left[\sum_{a=1}^{\mathtt{q}-1}(\beta'_a-\beta_a)L_{KT_a}\right]
\label{eq:rho of q partite q component}
\end{align}
Now we add another key torus $T_{\mathtt{q}}$ to the first party.
The partition of the system becomes
\begin{align}
\mathcal{A}_1=KT_\mathtt{q}\,,\qquad\mathcal{A}_2=T_1\,,\qquad\cdots\,,\qquad\mathcal{A}_\mathtt{q}=T_{\mathtt{q}-1}\,,
\end{align}
and the number of components becomes $\mathtt{q}+1$. The density operator of the whole $\mathtt{q}+1$ components is
\begin{align}
\rho^{\alpha'\beta'_1\beta'_2\cdots\beta'_{\mathtt{q}-1}\beta'_\mathtt{q}}_{\alpha\beta_1\beta_2\cdots\beta_{\mathtt{q}-1}\beta_\mathtt{q}}=\frac{1}{k^{\mathtt{q}+1}}\exp{\left(\frac{2\pi i}{k}\sum_{a=1}^{\mathtt{q}}L_{KT_a}\left(\alpha'\beta'_a-\alpha\beta_a\right)\right)}
\end{align}
where the additional index $\beta_\mathtt{q}$ corresponds to the key torus $T_\mathtt{q}$. Consequently, the exponent of the prefactor $(1/k)$ increases by one. However, tracing out the first party that contains the key-ring torus and the key torus $T_\mathtt{q}$, the traced indices satisfy $\beta_\mathtt{q}'=\beta_\mathtt{q}$. Therefore, $(\beta_\mathtt{q}'-\beta_\mathtt{q})$ vanishes. We thus obtain 
\begin{align}
&(\rho_{\mathcal{A}_2\mathcal{A}_3\cdots\mathcal{A}_\mathtt{q}})^{\beta'_1\beta'_2\cdots\beta'_{\mathtt{q}-1}}_{\beta_1\beta_2\cdots\beta_{\mathtt{q}-1}}
=\rho^{\alpha\beta'_1\beta'_2\cdots\beta'_{\mathtt{q}-1}\beta_\mathtt{q}}_{\alpha\beta_1\beta_2\cdots\beta_{\mathtt{q}-1}\beta_\mathtt{q}}\notag\\
&=\sum_{\beta_\mathtt{q}=0}^{k-1}\frac{1}{k^{(\mathtt{q}+1)-1}}\eta\left[
\sum_{a=1}^{\mathtt{q}-1}(\beta'_a-\beta_a)L_{KT_a}\right]
=\frac{1}{k^{\mathtt{q}-1}}\eta\left[\sum_{a=1}^{\mathtt{q}-1}(\beta'_a-\beta_a)L_{KT_a}\right]
\end{align}
which is exactly the same as \eqref{eq:rho of q partite q component}. Therefore, constructing $n^{\mathtt{q}-1}$ replicas and performing the corresponding contractions, both reduced density operators give rise to exactly the same $\mathtt{q}$-partite partition function. Repeating the same process, one can readily show that adding any number of key tori to the key-ring party $\mathcal A_1$ leaves the $\mathtt{q}$-partite partition function unchanged.

\subsubsection{Adding key tori to a non-key-ring party $\mathcal A_i$ with $i\geq 2$}
\label{subsubsec:Adding key tori to a non-key-ring party}
The second feature is that adding key tori to a party that does not contain the key-ring modifies the $\mathtt{q}$-partite partition function. However, the exponent of the prefactor $(1/k)$ depends only on the number of parties $\mathtt{q}$ we are computing and therefore remains unchanged. Moreover, the structure of the partition function is preserved; we simply replace the linking number between the key-ring and the original key torus by the effective linking number between the key-ring and the corresponding party treated as a whole.

To illustrate this observation, we take the three-component key-ring state ($KT_1T_2$) with the partition
\begin{align}
\mathcal{A}_1=K\,,\qquad\mathcal{A}_2=T_1\,,\qquad\mathcal{A}_3=T_2
\end{align}
as an example. Tracing out the first party containing the key-ring torus, the reduced density operator given by \eqref{eq:rho of key-ring} is
\begin{align}
(\rho_{\mathcal{A}_2\mathcal{A}_3})^{\beta_1'\beta_2'}_{\beta_1\beta_2}=(\rho_{T_1T_2})^{\beta_1'\beta_2'}_{\beta_1\beta_2}=\frac{1}{k^2}\eta\left[(\beta_1'-\beta_1)L_{KT_1}+(\beta_2'-\beta_2)L_{KT_2}\right]
\end{align}
The tripartite partition function corresponding to this reduced density operator is computed in Sec.~\ref{subsec:tripartite partition function for key-ring}.

We then add another key torus ($T_3$) to a party that does not contain the key-ring (namely, $\mathcal{A}_3$), thereby obtaining the four-component key-ring state ($KT_1T_2T_3$) with the new partition
\begin{align}
\mathcal{A}_1=K\,,\qquad\mathcal{A}_2=T_1\,,\qquad\mathcal{A}_3=T_2T_3.
\end{align}
Again, tracing out the first party, the new reduced density operator given by \eqref{eq:rho of key-ring} becomes
\begin{align}
(\rho_{\mathcal{A}_2\mathcal{A}_3})^{\beta_1'\beta_2'\beta_3'}_{\beta_1\beta_2\beta_3}=(\rho_{T_1T_2T_3})^{\beta_1'\beta_2'\beta_3'}_{\beta_1\beta_2\beta_3}=\frac{1}{k^3}\eta\left[(\beta_1'-\beta_1)L_{KT_1}+(\beta_2'-\beta_2)L_{KT_2}+(\beta_3'-\beta_3)L_{KT_3}\right].
\end{align}
Applying the identities in \eqref{eta constraint properties} as in Sec.~\ref{subsec:tripartite partition function for key-ring} and using the counting formula proven in Appendix~\ref{app:one_eta_counting},
the corresponding tripartite partition function is
\begin{equation}
\scalebox{0.95}{$
\begin{aligned}
\hspace{-1mm}\mathcal{Z}^{(3)}_n(K:T_1:T_2T_3)=&\frac{k^{3n}}{k^{3n^2}}\underbrace{(k^2\cdot(k,L_{KT_1},L_{KT_2},L_{KT_3}))^{(n-1)^2}}_{\text{$(n-1)\times(n-1)$ block}}\underbrace{(k,L_{KT_1})^{n-1}}_{\text{$(n-1)$ boundary}}\underbrace{(k\cdot(k,L_{KT_2},L_{KT_3}))^{n-1}}_{\text{$(n-1)$ boundary}}\\
=&\frac{1}{k^{n^2-1}}(k,L_{KT_1},L_{KT_2},L_{KT_3})^{(n-1)^2}(k,L_{KT_1})^{n-1}(k,L_{KT_2},L_{KT_3})^{n-1}
\end{aligned}$}
\end{equation}
where the factor of $k^{3n}$ comes from summing over the unconstrained reference indices. We also give an explicit illustration of the replica contraction and the reduction of the constraints in Appendix~\ref{Appendix_effective_linking_tripartite}.

Comparing with \eqref{eq:Z^3_n our key} in Sec.~\ref{subsec:tripartite partition function for key-ring}, we find that, although each reduced density operator now contributes a factor of ($1/k^3$) instead of ($1/k^2$), the final exponent of the prefactor ($1/k$) remains unchanged after summing over all indices. Moreover, since the additional linking number $L_{KT_3}$ always appears together with $L_{KT_2}$, we can collect them together as follows.
\begin{equation}
\mathcal{Z}^{(3)}_n(K:T_1:T_2T_3)
=\frac{1}{k^{n^2-1}}(k,L_{KT_1},\textcolor{magenta}{(L_{KT_2},L_{KT_3})})^{(n-1)^2}(k,L_{KT_1})^{n-1}(k,\textcolor{magenta}{(L_{KT_2},L_{KT_3})})^{n-1}
\end{equation}
Comparing with \eqref{eq:Z^3_n our key}, the only effect is the replacement
\begin{align}
L_{KT_2}\mapsto(L_{KT_2},L_{KT_3})
\end{align}
This observation motivates the introduction of an effective linking
number, which will be defined in Sec.~\ref{effectivelinkingn}.

For generic $n$ and $\mathtt{q}$, we first consider the $N=\mathtt{q}$ key-ring state ($KT_1T_2\cdots T_{\mathtt{q}-1}$). The corresponding $\mathtt{q}$-partite partition function is given by \eqref{eq:q=N key partition} whose prefactor is
\begin{align}
\frac{1}{k^{n^{\mathtt{q-1}}-1}}
\end{align}
After adding one key torus to a party that does not contain the key-ring, the reduced density operator acquires one additional factor of $(1/k)$, since the total number of tori increases from $\mathtt{q}$ to $(\mathtt{q}+1)$. Since there are $n^{\mathtt{q}-1}$ replicas, each contributing an additional factor of $(1/k)$, the total extra factor is
\begin{align}
\frac{1}{k^{n^{\mathtt{q}-1}}}
\label{eq:extra denominator}
\end{align}
However, $(n-1)n^{\mathtt{q}-2}$ $\eta$-constraint functions acquire one additional linking number. By the assignment-counting formula derived in Appendix~\ref{app:one_eta_counting}, each additional linking number contributes one extra factor of $k$. Moreover, the additional key torus introduces $n^{\mathtt{q}-2}$ additional unconstrained reference indices among the $n^{\mathtt{q}-1}$ replicas. Summing over these additional reference indices contributes another factor of $k^{n^{\mathtt{q}-2}}$. Therefore, we obtain a factor of
\begin{align}
k^{(n-1)n^{\mathtt{q}-2}}k^{n^{\mathtt{q}-2}}=k^{n^{\mathtt{q}-1}}
\end{align}
which perfectly cancels the extra factor \eqref{eq:extra denominator}. Therefore, we conclude that, for generic $n$ and $\mathtt{q}$, adding a key torus to a party that does not contain the key-ring does not change the exponent of the prefactor $(1/k)$. Furthermore, as illustrated explicitly for $n=3$ and $\mathtt{q}=3$, the linking number between the additional key torus and the key-ring always appears in the greatest common divisors together with the linking numbers associated with the other tori in the same party. Hence, the overall structure of the partition function remains unchanged. The only modification is that, whenever the original linking number appears as an argument of a greatest common divisor, it should be replaced by the greatest common divisor of all linking numbers associated with the key tori belonging to that party.
\subsubsection{Effective linking number for lower partite partition function}
\label{effectivelinkingn}
From the discussion in Sec.~\ref{subsubsec:Adding key tori to the key-ring party} and Sec.~\ref{subsubsec:Adding key tori to a non-key-ring party}, we define the effective linking number between the party containing the key-ring torus and another party ($\mathcal{A}_i\,,i\neq1$) as
\begin{align}
L_{\mathcal{A}_1\mathcal{A}_i}=\mbox{gcd  of all $L_{KT_a}$} \,,\ T_a\in\mathcal{A}_i\,, \quad i\neq1 
\end{align}
We emphasize that the effective linking number defined above is specific to key-ring states. In a general link state, the effective linking between two parties could receive contributions from all pairwise linking numbers between the component tori in the two parties, and therefore cannot, in general, be written as such a simple greatest common divisor. With this definition, the $\mathtt{q}$-partite partition function for an arbitrary $N$-component key-ring state can be obtained directly from the $\mathtt{q}=N$ expression \eqref{eq:q=N key partition} by replacing the original linking numbers between tori with the corresponding effective linking numbers between parties.
\begin{equation}
\mathcal Z_n^{(\mathtt q)}
(\mathcal{A}_1:\mathcal{A}_2:\cdots :\mathcal{A}_{\mathtt{q}})
=
\frac{1}{k^{n^{\mathtt q-1}-1}}
\prod_{\mathtt D=1}^{\mathtt q-1}
\prod_{\substack{
2\leq i_1<\cdots<i_{\mathtt D}\leq\mathtt q
}}
\left(
k,L_{\mathcal{A}_1\mathcal{A}_{i_1}},\ldots,L_{\mathcal{A}_1\mathcal{A}_{i_{\mathtt D}}}
\right)^{(n-1)^{\mathtt D}}
\label{eq:q<N key partition}
\end{equation}
For example, consider a seven-component key-ring state ($K, T_1, T_2, T_3, T_4, T_5, T_6$) with the partition
\begin{align}
\mathcal{A}_1=KT_1\,,\qquad\mathcal{A}_2=T_2T_3\,,\qquad\mathcal{A}_3=T_4T_5T_6
\end{align}
The corresponding effective linking numbers are
\begin{align}
L_{\mathcal{A}_1\textcolor{magenta}{\mathcal{A}_2}}=(L_{K\textcolor{magenta}{T_2}},L_{K\textcolor{magenta}{T_3}})\,,\qquad
L_{\mathcal{A}_1\textcolor{magenta}{\mathcal{A}_3}}=(L_{K\textcolor{magenta}{T_4}},L_{K\textcolor{magenta}{T_5}},L_{K\textcolor{magenta}{T_6}})
\end{align}
and therefore the tripartite partition function is
\begin{align}
&\mathcal{Z}^{(3)}_n(\mathcal{A}_1:\mathcal{A}_2:\mathcal{A}_3)  =\mathcal{Z}^{(3)}_n(KT_1:T_2T_3:T_4T_5T_6)\notag\\
&= \frac{1}{k^{n^{3-1}-1}}   \left((k,L_{\mathcal{A}_1\mathcal{A}_2})(k,L_{\mathcal{A}_1\mathcal{A}_3})\right)^{(n-1)}\left(k,L_{\mathcal{A}_1\mathcal{A}_2},L_{\mathcal{A}_1\mathcal{A}_3}\right)^{(n-1)^2}
\notag\\
&=  \frac{1}{k^{n^{3-1}-1}}  \left((k,L_{KT_2},L_{KT_3})(k,L_{KT_4},L_{KT_5},L_{KT_6})\right)^{(n-1)}\left(k,L_{KT_2},L_{KT_3},L_{KT_4},L_{KT_5},L_{KT_6}\right)^{(n-1)^2}
\end{align}

\subsection{Genuine Rényi multi-entropy for four-component key-ring states}
In this section, we compute the genuine Rényi multi-entropy for a four-component key-ring state ($K, T_1, T_2, T_3$), where $K$ serves as the key-ring, and show its relation to $I_3$. To compute this, we need all partite partition functions for the four-component key-ring state.

The quadripartite partition function is obtained directly from \eqref{eq:q=N key partition} in Sec.~\ref{sub:q partite partition fun for keyring},
\begin{align}
\mathcal{Z}^{(4)}_n[1:1:1:1]=&\mathcal{Z}^{(4)}_n(K:T_1:T_2:T_3)\notag\\=&\frac{1}{k^{n^3-1}}\left((k,L_{KT_1})(k,L_{KT_2})(k,L_{KT_3})\right)^{n-1}\notag\\&\left((k,L_{KT_1},L_{KT_2})(k,L_{KT_2},L_{KT_3})(k,L_{KT_3},L_{KT_1})\right)^{(n-1)^2}\notag\\
&(k,L_{KT_1},L_{KT_2},L_{KT_3})^{(n-1)^3}.
\label{eq:Z^4_n key-ring summary}
\end{align}

The tripartite partition functions $(2:1:1)$ can be classified into two categories according to whether the key-ring $K$ forms a subsystem by itself. In the first category, the key-ring $K$ is treated alone as one party by itself. From \eqref{eq:q<N key partition}, a representative example is
\begin{align}
\mathcal{Z}^{(3)}_n(K:T_1:T_2T_3)=\frac{1}{k^{n^2-1}}(k,L_{KT_1})^{n-1}(k,L_{KT_2},L_{KT_3})^{n-1}(k,L_{KT_1},L_{KT_2},L_{KT_3})^{(n-1)^2}
\end{align}
The remaining two partition functions in this category are obtained by permuting the labels $T_1$, $T_2$, and $T_3$ (see \ref{app:G.1}). In the second category, the key-ring $K$ is treated as belonging to a two-component subsystem. From \eqref{eq:q<N key partition}, a representative example is
\begin{align}
\mathcal{Z}^{(3)}_n(KT_1:T_2:T_3)
=\frac{1}{k^{n^2-1}}(k,L_{KT_2})^{n-1}(k,L_{KT_3})^{n-1}(k,L_{KT_2},L_{KT_3})^{(n-1)^2}
\label{eq:Z^3_n(T_2:T_3:KT_1) key-ring summary}
\end{align}
The remaining two partition functions are again obtained by permuting the labels $T_1$, $T_2$, and $T_3$ (see \ref{app:G.1}). Multiplying all six partition functions together, namely the three partition functions from each category, yields
\begin{align}
\mathcal{Z}^{(3)}_n[2:1:1]
=\, &\mathcal{Z}^{(3)}_n(K:T_1:T_2T_3)\mathcal{Z}^{(3)}_n(K:T_2:T_1T_3)\mathcal{Z}^{(3)}_n(K:T_3:T_1T_2)\notag\\
&\quad \mathcal{Z}^{(3)}_n(T_1:T_3:KT_2)\mathcal{Z}^{(3)}_n(T_1:T_2:KT_3)\mathcal{Z}^{(3)}_n(T_2:T_3:KT_1)\notag\\
=&\frac{1}{k^{6(n^2-1)}}\left((k,L_{KT_1})(k,L_{KT_2})(k,L_{KT_3})\right)^{3(n-1)}\notag\\
&\quad  \times \left((k,L_{KT_1},L_{KT_2})(k,L_{KT_2},L_{KT_3})(k,L_{KT_3},L_{KT_1})\right)^{(n-1)+(n-1)^2} \notag\\
& \quad  \times (k,L_{KT_1},L_{KT_2},L_{KT_3})^{3(n-1)^2}  
\label{eq:Z^3_n(1:1:2) product key-ring summary}
\end{align}
Repeating the above analysis for the remaining partition types yields
\begin{gather}
\mathcal{Z}^{(2)}_n[3:1]\,,\qquad\mathcal{Z}^{(2)}_n[2:2]
\end{gather}
All detailed derivations for bipartite partition functions are shown in \ref{app:G.1}.

Thus, from \eqref{eq:GM4-definition}, the genuine R\'enyi multi-entropy for this four-component key-ring state with $a=0$ is
\begin{align}\label{eq:GM^4_n in terms of Z for key ring state}
\left. \GM[4]_n\right|_{a=0}=\, &S^{(4)}_n[1:1:1:1]
-\frac{1}{3}S^{(3)}_n[2:1:1]+\frac{1}{3}S^{(2)}_n[3:1]\notag\\
=&- \frac{1}{n-1}\frac{1}{n^2}\log\left(\frac{\mathcal{Z}^{(4)}_n[1:1:1:1]\left(\mathcal{Z}^{(2)}_n[3:1]\right)^{n^2/3}}{\left(\mathcal{Z}^{(3)}_n[2:1:1]\right)^{n/3}}\right)\notag\\
=&-\frac{3-3n+n^2}{3n^2} \log \left(\frac{(k,L_{KT_1})(k,L_{KT_2})(k,L_{KT_3})(k,L_{KT_1},L_{KT_2},L_{KT_3})}{k\cdot(k,L_{KT_1},L_{KT_2})(k,L_{KT_2},L_{KT_3})(k,L_{KT_3},L_{KT_1})}\right)
\end{align}
while the $I_{3,n}$ is
\begin{align}
I_{3,n}&
=S^{(2)}_n[3:1]-S^{(2)}_n[2:2]=- \frac{1}{n-1} \log \left(\frac{\mathcal{Z}^{(2)}_n[3:1]}{\mathcal{Z}^{(2)}_n[2:2]}\right)\notag\\
&=- \log \left(\frac{(k,L_{KT_1},L_{KT_2},L_{KT_3})(k,L_{KT_1})(k,L_{KT_2})(k,L_{KT_3})}{k\cdot (k,L_{KT_1},L_{KT_2})(k,L_{KT_2},L_{KT_3})(k,L_{KT_3},L_{KT_1})}\right)
\label{I3nkeyringcase}
\end{align}
Thus, we obtain the following relation
\begin{align}
\left. \GM^{(4)}_n\right|_{a=0}
=&-\frac{3-3n+n^2}{3n^2} \log \left(\frac{(k,L_{KT_1})(k,L_{KT_2})(k,L_{KT_3})(k,L_{KT_1},L_{KT_2},L_{KT_3})}{k\cdot(k,L_{KT_1},L_{KT_2})(k,L_{KT_2},L_{KT_3})(k,L_{KT_3},L_{KT_1})}\right)\notag \\
=&\, \frac{n^2-3n+3}{3n^2} I_{3, n} \,.
\end{align}
In general, we have 
\begin{align}\label{eq:GMI3-key-q4}
\boxed{
\left. \GM^{(4)}_n \right|_{a}=\left. \GM^{(4)}_n\right|_{a=0}-a I_{3,n}=\left(\frac{n^2-3n+3}{3n^2}-a\right) I_{3,n} \,.
}
\end{align}
For key-ring states, \eqref{eq:GMI3-key-q4} holds for arbitrary $n$. Since we restrict our analysis to key-ring states rather than general link states, some additional multipartite contributions may already vanish as a consequence of this restriction.

\subsection{Genuine Rényi multi-entropy for five-component key-ring states}
In this section, we compute the genuine Rényi multi-entropy for a five-component key-ring state ($K, T_1, T_2, T_3, T_4$), where $K$ serves as the key-ring, and show its relation to $\partial_{b}{\GM^{(5)}_n}$. To compute this, we need all partite partition functions for the five-component key-ring state.

The quintipartite partition function is obtained directly from \eqref{eq:q=N key partition} in Sec.~\ref{sub:q partite partition fun for keyring},
\begin{align}
&\mathcal{Z}^{(5)}_n[1:1:1:1:1]=\mathcal{Z}^{(5)}_n(K:T_1:T_2:T_3:T_4)\notag\\
& =\frac{1}{k^{n^4-1}}\left((k,L_{KT_1})(k,L_{KT_2})(k,L_{KT_3})(k,L_{KT_4})\right)^{n-1} \\
& \hspace{-5mm}\times \left((k,L_{KT_1},L_{KT_2})(k,L_{KT_1},L_{KT_3})(k,L_{KT_1},L_{KT_4})(k,L_{KT_2},L_{KT_3})(k,L_{KT_2},L_{KT_4})(k,L_{KT_3},L_{KT_4})\right)^{(n-1)^2}\notag\\
& \quad\times \left((k,L_{KT_1},L_{KT_2},L_{KT_3})(k,L_{KT_4},L_{KT_1},L_{KT_2})(k,L_{KT_3},L_{KT_4},L_{KT_1})(k,L_{KT_2},L_{KT_3},L_{KT_4})\right)^{(n-1)^3} \notag\\
& \quad \times (k,L_{KT_1},L_{KT_2},L_{KT_3},L_{KT_4})^{(n-1)^4} .  \notag
\end{align}

The quadripartite partition functions $(2:1:1:1)$ can be classified into two categories according to whether the key-ring $K$ forms a subsystem by itself. In the first category, the key-ring $K$ is treated alone as one party by itself. From \eqref{eq:q<N key partition}, a representative example is
\begin{align}
&\mathcal{Z}^{(4)}_n(K:T_1:T_2:T_3T_4)\notag\\&=\frac{1}{k^{n^3-1}}\left((k,L_{KT_1})(k,L_{KT_2})\right)^{(n-1)}(k,L_{KT_3},L_{KT_4})^{(n-1)}\notag\\ 
& \quad\times(k,L_{KT_1},L_{KT_2})^{(n-1)^2}\left((k,L_{KT_1},L_{KT_3},L_{KT_4})(k,L_{KT_2},L_{KT_3},L_{KT_4})\right)^{(n-1)^2}\notag\\
& \quad\times(k,L_{KT_1},L_{KT_2},L_{KT_3},L_{KT_4})^{(n-1)^3} .
\label{eq367}
\end{align}
The remaining five partition functions in this category are obtained by permuting the labels $T_1$, $T_2$, $T_3$, and $T_4$ (see \ref{app:G.2}). In the second category, the key-ring $K$ is treated as belonging to a two-component subsystem. From \eqref{eq:q<N key partition}, a representative example is
\begin{align}
&\mathcal{Z}^{(4)}_n(KT_1:T_2:T_3:T_4) \notag\\
&=\frac{1}{k^{n^3-1}}\left((k,L_{KT_2})(k,L_{KT_3})(k,L_{KT_4})\right)^{(n-1)}\notag\\
& \quad\times\left((k,L_{KT_2},L_{KT_4})(k,L_{KT_2},L_{KT_3})(k,L_{KT_3},L_{KT_4})\right)^{(n-1)^2}(k,L_{KT_2},L_{KT_3},L_{KT_4})^{(n-1)^3}
\label{eq368}
\end{align}
The remaining three partition functions are again obtained by permuting the labels $T_1$, $T_2$, $T_3$, and $T_4$ (see \ref{app:G.2}). Multiplying all ten partition functions together, namely the six partition functions of the first category and the four of the second category, yields
\begin{align}
 & \mathcal{Z}^{(4)}_n[2:1:1:1]\notag\\
& =\frac{1}{k^{10({n^3-1})}}\left((k,L_{KT_1})(k,L_{KT_2})(k,L_{KT_3})(k,L_{KT_4})\right)^{6(n-1)}\notag\\
& \quad\times \left((k,L_{KT_1},L_{KT_2})(k,L_{KT_1},L_{KT_3})(k,L_{KT_1},L_{KT_4}) \right. \notag \\ 
& \qquad \qquad \qquad \qquad \times \left. (k,L_{KT_2},L_{KT_3})(k,L_{KT_2},L_{KT_4})(k,L_{KT_3},L_{KT_4})\right)^{3(n-1)^2+(n-1)}\notag\\
& \quad\times \left((k,L_{KT_1},L_{KT_2},L_{KT_3})(k,L_{KT_4},L_{KT_1},L_{KT_2}) \right. \notag \\& \left. \qquad \qquad \qquad \qquad \times  (k,L_{KT_3},L_{KT_4},L_{KT_1})(k,L_{KT_2},L_{KT_3},L_{KT_4})\right)^{3(n-1)^2+(n-1)^3 }\notag\\
& \quad\times (k,L_{KT_1},L_{KT_2},L_{KT_3},L_{KT_4})^{6(n-1)^3}
\label{eq369}
\end{align}

The tripartite partition functions of type $(2:2:1)$ are likewise divided into two categories according to whether the key-ring $K$ forms a subsystem by itself. In the first category, the key-ring $K$ is treated alone as one party by itself. From \eqref{eq:q<N key partition}, a representative example is
\begin{align}
&\mathcal{Z}^{(3)}_n(K:T_1T_2:T_3T_4)\notag\\
&=\frac{1}{k^{n^2-1}} \left((k,L_{KT_1},L_{KT_2})(k,L_{KT_3},L_{KT_4})\right)^{(n-1)}(k,L_{KT_1},L_{KT_2},L_{KT_3},L_{KT_4})^{(n-1)^2}
\label{eq370}
\end{align}
The remaining two partition functions in this category are obtained by permuting the labels $T_1$, $T_2$, $T_3$, and $T_4$ (see \ref{app:G.2}). In the second category, the key-ring $K$ is treated as belonging to a two-component subsystem. From \eqref{eq:q<N key partition}, a representative example is
\begin{align}
&\mathcal{Z}^{(3)}_n(KT_1:T_2T_3:T_4)\notag\\
&=\frac{1}{k^{n^2-1}} (k,L_{KT_4})^{(n-1)}
(k,L_{KT_2},L_{KT_3})^{(n-1)}(k,L_{KT_2},L_{KT_3},L_{KT_4})^{(n-1)^2}
\label{eq371}
\end{align}
The remaining eleven partition functions are again obtained by permuting the labels $T_1$, $T_2$, $T_3$, and $T_4$ (see \ref{app:G.2}). Multiplying all fifteen partition functions together, namely the three partition functions of the first category and the twelve of the second category, yields
\begin{align}
&\mathcal{Z}^{(3)}_n[2:2:1]\notag\\
&=\frac{1}{k^{15(n^2-1)}}
\left((k,L_{KT_1})(k,L_{KT_2})(k,L_{KT_3})(k,L_{KT_4})\right)^{3(n-1)}\notag\\
& \hspace{-2mm}\times \left((k,L_{KT_1},L_{KT_2})(k,L_{KT_1},L_{KT_3})(k,L_{KT_1},L_{KT_4})(k,L_{KT_2},L_{KT_3})(k,L_{KT_2},L_{KT_4})(k,L_{KT_3},L_{KT_4})\right)^{3(n-1)}\notag\\
& \quad\times  \left((k,L_{KT_1},L_{KT_2},L_{KT_3})(k,L_{KT_4},L_{KT_1},L_{KT_2})(k,L_{KT_3},L_{KT_4},L_{KT_1})(k,L_{KT_2},L_{KT_3},L_{KT_4})\right)^{3(n-1)^2}\notag\\
& \quad\times  (k,L_{KT_1},L_{KT_2},L_{KT_3},L_{KT_4})^{3(n-1)^2} \,.
\label{eq372}
\end{align}

Repeating the above analysis for the remaining partition types yields
\begin{gather}
\mathcal{Z}^{(3)}_n[3:1:1]\,,\qquad\mathcal{Z}^{(2)}_n[3:2]\,,\qquad\mathcal{Z}^{(2)}_n[4:1]
\label{rest of the z}
\end{gather}
All detailed derivations for bipartite partition functions are shown in \ref{app:G.2}.

Thus, from \eqref{eq:GM5-definition}, the genuine R\'enyi multi-entropy for this five-component key-ring state with $b=0$ is
\begin{align}\label{GM_5 computation for q=5}
\left.\GM^{(5)}_n\right|_{b=0}=& S^{(5)}_n[1:1:1:1:1]- \frac{1}{4}S^{(4)}_n[2:1:1:1]
+\frac{1}{10}S^{(3)}_n[2:2:1]\notag\\
& +\frac{1}{20}S^{(3)}_n[3:1:1]-\frac{1}{20}S^{(2)}_n[3:2]\notag\\
=&-\frac{1}{n-1}\frac{1}{n^3}\log\left(\frac{\mathcal{Z}^{(5)}_n[1:1:1:1:1]\left(\mathcal{Z}^{(3)}_n[2:2:1]\right)^{n^2/10}\left(\mathcal{Z}^{(3)}_n[3:1:1]\right)^{n^2/20}}{\left(\mathcal{Z}^{(4)}_n[2:1:1:1]\right)^{n/4}\left(\mathcal{Z}^{(2)}_n[3:2]\right)^{n^3/20}}\right)\notag\\
=&-\frac{(n-1)(n-2)}{2n^3}\log\left[k^{-1}\left((k,L_{KT_1})(k,L_{KT_2})(k,L_{KT_3})(k,L_{KT_4})\right)\right.\notag \\
&\scalebox{0.9}{$\left.\left((k,L_{KT_1},L_{KT_2})(k,L_{KT_1},L_{KT_3})(k,L_{KT_1},L_{KT_4})(k,L_{KT_2},L_{KT_3})(k,L_{KT_2},L_{KT_4})(k,L_{KT_3},L_{KT_4})\right)^{-1}\right.$}\notag\\
&\scalebox{0.9}{$\left.\left((k,L_{KT_1},L_{KT_2},L_{KT_3})(k,L_{KT_4},L_{KT_1},L_{KT_2})(k,L_{KT_3},L_{KT_4},L_{KT_1})(k,L_{KT_2},L_{KT_3},L_{KT_4})\right)\right.$}\notag\\
&\left.(k,L_{KT_1},L_{KT_2},L_{KT_3},L_{KT_4})^{-1}\right]
\end{align}
while the derivative of $\GM^{(5)}_n$ with respect to $b$ is
\begin{align}
\partial_b\GM^{(5)}_n=& \frac{2}{5}S^{(3)}_n[2:2:1]- \frac{4}{5}S^{(3)}_n[3:1:1]
-\frac{1}{5}S^{(2)}_n[3:2]+S^{(2)}_n[4:1]\notag\\
=&- \frac{1}{n-1}\frac{1}{n}\log\left(\frac{\left(\mathcal{Z}^{(3)}_n[2:2:1]\right)^{2/5}\left(\mathcal{Z}^{(2)}_n[4:1]\right)^n}{\left(\mathcal{Z}^{(3)}_n[3:1:1]\right)^{4/5}\left(\mathcal{Z}^{(2)}_n[3:2]\right)^{n/5}}\right)\notag\\
=&-\frac{n-2}{n}\log\left[k^{-1}\left((k,L_{KT_1})(k,L_{KT_2})(k,L_{KT_3})(k,L_{KT_4})\right)\right.\notag \\
&\scalebox{0.9}{$\left.\left((k,L_{KT_1},L_{KT_2})(k,L_{KT_1},L_{KT_3})(k,L_{KT_1},L_{KT_4})(k,L_{KT_2},L_{KT_3})(k,L_{KT_2},L_{KT_4})(k,L_{KT_3},L_{KT_4})\right)^{-1}\right.$}\notag\\
&\scalebox{0.9}{$\left.\left((k,L_{KT_1},L_{KT_2},L_{KT_3})(k,L_{KT_4},L_{KT_1},L_{KT_2})(k,L_{KT_3},L_{KT_4},L_{KT_1})(k,L_{KT_2},L_{KT_3},L_{KT_4})\right)\right.$}\notag\\
&\left.(k,L_{KT_1},L_{KT_2},L_{KT_3},L_{KT_4})^{-1}\right]
\end{align}
Thus, we obtain the following relation
\begin{align}
\left. \GM^{(5)}_n\right|_{b=0}
=&-\frac{(n-2)(n-1)}{2n^3}\log\left[k^{-1}\left((k,L_{KT_1})(k,L_{KT_2})(k,L_{KT_3})(k,L_{KT_4})\right)\right.\notag \\
&\scalebox{0.9}{$\left.\left((k,L_{KT_1},L_{KT_2})(k,L_{KT_1},L_{KT_3})(k,L_{KT_1},L_{KT_4})(k,L_{KT_2},L_{KT_3})(k,L_{KT_2},L_{KT_4})(k,L_{KT_3},L_{KT_4})\right)^{-1}\right.$}\notag\\
&\scalebox{0.9}{$\left.\left((k,L_{KT_1},L_{KT_2},L_{KT_3})(k,L_{KT_4},L_{KT_1},L_{KT_2})(k,L_{KT_3},L_{KT_4},L_{KT_1})(k,L_{KT_2},L_{KT_3},L_{KT_4})\right)\right.$}\notag\\
&\left.(k,L_{KT_1},L_{KT_2},L_{KT_3},L_{KT_4})^{-1}\right]\notag\\
=&\, \frac{n-1}{2n^2}\partial_b\GM^{(5)}_n
\end{align}
In general, we have 
\begin{align}\label{eq:GMI3-key-q5}
\boxed{
\left. \GM^{(5)}_n \right|_{b}=\left. \GM^{(5)}_n\right|_{b=0}+b \, \partial_b\GM^{(5)}_n=\left(\frac{n-1}{2n^2}+b\right)\partial_b\GM^{(5)}_n \,.
}
\end{align}
For key-ring states, \eqref{eq:GMI3-key-q5} holds for arbitrary $n$. 

Equations~(\ref{eq:GMI3-key-q4}) and~(\ref{eq:GMI3-key-q5}), for $q=4$ and $q=5$, respectively, hold for arbitrary $n$ for the key-ring link state and constitute the main results of this section.

\section{\texorpdfstring{$\mathtt{q}=4$}{q=4} genuine multi-entropy}
\label{sec:numerics}

\subsection{From key-ring states to generic four-component link states}
\label{subsec:primary-target}

The main goal of this section is to determine whether the collapse relation \eqref{eq:GMI3-key-q4}
\begin{equation}
\left.\GM^{(4)}_n\right|_{a=0}
=
\frac{n^2-3n+3}{3n^2}\,I_{3,n},
\label{eq:collapse-test}
\end{equation}
which holds exactly for key-ring states as shown in
Sec.~\ref{sec:keyring}, continues to hold for generic four-component
link states. In particular, we turn on arbitrary pairwise linking numbers
among the three components $T_1,T_2,T_3$,
\begin{equation}
L_{T_1T_2},\qquad L_{T_1T_3},\qquad L_{T_2T_3},
\end{equation}
and ask when the resulting state contains multipartite information that
is not captured by $I_{3,n}$ through \eqref{eq:collapse-test}.

A particularly natural question is whether the answer depends on the
arithmetic structure of the Chern--Simons level $k$. For prime $k$, the
greatest common divisor $(k,L)$ of a linking number
$L\in\{0,\ldots,k-1\}$ with $k$ can take only two values,
\begin{equation}
(k,L)=
\begin{cases}
k, & L=0,\\
1, & L\neq0.
\end{cases}
\end{equation}
Consequently, the bipartite entropy associated with a single linking
number is either $0$ or $\log k$, with no intermediate value. For
composite $k$, by contrast, additional possibilities arise. The
simplest example is $k=4$, for which
\begin{equation}
(4,2)=2,
\end{equation}
and hence a linking number $L=2$ produces the intermediate bipartite
entropy $S=\log2$. Algebraically, this reflects the fact that
$\mathbb Z_4$ is not a field and contains the zero divisor
\begin{equation}
2\times2\equiv0\pmod4.
\end{equation}

This makes $k=4$ a natural first test of whether composite-level
arithmetic modifies the collapse pattern. The graph-state/stabilizer
identification of Sec.~\ref{sec:graphstates} implies that the Abelian
link states considered here are stabilizer states, but by itself does
not establish that the $n=\mathtt q$ threshold pattern found previously
for qubit stabilizer states continues to hold for general $k$. In
particular, the $n=2$ collapse was first observed in
\cite{Iizuka:2025pqq}, and the $n=3$ collapse for qubit stabilizer
states was subsequently found in \cite{Akella:2026rbe}. It is therefore
important to test directly what survives once $k>2$.

As we will see below, however, the resulting structure is considerably
richer than this initial prime-versus-composite motivation might
suggest. In particular, the breakdown is not controlled simply by the
presence of zero divisors. Instead, its occurrence depends nontrivially
on both the R\'enyi index $n$ and the arithmetic structure of the level
$k$. We first study the representative levels $k=2,3,4$, which expose
the basic mechanisms, and then extend the numerical scan to
$2\leq k\leq24$.

Before turning to generic link states, we benchmark our numerical
implementation against the exact key-ring results of
Sec.~\ref{sec:keyring}. For key-ring configurations,
\begin{equation}
L_{T_1T_2}=L_{T_1T_3}=L_{T_2T_3}=0,
\end{equation}
the relevant partition functions and genuine multi-entropies are known
analytically for arbitrary $n$ and $k$. This provides a direct check of
the replica contractions, index assignments, and normalization used in
the numerical calculation.

For several choices of the key-ring linking numbers
$L_{KT_1},L_{KT_2},L_{KT_3}$, we evaluate the quadripartite partition
function
\begin{equation}
\mathcal Z_n^{(4)}(K:T_1:T_2:T_3)
\end{equation}
directly from the link-state wavefunction \eqref{eq:U1-link-state} and
compare it with the closed-form expression
\eqref{eq:Z^4_n key-ring summary}. We also construct
$\left.\GM^{(4)}_n\right|_{a=0}$ from the numerically evaluated
lower-partite partition functions and verify the exact key-ring relation
\begin{equation}
\left.\GM^{(4)}_n\right|_{a=0}
=
\frac{n^2-3n+3}{3n^2}\,I_{3,n}.
\end{equation}

We numerically evaluate $\left.\GM^{(4)}_n\right|_{a=0}$ and $I_{3,n}$
for $n=2,3,4,5$ and $k=2,3,\ldots,9$. As a benchmark, we choose the
key-ring configuration
\begin{equation}
L_{KT_1}=L_{KT_2}=L_{KT_3}=1,
\qquad
L_{T_1T_2}=L_{T_1T_3}=L_{T_2T_3}=0.
\end{equation}
For this configuration, Eq.~\eqref{I3nkeyringcase} gives
\begin{equation}
I_{3,n}=\log k,
\end{equation}
and hence the exact key-ring collapse relation implies
\begin{equation}
\left.\GM^{(4)}_n\right|_{a=0}
=
\frac{n^2-3n+3}{3n^2}\log k.
\end{equation}
Our numerical evaluation reproduces these results for all the values of
$n$ and $k$ listed above, providing a nontrivial validation of the
replica contractions, index assignments, and normalization used in the
numerical pipeline.

To quantify the breakdown of the key-ring collapse relation away from
the key-ring class, we define
\begin{equation}
\Delta_n(L)
\equiv
\left.\GM^{(4)}_n\right|_{a=0}
-
\frac{n^2-3n+3}{3n^2}\,I_{3,n},
\label{eq:Delta-definition}
\end{equation}
where $L$ collectively denotes the six independent pairwise linking
numbers of the four-component link. Thus, $\Delta_n(L)=0$ indicates
that the key-ring collapse relation continues to hold, whereas
$\Delta_n(L)\neq0$ signals its breakdown.

We proceed as follows. Sec.~\ref{subsec:k2-deviation} uses the
well-understood qubit case $k=2$ as a controlled baseline for
characterizing the onset of the breakdown. Sec.~\ref{subsec:k3-comparison}
then turns to the prime level $k=3$, where a qualitatively new violation
already appears at $n=3$. Sec.~\ref{subsec:k4-main} examines the
composite level $k=4$ and tests whether the breakdown is controlled by
the zero-divisor structure of $\mathbb Z_4$. Finally,
Sec.~\ref{subsec:numerics-summary} extends the analysis to the full scan
over $2\leq k\leq24$ and identifies the resulting dependence on the
R\'enyi index and the arithmetic structure of $k$.

\subsection{\texorpdfstring{$k=2$}{k=2}: collapse and breakdown at the threshold}
\label{subsec:k2-deviation}
For $k=2$, the link state is a qubit graph state, and the collapse of $\GM^{(4)}_n$ onto $I_{3,n}$ for $n<\mathtt q$ is precisely the $n=\mathtt q$ threshold pattern established numerically for general qubit stabilizer states -- the $n=2$ case in \cite{Iizuka:2025pqq} and the $n=3$ case in \cite{Akella:2026rbe} -- independently of the key-ring restriction. Numerically reproducing the collapse at $n=2,3$ for $k=2$ is therefore a useful cross-check of our pipeline, but not new physics. What Refs.~\cite{Akella:2026rbe} do not address is the
detailed behavior of the breakdown at $n=4$: this is the genuinely new
content at $k=2$, and it is here that we use the qubit case to build
intuition before turning to higher levels $k\geq3$.

Concretely, we first fix
\begin{equation}
L_{KT_1}=L_{KT_2}=L_{KT_3}=1,
\end{equation}
and consider the $2^3=8$ assignments of the remaining linking numbers
$\{L_{T_1T_2},L_{T_1T_3},L_{T_2T_3}\}$.
We organize these configurations by the number
$m=0,1,\ldots,3$ of nonzero linking numbers among
$\{L_{T_1T_2},L_{T_1T_3},L_{T_2T_3}\}$. For each $m$, we evaluate the deviation
\begin{equation}
\Delta(L)\equiv\left.\GM^{(4)}_n\right|_{n=4}-\frac{n^2-3n+3}{3n^2}\bigg|_{n=4} I_{3,4}
\end{equation}
across all configurations at that value of $m$, and examine both its typical magnitude and whether $\Delta(L)=0$ persists for some nonzero $m$ (i.e.\ whether the collapse is protected by some residual structure even away from the strict key-ring point) or whether $\Delta(L)\neq0$ immediately once $m\geq1$.

This provides a controlled, low-cost characterization of the breakdown
at the simplest value of $k$, and helps identify which qualitative
features already appear at $k=2$ and which emerge only for $k\geq3$,
including those associated with the richer arithmetic structure of
higher levels.

For $k=2$, the numerical results are summarized in Table~\ref{tab:k2-table}. Using the permutation symmetry among $T_1,T_2,T_3$, we can choose four inequivalent configurations $(L_{T_1T_2},L_{T_1T_3},L_{T_2T_3})=(0,0,0)$, $(1,0,0)$, $(1,1,0)$, and $(1,1,1)$, with multiplicities $1,3,3,1$, respectively. Here, the  $(0,0,0)$ corresponds to the strict key-ring point. As expected, the relation holds for all configurations at $n=2,3$. At $n=4$, however, it is violated for the two intermediate classes, while it remains exact for both $(0,0,0)$ and $(1,1,1)$. The same pattern persists at $n=5$, showing that the breakdown is not simply controlled by the number of nonzero triangle linking numbers.

\begin{table}[t]
\centering
\begin{tabular}{c c c c c c}
\toprule
$\left(L_{T_1T_2},L_{T_1T_3},L_{T_2T_3}\right)$ & multiplicity & $n=2$ & $n=3$ & $n=4$ & $n=5$ \\ 
\midrule
$(0,0,0)$ & $1$ & $0$ & $0$ & $0$ & $0$ \\ 
$(1,0,0)$ & $3$ & $0$ & $0$ & $\frac{1}{24}$ & $\frac{2}{25}$ \\ 
$(1,1,0)$ & $3$ & $0$ & $0$ & $\frac{1}{24}$ & $\frac{2}{25}$ \\ 
$(1,1,1)$ & $1$ & $0$ & $0$ & $0$ & $0$ \\ 
\bottomrule
\end{tabular}
\caption{
Numerical results for the $k=2$ key-ring case with additional linkings. The entries show $\Delta_n/\log 2$, with $\Delta_n\equiv \GM_n^{(4)}-\frac{n^2-3n+3}{3n^2}I_{3,n}$. The three linking numbers are those among $T_1,T_2,T_3$, after using their permutation symmetry to choose a canonical representative.
}
\label{tab:k2-table}
\end{table}

This complementary pattern -- exact collapse at both $m=0$ and $m=3$, violation at $m=1,2$ -- follows from the local complementation symmetry of qubit graph states. In the present setup $L_{KT_1}=L_{KT_2}=L_{KT_3}=1$, so the neighborhood of $K$ in the graph is exactly $\{T_1,T_2,T_3\}$. Local complementation at the vertex $K$ toggles all edges among its neighbors\footnote{More concretely, by writing out the wave function eq.~\eqref{Thewavefn} explicitly for $k=2$, $N=4$ with $L_{KT_1}=L_{KT_2}=L_{KT_3}=1$, one finds $U_K|\mathcal{L}(L_{T_1T_2},L_{T_1T_3},L_{T_2T_3})\rangle = |\mathcal{L}(1-L_{T_1T_2},1-L_{T_1T_3},1-L_{T_2T_3})\rangle$ with $U_K=R_X|_K\otimes S^\dagger|_{T_1}\otimes S^\dagger|_{T_2}\otimes S^\dagger|_{T_3}$, where $R_X=\frac12 \scalebox{0.8}{$\begin{pmatrix}1+i&1-i\\1-i&1+i\end{pmatrix}$}$ ($=\sqrt{X}$, with $R_X^2=X$) and $S^\dagger=\mathrm{diag}(1,-i)=\sqrt{Z}^{-1}$.}, {\it i.e.}, simultaneously flips $(L_{T_1T_2},L_{T_1T_3},L_{T_2T_3})\to(1-L_{T_1T_2},\,1-L_{T_1T_3},\,1-L_{T_2T_3})$, and is known to act as a local Clifford (hence local unitary) transformation on the graph state~\cite{Hein:2004zjp, Nest:2004khg}. Since $\mathrm{GM}^{(4)}_n$ and $I_{3,n}$ are both invariant under local unitaries, $\Delta_n$ is invariant under this map. This sends $(0,0,0)\leftrightarrow(1,1,1)$, and the $(1,0,0)$-type class to the $(1,1,0)$-type class, explaining not only why both extremes are protected but also why the two intermediate classes share \emph{identical} values of $\Delta_4$ and $\Delta_5$ in Table~\ref{tab:k2-table}, rather than merely both being nonzero. We emphasize that this argument is specific to $k=2$: whether an analogous protection mechanism exists for qudit ($k\geq3$) graph/stabilizer states is left open.

We next relax the above condition and consider all $2^6=64$ assignments of the six linking numbers. We identify configurations related by the $S_4$ subsystem permutations, local unit scalings, and global complex conjugation, which leave the multi-entropies and hence the genuine multi-entropy unchanged, as explained in Appendix~\ref{app:linking-equivalence}. Grouping configurations related by these transformations into equivalence classes, the 64 labelled configurations reduce to 11 inequivalent representatives.

For all 11 representatives, $\Delta_n=0$ at $n=2,3$. The first violations appear at $n=4$, as summarized in Table~\ref{tab:k2-all-linkings}. The same four inequivalent classes remain violating at $n=5$. Taking their multiplicities into account, $33$ of the $64$ labelled configurations violate the relation, corresponding to a violation rate of $51.56\%$.

This exact collapse at $n=2,3$, holding without exception across all
64 configurations, is consistent with the qubit-stabilizer results of
Refs.~\cite{Iizuka:2025pqq,Akella:2026rbe}; the genuinely new content
of the generic-link analysis first appears at $n=4$.

\begin{table}[t]
\centering
\begin{tabular}{c c c c c c}
\hline
$\left(L_{KT_1},L_{KT_2},L_{KT_3},L_{T_1T_2},L_{T_1T_3},L_{T_2T_3}\right)$
& multiplicity & $n=2$ & $n=3$ & $n=4$ & $n=5$ \\
\hline
$(0,0,0,0,0,0)$ & $1$  & $0$ & $0$ & $0$ & $0$ \\
$(1,0,0,0,0,0)$ & $6$  & $0$ & $0$ & $0$ & $0$ \\
$(1,1,0,0,0,0)$ & $12$ & $0$ & $0$ & $0$ & $0$ \\
$(1,0,0,0,0,1)$ & $3$  & $0$ & $0$ & $0$ & $0$ \\
$(1,1,1,0,0,0)$ & $4$  & $0$ & $0$ & $0$ & $0$ \\
$(1,1,0,1,0,0)$ & $4$  & $0$ & $0$ & $0$ & $0$ \\
$(1,1,0,0,1,0)$ & $12$ & $0$ & $0$ & $\frac{1}{24}$ & $\frac{2}{25}$ \\
$(1,1,1,1,0,0)$ & $12$ & $0$ & $0$ & $\frac{1}{24}$ & $\frac{2}{25}$ \\
$(1,1,0,0,1,1)$ & $3$  & $0$ & $0$ & $\frac{1}{24}$ & $\frac{2}{25}$ \\
$(1,1,1,1,1,0)$ & $6$  & $0$ & $0$ & $\frac{1}{24}$ & $\frac{2}{25}$ \\
$(1,1,1,1,1,1)$ & $1$  & $0$ & $0$ & $0$ & $0$ \\
\hline
\multicolumn{2}{c}{Violation rate}
& $0\%$
& $0\%$
& $51.56\%$
& $51.56\%$ \\
\hline
\end{tabular}
\caption{
Numerical results for all inequivalent $k=2$ linking-number representatives.
The entries show $\Delta_n/\log 2$ for $n=2,3,4,5$.
For each equivalence class, we choose the representative with entries equal to $1$ placed as far to the left as possible in the ordering shown above.
The multiplicities sum to $2^6=64$.
The relation holds for all configurations at $n=2,3$, while violations first appear at $n=4$.
}
\label{tab:k2-all-linkings}
\end{table}

For each of the four violating classes, the $n=4$ values are
\begin{equation}
\GM_4^{(4)}=-\frac{5}{48}\log 2,\qquad
I_{3,4}=-\log 2,\qquad
\Delta_4=\frac{1}{24}\log 2,
\end{equation}
while at $n=5$ they become
\begin{equation}
\GM_5^{(4)}=-\frac{7}{75}\log 2,\qquad
I_{3,5}=-\log 2,\qquad
\Delta_5=\frac{2}{25}\log 2.
\end{equation}
Thus, for $k=2$, the breakdown begins at the threshold $n=4$ and is not restricted to the key-ring case with additional linkings.

\subsection{\texorpdfstring{$k=3$}{k=3}: breakdown below the qubit threshold}
\label{subsec:k3-comparison}
For prime $k$, the greatest common divisor $(k,L)$ of a linking number
$L\in\{0,1,\ldots,k-1\}$ with $k$ can only take two values,
\begin{equation}
(k,L)=
\begin{cases}
k, & L=0,\\
1, & L\neq0,
\end{cases}
\end{equation}
so that every bipartite entanglement entropy
$S=\log\big(k/(k,L)\big)$ is either $0$ (unentangled) or $\log k$
(maximally entangled); no intermediate value occurs. Since $k=3$ is
again prime, $\mathbb Z_3$ is a field, and the
stabilizer underlying
Sec.~\ref{sec:graphstates} extends naturally from $k=2$. One might
therefore expect the qualitative pattern found for qubit stabilizer
states in \cite{Akella:2026rbe} to persist at $k=3$, although no general proof of the
$n=\mathtt q$ threshold is known for qudit stabilizer states.

We test this expectation by repeating the scan of
Sec.~\ref{subsec:k2-deviation} for all $3^6=729$ configurations of the
linking matrix. The result reveals the first qualitative surprise:
while the collapse remains exact for every configuration at $n=2$,
it already fails for some configurations at $n=3$. Thus the qubit
threshold pattern does not extend even to the next prime-dimensional
case. This shows that neither primality of $k$ nor the stabilizer
structure alone is sufficient to protect the collapse up to
$n=\mathtt q-1$.

After quotienting by the $S_4$ subsystem permutations, local unit scalings, and global complex conjugation, which leave the multi-entropies and the genuine multi-entropy invariant as explained in Appendix~\ref{app:linking-equivalence}, the $3^6=729$ labelled configurations reduce to 14 inequivalent representatives.
For a simple visual illustration of some of these equivalence classes, see
Fig.~\ref{fig:linking-equivalence-examples}.

The numerical results for these representatives are summarized in Table~\ref{tab:k3-all-linkings}. 
Here we again use
\begin{equation}
\Delta_n\equiv \GM_n^{(4)}-\frac{n^2-3n+3}{3n^2}I_{3,n}.
\end{equation}
The relation holds for all $729$ configurations at $n=2$, but already fails for $120$ configurations at $n=3$, in contrast to the qubit case $k=2$. At $n=4$ and $n=5$, the numbers of configurations satisfying the relation are $153$ and $273$, respectively. Thus, the $k=3$ results show that the breakdown can occur already at $n=3$.

\begin{table}[t]
\centering
\begin{tabular}{c c c c c c}
\toprule
$\left(L_{KT_1},L_{KT_2},L_{KT_3},L_{T_1T_2},L_{T_1T_3},L_{T_2T_3}\right)$ & multiplicity & $n=2$ & $n=3$ & $n=4$ & $n=5$ \\
\midrule
$(0,0,0,0,0,0)$ & $1$   & $0$ & $0$ & $0$ & $0$ \\
$(1,0,0,0,0,0)$ & $12$  & $0$ & $0$ & $0$ & $0$ \\
$(1,1,0,0,0,0)$ & $48$  & $0$ & $0$ & $0$ & $0$ \\
$(1,0,0,0,0,1)$ & $12$  & $0$ & $0$ & $0$ & $0$ \\
$(1,1,0,0,1,0)$ & $96$  & $0$ & $0$ & $\frac{1}{24}$ & $\frac{2}{25}$ \\
$(1,1,1,0,0,0)$ & $32$  & $0$ & $0$ & $0$ & $0$ \\
$(1,1,0,1,0,0)$ & $32$  & $0$ & $0$ & $0$ & $0$ \\
$(1,1,1,1,0,0)$ & $192$ & $0$ & $0$ & $\frac{1}{24}$ & $\frac{2}{25}$ \\
$(1,1,0,0,1,1)$ & $24$  & $0$ & $0$ & $\frac{1}{24}$ & $\frac{2}{25}$ \\
$(1,1,0,0,1,2)$ & $24$  & $0$ & $-\frac{2}{9}$ & $-\frac{1}{8}$ & $0$ \\
$(1,1,1,1,1,0)$ & $96$  & $0$ & $0$ & $\frac{1}{24}$ & $\frac{2}{25}$ \\
$(1,1,1,1,0,2)$ & $96$  & $0$ & $-\frac{2}{9}$ & $-\frac{1}{8}$ & $0$ \\
$(1,1,1,1,1,1)$ & $16$  & $0$ & $0$ & $0$ & $0$ \\
$(1,1,1,1,1,2)$ & $48$  & $0$ & $0$ & $\frac{1}{24}$ & $\frac{2}{25}$ \\
\midrule
\multicolumn{2}{c}{Violation rate}
& $0\%$
& $16.46\%$
& $79.01\%$
& $62.55\%$ \\
\bottomrule
\end{tabular}
\caption{
Numerical results for all inequivalent linking-number representatives at $k=3$. The entries show $\Delta_n/\log 3$; hence a zero denotes an exact collapse onto $I_{3,n}$.
Within each equivalence class, we choose a representative in which entries equal to $1$ are placed as far to the left as allowed by the symmetry transformations. The multiplicity gives the number of labelled configurations represented by each row, and the multiplicities sum to $3^6=729$.
The last row gives the fraction of all labelled configurations for which the relation is violated.
}
\label{tab:k3-all-linkings}
\end{table}

The likely reason is that the structural argument of Sec.~\ref{subsec:primary-target} -- the absence of intermediate values $(k,L)\notin\{1,k\}$ for prime $k$ -- constrains only \emph{bipartite} entanglement. It gives no reason to expect the collapse of the genuinely \emph{multipartite} quantities $\mathrm{GM}^{(4)}_n$ and $I_{3,n}$ to survive at $n=3$, so the primality of $k$ alone does not guarantee protection there.

\subsection{\texorpdfstring{$k=4$}{k=4}: probing the role of zero divisors}
\label{subsec:k4-main}

We next turn to $k=4$, the simplest composite level with a nontrivial
zero divisor. This provides a natural setting in which to test whether
the breakdown of the collapse relation is controlled by arithmetic
features specific to composite $\mathbb Z_k$. We perform a full scan over
all $4^6=4096$ configurations of the linking matrix
$\{L_{KT_1},L_{KT_2},L_{KT_3},L_{T_1T_2},L_{T_1T_3},L_{T_2T_3}\}$,
for $n=2,3,4,5$. In addition to the number of nonzero linking numbers,
we track, for each configuration, the number
$p\in\{0,\ldots,6\}$ of links taking the zero-divisor value
$L=2\pmod4$, as opposed to the invertible values $L=1,3$.

The main questions are:
\begin{itemize}
\item[(i)] Does the collapse
\begin{equation}
\left.\GM^{(4)}_n\right|_{a=0}
=
\frac{n^2-3n+3}{3n^2}I_{3,n}
\end{equation}
continue to hold at $n=2,3$, as it does for $k=2$, or does the
composite nature of $k=4$ already induce an earlier breakdown?

\item[(ii)] Once the collapse breaks down, is the violation correlated
with the presence or number of zero-divisor linking numbers
$L=2\pmod4$?

\item[(iii)] How do the magnitude and configuration dependence of the
deviation $\Delta_n$ compare with the $k=2$ case?
\end{itemize}

We first answer question (i).
After quotienting by the $S_4$ subsystem permutations, local unit scalings, and global complex conjugation, which leave the multi-entropies and the genuine multi-entropy invariant as explained in Appendix~\ref{app:linking-equivalence}, the $4^6=4096$ labelled configurations reduce to 72 inequivalent representatives. The numerical results are summarized in Table~\ref{tab:k4-summary}, while the results for all inequivalent representatives are listed in Table~\ref{tab:k4-all-linkings-full} in Appendix~\ref{app:Numerical-table-k=4}. Thus, we conclude that the
collapse relation holds for all configurations at $n=2$ and $n=3$. At both $n=4$ and $n=5$, it holds for 25 of the 72 inequivalent representatives, corresponding to $583$ of the $4096$ labelled configurations, while the remaining $3513$ configurations violate the relation.

\begin{table}[ht]
\centering
\begin{tabular}{c c c c}
\toprule
$n$ & $\Delta_n=0$ representatives & $\Delta_n=0$ linking-number configurations & violation rate \\
\midrule
$2$ & $72/72$ & $4096/4096$ & $0\%$ \\
$3$ & $72/72$ & $4096/4096$ & $0\%$ \\
$4$ & $25/72$ & $583/4096$ & $85.77\%$ \\
$5$ & $25/72$ & $583/4096$ & $85.77\%$ \\
\bottomrule
\end{tabular}
\caption{
Summary of the numerical evaluation over all $4^6=4096$ linking-number configurations at $k=4$. The second column counts inequivalent representatives with $\Delta_n=0$, while the third column restores their multiplicities and counts the corresponding linking-number configurations.
}
\label{tab:k4-summary}
\end{table}

We now turn to question (ii). For each configuration, we count the
number $p\in\{0,\ldots,6\}$ of linking numbers equal to the
zero-divisor value $L=2\pmod4$, and examine whether the $n=4$
violation rate is correlated with $p$. The result is shown in
Table~\ref{tab:k4-zerodivisor-correlation}.

\begin{table}[t]
\centering
\begin{tabular}{c c c c}
\toprule
$p$ & total configurations & violating configurations & violation rate \\
\midrule
$0$ & $729$  & $576$  & $79.01\%$ \\
$1$ & $1458$ & $1296$ & $88.89\%$ \\
$2$ & $1215$ & $1080$ & $88.89\%$ \\
$3$ & $540$  & $420$  & $77.78\%$ \\
$4$ & $135$  & $123$  & $91.11\%$ \\
$5$ & $18$   & $18$   & $100\%$ \\
$6$ & $1$    & $0$    & $0\%$ \\
\bottomrule
\end{tabular}
\caption{Correlation between the number $p$ of zero-divisor linking numbers ($L=2\bmod4$) and the $n=4$ violation rate, weighted by multiplicity. The total number of violating configurations, $576+1296+1080+420+123+18+0=3513$, agrees with Table~\ref{tab:k4-summary}.}
\label{tab:k4-zerodivisor-correlation}
\end{table}

The data in Table~\ref{tab:k4-zerodivisor-correlation} show that
zero-divisor linking numbers are \emph{not} responsible for the breakdown in
any simple way. Even for $p=0$, where no linking number takes the
zero-divisor value $L=2$, $79.01\%$ of the configurations violate the
collapse relation. Moreover, the violation rate is not monotonic in
$p$. In particular, the unique configuration with $p=6$,
$(2,2,2,2,2,2)$, satisfies the collapse exactly. Therefore, neither the
presence nor the number of zero-divisor linking numbers controls the
breakdown. The relevant structure must instead depend on the full
multipartite linking pattern.

Turning to question (iii), whereas the $k=2$ analysis of Sec.~\ref{subsec:k2-deviation} found a single deviation value $\Delta_4=\frac{1}{24}\log2$ across all violating configurations, the $k=4$ data in Table~\ref{tab:k4-all-linkings-full} exhibit four distinct nonzero values, $\Delta_4/\log4\in\left\{\frac{1}{48},\frac{1}{24},-\frac{1}{24},-\frac{3}{32}\right\}$, i.e.\ $\Delta_4/\log2\in\left\{\frac{1}{24},\frac{1}{12},-\frac{1}{12},-\frac{3}{16}\right\}$ once expressed in the same $\log2$ units as the $k=2$ case. Two of these exceed the magnitude of the sole $k=2$ value, most notably $-\frac{3}{16}$, which is more than seven times larger. Thus the transition from $k=2$ to the composite level $k=4$ not only increases the fraction of violating configurations but also diversifies and, in some cases, amplifies the size of the breakdown itself -- a richer structure than a single zero-divisor count $p$ alone can capture.

\subsection{\texorpdfstring{Full numerical scan over $2\leq k\leq24$}{Full numerical scan over 2 <= k <= 24}}
\label{subsec:numerics-summary}

Table~\ref{tab:numerical-violation-summary} summarizes the full numerical scans for $2\leq k\leq 24$ and $2\leq n\leq 5$, showing the fraction of the $k^6$ labelled linking-number configurations for which $\Delta_n\neq 0$. The relation holds for all scanned values of $k$ at $n=2$. At $n=3$, violations occur only for $k$ divisible by $3$. Within the present scan, the violation rate is $16.46\%$ for $k=3,6,12,15,21,24$ and $62.74\%$ for $k=9,18$. At both $n=4$ and $n=5$, violations occur for every scanned value of $k$, with rates that vary substantially with $k$.

\begin{table}[t]
\centering
\begin{tabular}{c c c c c}
\hline
$k$ & $n=2$ & $n=3$ & $n=4$ & $n=5$ \\
\hline
$2$  & $0\%$ & $0\%$     & $51.56\%$ & $51.56\%$ \\
$3$  & $0\%$ & $16.46\%$ & $79.01\%$ & $62.55\%$ \\
$4$  & $0\%$ & $0\%$     & $85.77\%$ & $85.77\%$ \\
$5$  & $0\%$ & $0\%$     & $93.39\%$ & $93.39\%$ \\
$6$  & $0\%$ & $16.46\%$ & $89.83\%$ & $81.86\%$ \\
$7$  & $0\%$ & $0\%$     & $73.81\%$ & $39.11\%$ \\
$8$  & $0\%$ & $0\%$     & $96.22\%$ & $96.22\%$ \\
$9$  & $0\%$ & $62.74\%$ & $97.44\%$ & $77.05\%$ \\
$10$ & $0\%$ & $0\%$     & $96.80\%$ & $96.80\%$ \\
$11$ & $0\%$ & $0\%$     & $50.46\%$ & $74.68\%$ \\
$12$ & $0\%$ & $16.46\%$ & $97.01\%$ & $94.67\%$ \\
$13$ & $0\%$ & $0\%$     & $43.39\%$ & $22.45\%$ \\
$14$ & $0\%$ & $0\%$     & $87.31\%$ & $70.50\%$ \\
$15$ & $0\%$ & $16.46\%$ & $98.61\%$ & $97.52\%$ \\
$16$ & $0\%$ & $0\%$     & $98.25\%$ & $99.02\%$ \\
$17$ & $0\%$ & $0\%$     & $33.80\%$ & $17.36\%$ \\
$18$ & $0\%$ & $62.74\%$ & $98.76\%$ & $88.88\%$ \\
$19$ & $0\%$ & $0\%$     & $30.42\%$ & $45.26\%$ \\
$20$ & $0\%$ & $0\%$     & $99.06\%$ & $99.06\%$ \\
$21$ & $0\%$ & $16.46\%$ & $94.50\%$ & $77.20\%$ \\
$22$ & $0\%$ & $0\%$     & $76.01\%$ & $87.74\%$ \\
$23$ & $0\%$ & $0\%$     & $25.33\%$ & $12.93\%$ \\
$24$ & $0\%$ & $16.46\%$ & $99.21\%$ & $98.58\%$ \\
\hline
\end{tabular}
\caption{
Violation rates, defined as the fractions of the $k^6$ labelled linking-number configurations with $\Delta_n\neq 0$, for the full numerical scans with $2\leq k\leq 24$ and $2\leq n\leq 5$.
}
\label{tab:numerical-violation-summary}
\end{table}

To illustrate the characteristic $n=3$ behavior more explicitly, we focus on the violating representatives for $k\equiv 0 \pmod{3}$ within the scanned range. Among the violating representatives, the maximal number of vanishing linking numbers is two for every such $k$. We therefore focus on this minimal-support sector. A common family of violating representatives can be chosen as
\begin{equation}
\left(L_{KT_1},L_{KT_2},L_{KT_3},L_{T_1T_2},L_{T_1T_3},L_{T_2T_3}\right)
=
\left(\frac{k}{3},\frac{k}{3},0,0,\frac{k}{3},\frac{2k}{3}\right),
\qquad k\equiv 0 \pmod{3}.
\end{equation}
Each representative contains exactly two vanishing linking numbers, while the four nonzero links form the same four-cycle support. Table~\ref{tab:n3-multiple-three-example} lists one such representative for each scanned value of $k$ divisible by $3$. These are illustrative examples and do not exhaust all violating configurations.

\begin{table}[t]
\centering
\begin{tabular}{c c c c c}
\hline
$k$ & one minimal-support violating representative & $\GM_3^{(4)}$ & $I_{3,3}$ & $\Delta_3$ \\
\hline
$3$  & $(1,1,0,0,1,2)$   & $-\frac{4}{9}\log 3$ & $-2\log 3$ & $-\frac{2}{9}\log 3$ \\
$6$  & $(2,2,0,0,2,4)$   & $-\frac{4}{9}\log 3$ & $-2\log 3$ & $-\frac{2}{9}\log 3$ \\
$9$  & $(3,3,0,0,3,6)$   & $-\frac{4}{9}\log 3$ & $-2\log 3$ & $-\frac{2}{9}\log 3$ \\
$12$ & $(4,4,0,0,4,8)$   & $-\frac{4}{9}\log 3$ & $-2\log 3$ & $-\frac{2}{9}\log 3$ \\
$15$ & $(5,5,0,0,5,10)$  & $-\frac{4}{9}\log 3$ & $-2\log 3$ & $-\frac{2}{9}\log 3$ \\
$18$ & $(6,6,0,0,6,12)$  & $-\frac{4}{9}\log 3$ & $-2\log 3$ & $-\frac{2}{9}\log 3$ \\
$21$ & $(7,7,0,0,7,14)$  & $-\frac{4}{9}\log 3$ & $-2\log 3$ & $-\frac{2}{9}\log 3$ \\
$24$ & $(8,8,0,0,8,16)$  & $-\frac{4}{9}\log 3$ & $-2\log 3$ & $-\frac{2}{9}\log 3$ \\
\hline
\end{tabular}
\caption{
Illustrative minimal-support violations at $n=3$ for the scanned values of $k$ divisible by $3$. The second column gives representatives in the ordering $\left(L_{KT_1},L_{KT_2},L_{KT_3},L_{T_1T_2},L_{T_1T_3},L_{T_2T_3}\right)$. Each example has exactly two vanishing linking numbers and the same four-cycle support. These examples do not enumerate all violating configurations.
}
\label{tab:n3-multiple-three-example}
\end{table}

The identical values along the family in
Table~\ref{tab:n3-multiple-three-example} are not accidental.
Let $k=3m$ and consider linking numbers of the form
\begin{equation}
L_{ij}=m\,\ell_{ij},
\qquad
\ell_{ij}\in\mathbb Z_3.
\end{equation}
The phase of the link state then becomes
\begin{equation}
\exp\left[
\frac{2\pi i}{k}\sum_{i<j}L_{ij}\alpha_i\alpha_j
\right]
=
\exp\left[
\frac{2\pi i}{3}\sum_{i<j}\ell_{ij}\alpha_i\alpha_j
\right],
\end{equation}
which depends only on $\alpha_i\bmod 3$. Writing
\begin{equation}
\alpha_i=r_i+3s_i,
\qquad
r_i\in\mathbb Z_3,
\qquad
s_i=0,\ldots,m-1,
\end{equation}
and performing the corresponding local basis relabeling, the link state
factorizes as
\begin{equation}
\ket{\mathcal L}_{k=3m,\,L=m\ell}
\simeq
\ket{\mathcal L}_{k=3,\,\ell}
\otimes
\ket{+}_m^{\otimes 4},
\qquad
\ket{+}_m
=
\frac{1}{\sqrt m}\sum_{s=0}^{m-1}\ket{s},
\label{eq:k3-embedding}
\end{equation}
where $\simeq$ denotes equivalence under local basis transformations.
The additional $m$-dimensional factors are completely unentangled and
therefore do not contribute to the multi-entropies. Consequently,
\begin{equation}
\GM_n^{(4)}(3m,m\ell)
=
\GM_n^{(4)}(3,\ell),
\qquad
I_{3,n}(3m,m\ell)
=
I_{3,n}(3,\ell),
\end{equation}
and hence $\Delta_n(3m,m\ell)=\Delta_n(3,\ell)$.
This explains analytically why the family displayed in
Table~\ref{tab:n3-multiple-three-example} has the same values for
all scanned levels divisible by $3$.

A further striking pattern emerges from the full scan: within the scanned range $2\leq k\leq24$, the $n=3$ violation rate depends only on the 3-adic valuation $v_3(k)$, defined as the number of times $3$ appears as a factor in $k$ (equivalently, the largest integer $m$ such that $3^m$ divides $k$): it vanishes for $v_3(k)=0$, equals $16.46\%$ for $v_3(k)=1$ ($k=3,6,12,15,21,24$), and rises to $62.74\%$ for $v_3(k)=2$ ($k=9,18$).

\section{Discussion}
\label{sec:discussion}

In this paper, we first restricted ourselves to the special class of key-ring link states and showed that, for the cases $\mathtt q=4,5$, the $\mathtt q$-partite genuine multi-entropy collapses onto lower-partite quantities involving at most $(\mathtt q-2)$ parties. For $\mathtt q=4$, the genuine multi-entropy $\GM^{(4)}_n$ reduces exactly to the tripartite information $I_{3,n}$, while for
$\mathtt q=5$ the two genuine diagnostics are themselves proportional, so that $\GM^{(5)}_n$ is completely determined by $\partial_b\GM^{(5)}_n$, a combination of lower-partite 
multi-entropies. Remarkably, these collapse relations hold for arbitrary R\'enyi index $n$ and Chern--Simons level $k$ within the key-ring class. This is a consequence of the restricted structure of key-ring states, in which only one distinguished component links to the others; it holds for every $n$, with no threshold in sight.

Once we move beyond the key-ring class and turn on generic pairwise linking numbers among all $\mathtt q=4$ components, this exact, all-$n$ collapse no longer survives. Our numerical scan over Chern-Simons levels $2\leq k\leq24$ reveals a three-tier structure:
\begin{itemize}
\item $n=2$: \text{the collapse holds for every level examined}
\item $n=3$: \text{within the scanned range, the collapse holds if and only if } $3\nmid k$
\item $n=4, 5$: \text{the collapse is violated for every level examined}.
\end{itemize}
This refines, rather than simply invalidates, the qubit-stabilizer results of \cite{Iizuka:2025pqq,Akella:2026rbe}: those results are recovered as the special case $k=2$, but the pattern that replaces a single $n=\mathtt q$ threshold for general $k$ is considerably richer, and is controlled by the arithmetic of $k$ itself.

The mechanism behind the collapse is qualitatively different for $n=2$ versus $n>2$. The $n=2$ collapse, which holds without exception across the entire range $2\leq k\leq24$ and, at $k=2$, across all $2^6=64$ labelled configurations, rests on solid analytic footing. This case was established directly for $\mathtt q=4$ qubit stabilizer states in Ref.~\cite{Iizuka:2025pqq}; more generally, Ref.~\cite{Akella:2026xza} shows via an explicit counting argument that the $n=2$ Coxeter multi-invariant for tripartite stabilizer states reduces to a product of bipartite entanglement entropies, and conjectures that the same structure extends to general $\mathtt q$, consistent with the collapse found here. The $n=3$ and $n=4$ breakdowns, by contrast, resist a single unifying mechanism. The composite level $k=4$ was originally singled out because $\mathbb Z_4$ contains a zero divisor, yet the $n=3$ violation already appears at the \emph{prime} level $k=3$ (Sec.~\ref{subsec:k3-comparison}), and at $k=4$ itself the $n=4$ violation rate depends on the zero-divisor count $p$ only mildly and non-monotonically, apart from the protected point $p=6$, which we conjecture is related to $L=2$ being self-inverse under $L\to-L\bmod4$ (Sec.~\ref{subsec:k4-main}). The zero-divisor structure of $\mathbb Z_k$ is therefore not the operative mechanism at either $n=3$ or $n=4$; the obstruction instead reflects a genuinely multipartite structure not captured by bipartite arithmetic alone.

Several questions remain open. First, the $n=4,5$ violation rate as a function of $k$ in Table~\ref{tab:numerical-violation-summary} shows no analogue of the clean $v_3(k)$ dependence found at $n=3$: it varies non-monotonically with $k$, and large primes such as $k=13,17,19,23$ show comparatively low violation rates relative to neighboring composite values. Identifying the number-theoretic or representation-theoretic origin of this dependence, and its relation (if any) to the zero-divisor mechanism, is left for future work. Second, the exact cancellation we observe at $k=2$ between the $(0,0,0)$ and $(1,1,1)$ classes, and between the two intermediate classes, is explained by local complementation, a local Clifford symmetry of qubit graph states~\cite{Hein:2004zjp,Nest:2004khg}; whether an analogous symmetry protects special configurations for qudit ($k\geq3$) link states -- potentially including the $p=6$ protection noted above -- is not known.  Third, throughout this work we have focused on Abelian
$U(1)_k$ Chern--Simons theory, for which the link-state wave function
is sufficiently explicit to allow both analytic and systematic
numerical treatment. It would be very interesting to extend the present
analysis to non-Abelian Chern--Simons theories. While genuine
multi-entropy for non-Abelian link states has been investigated
numerically in Ref.~\cite{Yuan:2025dgx}, it remains an open question
whether analytic collapse relations, or analytic criteria for their
breakdown, can be established in the non-Abelian setting.

More broadly, these results suggest that the ``$n=\mathtt q$ threshold'' intuition built from qubit stabilizer states -- and used as an organizing principle in earlier genuine-multi-entropy studies such as the toric-code analysis of~\cite{Akella:2026rbe} -- should not be assumed to carry over unmodified to general qudit or link-state settings. Instead, the collapse of genuine multi-entropy onto lower-partite quantities, and its breakdown, appears to be governed by an interplay between the R\'enyi index $n$ and the arithmetic of the underlying level $k$, of which the $n=3$, $3\mid k$ obstruction identified here is the cleanest example.

\acknowledgments
The work of N.I. was supported in part 
by  NSTC of Taiwan Grant Number 114-2112-M-007-025-MY3, 
and 
by MEXT KAKENHI Grant-in-Aid for Transformative Research Areas A “Extreme Universe” No. 21H05184. 
The work of A.M. was supported by JSPS KAKENHI Grant Number JP26KJ0186. The authors used generative AI tools for limited language polishing.


\appendix

\section{Detailed expressions for $\GM[5]_n(A:B:C:D:E)$}
\label{detailGM5}

The linear combinations of $S_n^{(\mathtt{q})}$ in \eqref{eq:GM5-definition} are defined by
\begin{align}
S_n^{(5)}[1:1:1:1:1]
:=&\;S_n^{(5)}(A:B:C:D:E),\\
S_n^{(4)}[2:1:1:1]:=&\;S_n^{(4)}(AB:C:D:E)+S_n^{(4)}(AC:B:D:E)+S_n^{(4)}(AD:B:C:E)\notag\\
+&\;S_n^{(4)}(AE:B:C:D)
+S_n^{(4)}(BC:A:D:E)+S_n^{(4)}(BD:A:C:E)\notag\\
+&\;S_n^{(4)}(BE:A:C:D)+S_n^{(4)}(CD:A:B:E)+S_n^{(4)}(CE:A:B:D)\notag\\
+&\;S_n^{(4)}(DE:A:B:C),\\
S_n^{(3)}[2:2:1]:=&\;S_n^{(3)}(AB:CD:E)+S_n^{(3)}(AC:BD:E)+S_n^{(3)}(AD:BC:E)\notag\\
+&\;S_n^{(3)}(AB:CE:D)+S_n^{(3)}(AC:BE:D)+S_n^{(3)}(AE:BC:D)\notag\\
+&\;S_n^{(3)}(AB:DE:C)+S_n^{(3)}(AD:BE:C)+S_n^{(3)}(AE:BD:C)\notag\\
+&\;S_n^{(3)}(AC:DE:B)+S_n^{(3)}(AD:CE:B)+S_n^{(3)}(AE:CD:B)\notag\\
+&\;S_n^{(3)}(BC:DE:A)+S_n^{(3)}(BD:CE:A)+S_n^{(3)}(BE:CD:A)\\
S_n^{(3)}[3:1:1]:=&\;S_n^{(3)}(ABC:D:E)+S_n^{(3)}(ABD:C:E)+S_n^{(3)}(ABE:C:D)\notag\\
+&\;S_n^{(3)}(ACD:B:E)+S_n^{(3)}(ACE:B:D)+S_n^{(3)}(ADE:B:C)\notag\\
+&\;S_n^{(3)}(BCD:A:E)+S_n^{(3)}(BCE:A:D)+S_n^{(3)}(BDE:A:C)\notag\\
+&\;S_n^{(3)}(CDE:A:B),\\
S_n^{(2)}[3:2]:=&\;S_n^{(2)}(ABC:DE)+S_n^{(2)}(ABD:CE)+S_n^{(2)}(ABE:CD)\notag\\
+&\;S_n^{(2)}(ACD:BE)+S_n^{(2)}(ACE:BD)+S_n^{(2)}(ADE:BC)\notag\\
+&\;S_n^{(2)}(BCD:AE)+S_n^{(2)}(BCE:AD)+S_n^{(2)}(BDE:AC)\notag\\
+&\;S_n^{(2)}(CDE:AB),\\
S_n^{(2)}[4:1]:=&\;S_n^{(2)}(ABCD:E)+S_n^{(2)}(ABCE:D)+S_n^{(2)}(ABDE:C)\notag\\
+&\;S_n^{(2)}(ACDE:B)+S_n^{(2)}(BCDE:A).
\end{align}

\section{Solution counting for a single \texorpdfstring{$\eta$}{eta}-constraint}
\label{app:one_eta_counting}

In this appendix, we show that
\begin{equation}
\sum_{\alpha_1,\ldots,\alpha_D=0}^{k-1}
\eta\left[
\sum_{i=1}^{D}\alpha_iL_i
\right]
=
k^{D-1}(k,L_1,\ldots,L_D),
\label{eq:single-eta-counting}
\end{equation}
where $(k,L_1,\ldots,L_D)$ denotes the greatest common divisor of its arguments.

Consider the sum on the left-hand side of
\eqref{eq:single-eta-counting}. Since the $\eta$-constraint function
takes the value one when its argument vanishes modulo $k$, and zero
otherwise, this sum counts the number of assignments
\begin{equation}
(\alpha_1,\ldots,\alpha_D)\in(\mathbb Z_k)^D
\end{equation}
that satisfy
\begin{equation}
\sum_{i=1}^{D}\alpha_iL_i
\equiv0
\pmod{k}.
\label{eq:eta-congruence}
\end{equation}

To count these assignments, we define the map
\begin{equation}
f:(\mathbb Z_k)^D\longrightarrow\mathbb Z_k
\end{equation}
by
\begin{equation}
f(\alpha_1,\ldots,\alpha_D)
=
\sum_{i=1}^{D}\alpha_iL_i
\pmod{k}.
\label{eq:f-map}
\end{equation}
The solutions of \eqref{eq:eta-congruence} are precisely the elements
of the kernel of $f$. Therefore,
\begin{equation}
\sum_{\alpha_1,\ldots,\alpha_D=0}^{k-1}
\eta\left[
\sum_{i=1}^{D}\alpha_iL_i
\right]
=
|\ker f|.
\end{equation}

Since the domain contains $k^D$ elements, we have
\begin{equation}
|\ker f|
=
\frac{k^D}{|\operatorname{Im}f|}.
\label{eq:kernel-image-relation}
\end{equation}
Thus, it remains to determine the number of elements in the image of
$f$.

Let
\begin{equation}
g=(k,L_1,\ldots,L_D).
\end{equation}
Since $g$ divides every $L_i$, we can write
\begin{equation}
L_i=g\,l_i
\end{equation}
for some integers $l_i$. It then follows from \eqref{eq:f-map} that
\begin{equation}
f(\alpha_1,\ldots,\alpha_D)
=
g\sum_{i=1}^{D}\alpha_i l_i
\pmod{k}.
\end{equation}
Therefore, every element of the image of $f$ is a multiple of $g$
modulo $k$. Hence,
\begin{equation}
\operatorname{Im}f
\subseteq
\left\{
0,g,2g,\ldots,
\left(\frac{k}{g}-1\right)g
\right\}.
\label{eq:image-subset}
\end{equation}

We next show that every multiple of $g$ appearing on the right-hand side
of \eqref{eq:image-subset} actually belongs to the image.
By B\'ezout's identity, there exist integers
$\lambda_0,\lambda_1,\ldots,\lambda_D$ such that
\begin{equation}
\lambda_0 k
+
\sum_{i=1}^{D}\lambda_iL_i
=
g.
\end{equation}
Reducing this relation modulo $k$, we obtain
\begin{equation}
\sum_{i=1}^{D}\lambda_iL_i
\equiv g
\pmod{k}.
\end{equation}
Thus, by choosing
\begin{equation}
\alpha_i\equiv\lambda_i\pmod{k},
\end{equation}
we find an element of the domain that is mapped to $g$. Therefore,
\begin{equation}
g\in\operatorname{Im}f.
\end{equation}

More generally, multiplying all $\alpha_i$ by an integer $z$ gives
\begin{equation}
f(z\alpha_1,\ldots,z\alpha_D)
\equiv zg
\pmod{k}.
\end{equation}
Hence all multiples
\begin{equation}
0,g,2g,\ldots,
\left(\frac{k}{g}-1\right)g
\end{equation}
belong to the image. These values are distinct modulo $k$, while the
next value satisfies
\begin{equation}
\frac{k}{g}g=k\equiv0\pmod{k}.
\end{equation}
Therefore,
\begin{equation}
\operatorname{Im}f
=
\left\{
0,g,2g,\ldots,
\left(\frac{k}{g}-1\right)g
\right\},
\end{equation}
and consequently
\begin{equation}
|\operatorname{Im}f|
=
\frac{k}{g}.
\end{equation}

Substituting this result into \eqref{eq:kernel-image-relation}, we
obtain
\begin{equation}
|\ker f|
=
\frac{k^D}{k/g}
=
k^{D-1}g
=
k^{D-1}(k,L_1,\ldots,L_D).
\end{equation}
This proves \eqref{eq:single-eta-counting}.

\section{Explicit evaluation of the \texorpdfstring{$\eta$}{eta}-constraints}
\label{app:3A}

\subsection{Tripartite case}
\label{Appendix_keyring_tripartite}

In this appendix, we explicitly illustrate how the $\eta$-constraints
appearing in the tripartite partition function are simplified by
repeated use of the identities in
\eqref{eta constraint properties}. This calculation justifies the
counting argument used in
Sec.~\ref{subsec:tripartite partition function for key-ring}.

Consider the three-component key-ring state associated with
$K,T_1,$ and $T_2$. From \eqref{eq:rho of key-ring}, the reduced
density matrix obtained by tracing out $K$ is
\begin{equation}
\left(\rho_{T_1T_2}\right)^{
\beta_1'\beta_2'
}_{
\beta_1\beta_2
}
=
\frac{1}{k^2}
\eta\left[
(\beta_1'-\beta_1)L_{KT_1}
+
(\beta_2'-\beta_2)L_{KT_2}
\right].
\label{eq:rho-K3-appendix}
\end{equation}

To make the contraction explicit, we consider $n=3$. The partition
function $\mathcal Z_3^{(3)}(K:T_1:T_2)$ contains nine copies of the
reduced density matrix. For notational convenience, we display these
copies in a $3\times3$ array:
\begin{equation}
\begin{split}
\mathcal Z_3^{(3)}(K:T_1:T_2)
={}&
\left(
\begin{array}{ccc}
\left(\rho_{T_1T_2}\right)^{
\beta_{1,(1)}\beta_{2,(1)}
}_{
\beta_{1,(2)}\beta_{2,(4)}
}
&
\left(\rho_{T_1T_2}\right)^{
\beta_{1,(2)}\beta_{2,(2)}
}_{
\beta_{1,(3)}\beta_{2,(5)}
}
&
\left(\rho_{T_1T_2}\right)^{
\beta_{1,(3)}\beta_{2,(3)}
}_{
\beta_{1,(1)}\beta_{2,(6)}
}
\\[4pt]
\left(\rho_{T_1T_2}\right)^{
\beta_{1,(4)}\beta_{2,(4)}
}_{
\beta_{1,(5)}\beta_{2,(7)}
}
&
\left(\rho_{T_1T_2}\right)^{
\beta_{1,(5)}\beta_{2,(5)}
}_{
\beta_{1,(6)}\beta_{2,(8)}
}
&
\left(\rho_{T_1T_2}\right)^{
\beta_{1,(6)}\beta_{2,(6)}
}_{
\beta_{1,(4)}\beta_{2,(9)}
}
\\[4pt]
\left(\rho_{T_1T_2}\right)^{
\beta_{1,(7)}\beta_{2,(7)}
}_{
\beta_{1,(8)}\beta_{2,(1)}
}
&
\left(\rho_{T_1T_2}\right)^{
\beta_{1,(8)}\beta_{2,(8)}
}_{
\beta_{1,(9)}\beta_{2,(2)}
}
&
\left(\rho_{T_1T_2}\right)^{
\beta_{1,(9)}\beta_{2,(9)}
}_{
\beta_{1,(7)}\beta_{2,(3)}
}
\end{array}
\right).
\end{split}
\label{eq:Z33-array-appendix}
\end{equation}
The entries in this array are multiplied together, and all repeated
indices are summed over according to the replica permutations. Each
copy contributes a factor $1/k^2$, so the overall normalization factor
is $1/k^{18}$.

We introduce the difference variables
\begin{equation}
\beta_{1,(i)(j)}
:=
\beta_{1,(i)}-\beta_{1,(j)},
\qquad
\beta_{2,(i)(j)}
:=
\beta_{2,(i)}-\beta_{2,(j)}.
\label{eq:beta-differences-appendix}
\end{equation}
After substituting \eqref{eq:rho-K3-appendix}, the nine
$\eta$-constraints are
\begin{equation}
\hspace{-8mm}
\left( \begin{array}{ccc}
\eta\left[
\beta_{1,(1)(2)}L_{KT_1}
+\beta_{2,(1)(4)}L_{KT_2}
\right]
&
\eta\left[
\beta_{1,(2)(3)}L_{KT_1}
+\beta_{2,(2)(5)}L_{KT_2}
\right]
&
\eta\left[
\beta_{1,(3)(1)}L_{KT_1}
+\beta_{2,(3)(6)}L_{KT_2}
\right]
\\[5pt]
\eta\left[
\beta_{1,(4)(5)}L_{KT_1}
+\beta_{2,(4)(7)}L_{KT_2}
\right]
&
\eta\left[
\beta_{1,(5)(6)}L_{KT_1}
+\beta_{2,(5)(8)}L_{KT_2}
\right]
&
\eta\left[
\beta_{1,(6)(4)}L_{KT_1}
+\beta_{2,(6)(9)}L_{KT_2}
\right]
\\[5pt]
\eta\left[
\beta_{1,(7)(8)}L_{KT_1}
+\beta_{2,(7)(1)}L_{KT_2}
\right]
&
\eta\left[
\beta_{1,(8)(9)}L_{KT_1}
+\beta_{2,(8)(2)}L_{KT_2}
\right]
&
\eta\left[
\beta_{1,(9)(7)}L_{KT_1}
+\beta_{2,(9)(3)}L_{KT_2}
\right].
\end{array}
\right)
\label{eq:eta-array-initial}
\end{equation}

We first simplify the constraints along each row. Using
\begin{equation}
\eta[x]\eta[y]
=
\eta[x]\eta[x+y],
\label{eqC5}
\end{equation}
together with
\begin{equation}
\beta_{1,(1)(2)}
+\beta_{1,(2)(3)}
+\beta_{1,(3)(1)}
=
0,
\end{equation}
the final constraint in the first row may be replaced by
\begin{equation}
\eta\left[
\left(
\beta_{2,(1)(4)}
+\beta_{2,(2)(5)}
+\beta_{2,(3)(6)}
\right)L_{KT_2}
\right].
\end{equation}
Applying the same operation to all three rows gives
\begin{equation}
\hspace{-10mm}
\left( \begin{array}{ccc}
\eta\left[
\beta_{1,(1)(2)}L_{KT_1}
+\beta_{2,(1)(4)}L_{KT_2}
\right]
&
\eta\left[
\beta_{1,(2)(3)}L_{KT_1}
+\beta_{2,(2)(5)}L_{KT_2}
\right]
&
\eta\left[
\left(
\beta_{2,(1)(4)}
+\beta_{2,(2)(5)}
+\beta_{2,(3)(6)}
\right)L_{KT_2}
\right]
\\[5pt]
\eta\left[
\beta_{1,(4)(5)}L_{KT_1}
+\beta_{2,(4)(7)}L_{KT_2}
\right]
&
\eta\left[
\beta_{1,(5)(6)}L_{KT_1}
+\beta_{2,(5)(8)}L_{KT_2}
\right]
&
\eta\left[
\left(
\beta_{2,(4)(7)}
+\beta_{2,(5)(8)}
+\beta_{2,(6)(9)}
\right)L_{KT_2}
\right]
\\[5pt]
\eta\left[
\beta_{1,(7)(8)}L_{KT_1}
+\beta_{2,(7)(1)}L_{KT_2}
\right]
&
\eta\left[
\beta_{1,(8)(9)}L_{KT_1}
+\beta_{2,(8)(2)}L_{KT_2}
\right]
&
\eta\left[
\left(
\beta_{2,(7)(1)}
+\beta_{2,(8)(2)}
+\beta_{2,(9)(3)}
\right)L_{KT_2}
\right].
\end{array}
\right)
\label{eq:eta-array-row-reduced}
\end{equation}

We next apply the same procedure along each column. The resulting
constraints can be written as
\begin{equation}
\hspace{-23mm}
\left( \begin{array}{ccc}
\eta\left[
\beta_{1,(1)(2)}L_{KT_1}
+\beta_{2,(1)(4)}L_{KT_2}
\right]
&
\eta\left[
\beta_{1,(2)(3)}L_{KT_1}
+\beta_{2,(2)(5)}L_{KT_2}
\right]
&
\eta\left[
\left(
\beta_{2,(1)(4)}
+\beta_{2,(2)(5)}
+\beta_{2,(3)(6)}
\right)L_{KT_2}
\right]
\\[5pt]
\eta\left[
\beta_{1,(4)(5)}L_{KT_1}
+\beta_{2,(4)(7)}L_{KT_2}
\right]
&
\eta\left[
\beta_{1,(5)(6)}L_{KT_1}
+\beta_{2,(5)(8)}L_{KT_2}
\right]
&
\eta\left[
\left(
\beta_{2,(4)(7)}
+\beta_{2,(5)(8)}
+\beta_{2,(6)(9)}
\right)L_{KT_2}
\right]
\\[5pt]
\eta\left[
\left(
\beta_{1,(1)(2)}
+\beta_{1,(4)(5)}
+\beta_{1,(7)(8)}
\right)L_{KT_1}
\right]
&
\eta\left[
\left(
\beta_{1,(2)(3)}
+\beta_{1,(5)(6)}
+\beta_{1,(8)(9)}
\right)L_{KT_1}
\right]
&
\eta[0].
\end{array}
\right)
\label{eq:eta-array-final}
\end{equation}

Equation~\eqref{eq:eta-array-final} makes the counting structure
transparent. For general $n$, the upper-left
$(n-1)\times(n-1)$ block contains $(n-1)^2$ constraints, each involving
one difference variable associated with $T_1$ and one associated with
$T_2$. This block therefore contains
\begin{equation}
2(n-1)^2
\end{equation}
independent difference variables.

The final row contains $n-1$ constraints involving only
$L_{KT_1}$, while the final column contains $n-1$ constraints involving
only $L_{KT_2}$. Together they contain
\begin{equation}
2(n-1)
\end{equation}
additional independent difference variables. The lower-right corner
gives the trivial constraint
\begin{equation}
\eta[0]=1.
\end{equation}

Since the original contraction contains $n^2$ indices associated with
$T_1$ and $n^2$ indices associated with $T_2$, the total number of
indices is $2n^2$. The number of indices appearing as independent
difference variables is
\begin{equation}
2(n-1)^2+2(n-1)=2n(n-1).
\end{equation}
The remaining
\begin{equation}
2n^2-2n(n-1)=2n
\end{equation}
reference indices can be summed over freely and therefore contribute a
factor of
\begin{equation}
k^{2n}.
\end{equation}

Using the counting formula derived in Appendix~\ref{app:one_eta_counting},
\begin{equation}
\sum_{\alpha_1,\ldots,\alpha_D=0}^{k-1}
\eta\left[
\sum_{i=1}^{D}\alpha_iL_i
\right]
=
k^{D-1}(k,L_1,\ldots,L_D),
\end{equation}
each constraint in the $(n-1)\times(n-1)$ bulk block contributes
\begin{equation}
k(k,L_{KT_1},L_{KT_2}).
\end{equation}
Each constraint in the final row contributes
\begin{equation}
(k,L_{KT_1}),
\end{equation}
and each constraint in the final column contributes
\begin{equation}
(k,L_{KT_2}).
\end{equation}

Combining these contributions with the normalization factor
$k^{-2n^2}$, we obtain
\begin{equation}
\begin{split}
\mathcal Z_n^{(3)}(K:T_1:T_2)
={}&
\frac{k^{2n}}{k^{2n^2}}
\left[
k(k,L_{KT_1},L_{KT_2})
\right]^{(n-1)^2}
\\
&\times
(k,L_{KT_1})^{n-1}
(k,L_{KT_2})^{n-1}
\\
={}&
\frac{1}{k^{n^2-1}}
(k,L_{KT_1})^{n-1}
(k,L_{KT_2})^{n-1}
(k,L_{KT_1},L_{KT_2})^{(n-1)^2},
\end{split}
\label{eq:Z3-key-ring-appendix}
\end{equation}
in agreement with \eqref{eq:Z^3_n our key}.

\subsection{Quadripartite case}
\label{Appendix_keyring_quadripartite}
In this appendix, we explicitly illustrate how the $\eta$-constraints
appearing in the quadripartite partition function are simplified by repeated use of the identities in
\eqref{eta constraint properties}. This calculation justifies the
counting argument used in
Sec.~\ref{subsec:q=N partition function for key-ring}.

Consider the four-component key-ring state associated with
$K,T_1,T_2,$ and $T_3$. From \eqref{eq:rho of key-ring}, the reduced
density matrix obtained by tracing out $K$ is
\begin{align}
(\rho_{T_1T_2T_3})^{\beta_1'\beta_2'\beta_3'}_{\beta_1\beta_2\beta_3}=\frac{1}{k^3}\eta\left[(\beta_1'-\beta_1)L_{KT_1}+(\beta_2'-\beta_2)L_{KT_2}+(\beta_3'-\beta_3)L_{KT_3}\right]
\end{align}
For the quadripartite partition function, we knew that there are $n^{4-1}$ replicas arranged as a ($n\times n\times n$)-cubic. We use different ( )-brackets to represent different layers on the third direction, and they are all multiplied together. Therefore, the $n=3$ quadripartite partition function is
\begin{gather}
\mathcal{Z}^{(4)}_3(K:T_1:T_2:T_3)=\notag\\
\begin{pmatrix}
(\rho_{T_1T_2T_3})^{\beta_{1,(1)}\beta_{2,(1)}\beta_{3,(1)}}_{\beta_{1,(2)}\beta_{2,(4)}\beta_{3,(10)}}&
(\rho_{T_1T_2T_3})^{\beta_{1,(2)}\beta_{2,(2)}\beta_{3,(2)}}_{\beta_{1,(3)}\beta_{2,(5)}\beta_{3,(11)}}&
(\rho_{T_1T_2T_3})^{\beta_{1,(3)}\beta_{2,(3)}\beta_{3,(3)}}_{\beta_{1,(1)}\beta_{2,(6)}\beta_{3,(12)}}\\
(\rho_{T_1T_2T_3})^{\beta_{1,(4)}\beta_{2,(4)}\beta_{3,(4)}}_{\beta_{1,(5)}\beta_{2,(7)}\beta_{3,(13)}}&
(\rho_{T_1T_2T_3})^{\beta_{1,(5)}\beta_{2,(5)}\beta_{3,(5)}}_{\beta_{1,(6)}\beta_{2,(8)}\beta_{3,(14)}}&
(\rho_{T_1T_2T_3})^{\beta_{1,(6)}\beta_{2,(6)}\beta_{3,(6)}}_{\beta_{1,(4)}\beta_{2,(9)}\beta_{3,(15)}}\\
(\rho_{T_1T_2T_3})^{\beta_{1,(7)}\beta_{2,(7)}\beta_{3,(7)}}_{\beta_{1,(8)}\beta_{2,(1)}\beta_{3,(16)}}&
(\rho_{T_1T_2T_3})^{\beta_{1,(8)}\beta_{2,(8)}\beta_{3,(8)}}_{\beta_{1,(9)}\beta_{2,(2)}\beta_{3,(17)}}&
(\rho_{T_1T_2T_3})^{\beta_{1,(9)}\beta_{2,(9)}\beta_{3,(9)}}_{\beta_{1,(7)}\beta_{2,(3)}\beta_{3,(18)}}
\end{pmatrix}\notag\\
\begin{pmatrix}
(\rho_{T_1T_2T_3})^{\beta_{1,(10)}\beta_{2,(10)}\beta_{3,(10)}}_{\beta_{1,(11)}\beta_{2,(13)}\beta_{3,(19)}}&
(\rho_{T_1T_2T_3})^{\beta_{1,(11)}\beta_{2,(11)}\beta_{3,(11)}}_{\beta_{1,(12)}\beta_{2,(14)}\beta_{3,(20)}}&
(\rho_{T_1T_2T_3})^{\beta_{1,(12)}\beta_{2,(12)}\beta_{3,(12)}}_{\beta_{1,(10)}\beta_{2,(15)}\beta_{3,(21)}}\\
(\rho_{T_1T_2T_3})^{\beta_{1,(13)}\beta_{2,(13)}\beta_{3,(13)}}_{\beta_{1,(14)}\beta_{2,(16)}\beta_{3,(22)}}&
(\rho_{T_1T_2T_3})^{\beta_{1,(14)}\beta_{2,(14)}\beta_{3,(14)}}_{\beta_{1,(15)}\beta_{2,(17)}\beta_{3,(23)}}&
(\rho_{T_1T_2T_3})^{\beta_{1,(15)}\beta_{2,(15)}\beta_{3,(15)}}_{\beta_{1,(14)}\beta_{2,(18)}\beta_{3,(24)}}\\
(\rho_{T_1T_2T_3})^{\beta_{1,(16)}\beta_{2,(16)}\beta_{3,(16)}}_{\beta_{1,(17)}\beta_{2,(10)}\beta_{3,(25)}}&
(\rho_{T_1T_2T_3})^{\beta_{1,(17)}\beta_{2,(17)}\beta_{3,(17)}}_{\beta_{1,(18)}\beta_{2,(11)}\beta_{3,(26)}}&
(\rho_{T_1T_2T_3})^{\beta_{1,(18)}\beta_{2,(18)}\beta_{3,(18)}}_{\beta_{1,(16)}\beta_{2,(12)}\beta_{3,(27)}}
\end{pmatrix}\notag\\
\begin{pmatrix}
(\rho_{T_1T_2T_3})^{\beta_{1,(19)}\beta_{2,(19)}\beta_{3,(19)}}_{\beta_{1,(20)}\beta_{2,(22)}\beta_{3,(1)}}&
(\rho_{T_1T_2T_3})^{\beta_{1,(20)}\beta_{2,(20)}\beta_{3,(20)}}_{\beta_{1,(21)}\beta_{2,(23)}\beta_{3,(2)}}&
(\rho_{T_1T_2T_3})^{\beta_{1,(21)}\beta_{2,(21)}\beta_{3,(21)}}_{\beta_{1,(19)}\beta_{2,(24)}\beta_{3,(3)}}\\
(\rho_{T_1T_2T_3})^{\beta_{1,(22)}\beta_{2,(22)}\beta_{3,(22)}}_{\beta_{1,(23)}\beta_{2,(25)}\beta_{3,(4)}}&
(\rho_{T_1T_2T_3})^{\beta_{1,(23)}\beta_{2,(23)}\beta_{3,(23)}}_{\beta_{1,(24)}\beta_{2,(26)}\beta_{3,(5)}}&
(\rho_{T_1T_2T_3})^{\beta_{1,(24)}\beta_{2,(24)}\beta_{3,(24)}}_{\beta_{1,(22)}\beta_{2,(27)}\beta_{3,(6)}}\\
(\rho_{T_1T_2T_3})^{\beta_{1,(25)}\beta_{2,(25)}\beta_{3,(25)}}_{\beta_{1,(26)}\beta_{2,(19)}\beta_{3,(7)}}&
(\rho_{T_1T_2T_3})^{\beta_{1,(26)}\beta_{2,(26)}\beta_{3,(26)}}_{\beta_{1,(27)}\beta_{2,(20)}\beta_{3,(8)}}&
(\rho_{T_1T_2T_3})^{\beta_{1,(27)}\beta_{2,(27)}\beta_{3,(27)}}_{\beta_{1,(25)}\beta_{2,(21)}\beta_{3,(9)}}
\end{pmatrix}\notag\\
\end{gather}
Now there are $n^{\mathtt{q}-1}\ (3^{4-1}=27)$ replicas, and each of them has a $1/k^{\mathtt{q}-1}\ (1/k^{4-1}=1/k^3)$ factor. The remaining $n^{\mathtt{q}-1}\ (3^{4-1}=27)$ $\eta$-constraint functions are
\begin{gather}
\scalebox{0.62}{$
\begin{pmatrix}
\eta\left[\beta_{1,(1)(2)}L_{KT_1}+\beta_{2,(1)(4)}L_{KT_2}+\beta_{3,(1)(10)}L_{KT_3}\right]&
\eta\left[\beta_{1,(2)(3)}L_{KT_1}+\beta_{2,(2)(5)}L_{KT_2}+\beta_{3,(2)(11)}L_{KT_3}\right]&
\eta\left[\beta_{1,(3)(1)}L_{KT_1}+\beta_{2,(3)(6)}L_{KT_2}+\beta_{3,(3)(12)}L_{KT_3}\right]\\
\eta\left[\beta_{1,(4)(5)}L_{KT_1}+\beta_{2,(4)(7)}L_{KT_2}+\beta_{3,(4)(13)}L_{KT_3}\right]&
\eta\left[\beta_{1,(5)(6)}L_{KT_1}+\beta_{2,(5)(8)}L_{KT_2}+\beta_{3,(5)(14)}L_{KT_3}\right]&
\eta\left[\beta_{1,(6)(4)}L_{KT_1}+\beta_{2,(6)(9)}L_{KT_2}+\beta_{3,(6)(15)}L_{KT_3}\right]\\
\eta\left[\beta_{1,(7)(8)}L_{KT_1}+\beta_{2,(7)(1)}L_{KT_2}+\beta_{3,(7)(16)}L_{KT_3}\right]&
\eta\left[\beta_{1,(8)(9)}L_{KT_1}+\beta_{2,(8)(2)}L_{KT_2}+\beta_{3,(8)(17)}L_{KT_3}\right]&
\eta\left[\beta_{1,(9)(7)}L_{KT_1}+\beta_{2,(9)(3)}L_{KT_2}+\beta_{3,(9)(18)}L_{KT_3}\right]\notag
\end{pmatrix}$}\notag\\
\scalebox{0.57}{$
\begin{pmatrix}
\eta\left[\beta_{1,(10)(11)}L_{KT_1}+\beta_{2,(10)(13)}L_{KT_2}+\beta_{3,(10)(19)}L_{KT_3}\right]&
\eta\left[\beta_{1,(11)(12)}L_{KT_1}+\beta_{2,(11)(14)}L_{KT_2}+\beta_{3,(11)(20)}L_{KT_3}\right]&
\eta\left[\beta_{1,(12)(10)}L_{KT_1}+\beta_{2,(12)(15)}L_{KT_2}+\beta_{3,(12)(21)}L_{KT_3}\right]\\
\eta\left[\beta_{1,(13)(14)}L_{KT_1}+\beta_{2,(13)(16)}L_{KT_2}+\beta_{3,(13)(22)}L_{KT_3}\right]&
\eta\left[\beta_{1,(14)(15)}L_{KT_1}+\beta_{2,(14)(17)}L_{KT_2}+\beta_{3,(14)(23)}L_{KT_3}\right]&
\eta\left[\beta_{1,(15)(14)}L_{KT_1}+\beta_{2,(15)(18)}L_{KT_2}+\beta_{3,(15)(24)}L_{KT_3}\right]\\
\eta\left[\beta_{1,(16)(17)}L_{KT_1}+\beta_{2,(16)(10)}L_{KT_2}+\beta_{3,(16)(25)}L_{KT_3}\right]&
\eta\left[\beta_{1,(17)(18)}L_{KT_1}+\beta_{2,(17)(11)}L_{KT_2}+\beta_{3,(17)(26)}L_{KT_3}\right]&
\eta\left[\beta_{1,(18)(16)}L_{KT_1}+\beta_{2,(18)(12)}L_{KT_2}+\beta_{3,(18)(27)}L_{KT_3}\right]\notag
\end{pmatrix}$}\notag\\
\scalebox{0.58}{$
\begin{pmatrix}
\eta\left[\beta_{1,(19)(20)}L_{KT_1}+\beta_{2,(19)(22)}L_{KT_2}+\beta_{3,(19)(1)}L_{KT_3}\right]&
\eta\left[\beta_{1,(20)(21)}L_{KT_1}+\beta_{2,(20)(23)}L_{KT_2}+\beta_{3,(20)(2)}L_{KT_3}\right]&
\eta\left[\beta_{1,(21)(19)}L_{KT_1}+\beta_{2,(21)(24)}L_{KT_2}+\beta_{3,(21)(3)}L_{KT_3}\right]\\
\eta\left[\beta_{1,(22)(23)}L_{KT_1}+\beta_{2,(22)(25)}L_{KT_2}+\beta_{3,(22)(4)}L_{KT_3}\right]&
\eta\left[\beta_{1,(23)(24)}L_{KT_1}+\beta_{2,(23)(26)}L_{KT_2}+\beta_{3,(23)(5)}L_{KT_3}\right]&
\eta\left[\beta_{1,(24)(22)}L_{KT_1}+\beta_{2,(24)(27)}L_{KT_2}+\beta_{3,(24)(6)}L_{KT_3}\right]\\
\eta\left[\beta_{1,(25)(26)}L_{KT_1}+\beta_{2,(25)(19)}L_{KT_2}+\beta_{3,(25)(7)}L_{KT_3}\right]&
\eta\left[\beta_{1,(26)(27)}L_{KT_1}+\beta_{2,(26)(20)}L_{KT_2}+\beta_{3,(26)(8)}L_{KT_3}\right]&
\eta\left[\beta_{1,(27)(25)}L_{KT_1}+\beta_{2,(27)(21)}L_{KT_2}+\beta_{3,(27)(9)}L_{KT_3}\right]\notag
\end{pmatrix}$}\notag
\end{gather}
Applying the same summing rule for the $\eta$-constraint function in Sec.~\ref{subsec:tripartite partition function for key-ring} to each row, column, and layer direction, we obtain
\begin{gather}
\scalebox{0.63}{$\begin{pmatrix}
\textcolor{red}{\eta\left[\beta_{1,(1)(2)}L_{KT_1}+\beta_{2,(1)(4)}L_{KT_2}+\beta_{3,(1)(10)}L_{KT_3}\right]}&
\textcolor{red}{\eta\left[\beta_{1,(2)(3)}L_{KT_1}+\beta_{2,(2)(5)}L_{KT_2}+\beta_{3,(2)(11)}L_{KT_3}\right]}&
\textcolor{blue}{\eta\left[\begin{aligned}
&(\beta_{2,(1)(4)}+\beta_{2,(2)(5)}+\beta_{2,(3)(6)})L_{KT_2}\\
&+(\beta_{3,(1)(10)}+\beta_{3,(2)(11)}+\beta_{3,(3)(12)})L_{KT_3}
\end{aligned}\right]}\\
\textcolor{red}{\eta\left[\beta_{1,(4)(5)}L_{KT_1}+\beta_{2,(4)(7)}L_{KT_2}+\beta_{3,(4)(13)}L_{KT_3}\right]}&
\textcolor{red}{\eta\left[\beta_{1,(5)(6)}L_{KT_1}+\beta_{2,(5)(8)}L_{KT_2}+\beta_{3,(5)(14)}L_{KT_3}\right]}&
\textcolor{blue}{\eta\left[\begin{aligned}
&(\beta_{2,(4)(7)}+\beta_{2,(5)(8)}+\beta_{2,(6)(9)})L_{KT_2}\\
&+(\beta_{3,(4)(13)}+\beta_{3,(5)(14)}+\beta_{3,(6)(15)})L_{KT_3}
\end{aligned}\right]}\\
\textcolor{blue}{\eta\left[\begin{aligned}
&(\beta_{1,(1)(2)}+\beta_{1,(4)(5)}+\beta_{1,(7)(8)})L_{KT_1}\\
&+(\beta_{3,(1)(10)}+\beta_{3,(4)(13)}+\beta_{3,(7)(16)})L_{KT_3}
\end{aligned}\right]}&
\textcolor{blue}{\eta\left[\begin{aligned}
&(\beta_{1,(2)(3)}+\beta_{1,(5)(6)}+\beta_{1,(8)(9)})L_{KT_1}\\
&+(\beta_{3,(2)(11)}+\beta_{3,(5)(14)}+\beta_{3,(8)(17)})L_{KT_3}
\end{aligned}\right]}&
\textcolor{green!60!black}{\eta\left[\begin{aligned}
(&\beta_{3,(1)(10)}+\beta_{3,(2)(11)}+\beta_{3,(3)(12)}\\
&+\beta_{3,(4)(13)}+\beta_{3,(5)(14)}+\beta_{3,(6)(15)}\\
&+\beta_{3,(7)(16)}+\beta_{3,(8)(17)}+\beta_{3,(9)(18)})L_{KT_3}
\end{aligned}\right]}
\end{pmatrix}\notag$}\\
\scalebox{0.585}{$\begin{pmatrix}
\textcolor{red}{\eta\left[\beta_{1,(10)(11)}L_{KT_1}+\beta_{2,(10)(13)}L_{KT_2}+\beta_{3,(10)(19)}L_{KT_3}\right]}&
\textcolor{red}{\eta\left[\beta_{1,(11)(12)}L_{KT_1}+\beta_{2,(11)(14)}L_{KT_2}+\beta_{3,(11)(20)}L_{KT_3}\right]}&
\textcolor{blue}{\eta\left[\begin{aligned}
&(\beta_{2,(10)(13)}+\beta_{2,(11)(14)}+\beta_{2,(12)(15)})L_{KT_2}\\
&+(\beta_{3,(10)(19)}+\beta_{3,(11)(20)}+\beta_{3,(12)(21)})L_{KT_3}
\end{aligned}\right]}\\
\textcolor{red}{\eta\left[\beta_{1,(13)(14)}L_{KT_1}+\beta_{2,(13)(16)}L_{KT_2}+\beta_{3,(13)(22)}L_{KT_3}\right]}&
\textcolor{red}{\eta\left[\beta_{1,(14)(15)}L_{KT_1}+\beta_{2,(14)(17)}L_{KT_2}+\beta_{3,(14)(23)}L_{KT_3}\right]}&
\textcolor{blue}{\eta\left[\begin{aligned}
&(\beta_{2,(13)(16)}+\beta_{2,(14)(17)}+\beta_{2,(15)(18)})L_{KT_2}\\
&+(\beta_{3,(13)(22)}+\beta_{3,(14)(23)}+\beta_{3,(15)(24)})L_{KT_3}
\end{aligned}\right]}\\
\textcolor{blue}{\eta\left[\begin{aligned}
&(\beta_{1,(10)(11)}+\beta_{1,(13)(14)}+\beta_{1,(16)(17)})L_{KT_1}\\
&+(\beta_{3,(10)(19)}+\beta_{3,(13)(22)}+\beta_{3,(16)(25)})L_{KT_3}
\end{aligned}\right]}&
\textcolor{blue}{\eta\left[\begin{aligned}
&(\beta_{1,(11)(12)}+\beta_{1,(14)(15)}+\beta_{1,(17)(18)})L_{KT_1}\\
&+(\beta_{3,(11)(20)}+\beta_{3,(14)(23)}+\beta_{3,(17)(26)})L_{KT_3}
\end{aligned}\right]}&
\textcolor{green!60!black}{\eta\left[\begin{aligned}
(&\beta_{3,(10)(19)}+\beta_{3,(11)(20)}+\beta_{3,(12)(21)}\\
&+\beta_{3,(13)(22)}+\beta_{3,(14)(23)}+\beta_{3,(15)(24)}\\
&+\beta_{3,(16)(25)}+\beta_{3,(17)(26)}+\beta_{3,(18)(27)})L_{KT_3}
\end{aligned}\right]}
\end{pmatrix}\notag$}\\
\scalebox{0.62}{$\begin{pmatrix}\label{final contraints in 4 partite}
\textcolor{blue}{\eta\left[\begin{aligned}
&(\beta_{1,(1)(2)}+\beta_{1,(10)(11)}+\beta_{1,(19)(20)})L_{KT_1}\\
&+(\beta_{2,(1)(4)}+\beta_{2,(10)(13)}+\beta_{2,(19)(22)})L_{KT_2}
\end{aligned}\right]}&
\textcolor{blue}{\eta\left[\begin{aligned}
&(\beta_{1,(2)(3)}+\beta_{1,(11)(12)}+\beta_{1,(20)(21)})L_{KT_1}\\
&+(\beta_{2,(2)(5)}+\beta_{2,(11)(14)}+\beta_{2,(20)(23)})L_{KT_2}
\end{aligned}\right]}&
\textcolor{green!60!black}{\eta\left[\begin{aligned}
(&\beta_{2,(1)(4)}+\beta_{2,(2)(5)}+\beta_{2,(3)(6)}\\
&+\beta_{2,(10)(13)}+\beta_{2,(11)(14)}+\beta_{2,(12)(15)}\\
&+\beta_{2,(19)(22)}+\beta_{2,(20)(23)}+\beta_{2,(21)(24)})L_{KT_2}
\end{aligned}\right]}\\
\textcolor{blue}{\eta\left[\begin{aligned}
&(\beta_{1,(4)(5)}+\beta_{1,(13)(14)}+\beta_{1,(22)(23)})L_{KT_1}\\
&+(\beta_{2,(4)(7)}+\beta_{2,(13)(16)}+\beta_{2,(22)(25)})L_{KT_2}
\end{aligned}\right]}&
\textcolor{blue}{\eta\left[\begin{aligned}
&(\beta_{1,(5)(6)}+\beta_{1,(14)(15)}+\beta_{1,(23)(24)})L_{KT_1}\\
&+(\beta_{2,(5)(8)}+\beta_{2,(14)(17)}+\beta_{2,(23)(26)})L_{KT_2}
\end{aligned}\right]}&
\textcolor{green!60!black}{\eta\left[\begin{aligned}
(&\beta_{2,(4)(7)}+\beta_{2,(5)(8)}+\beta_{2,(6)(9)}\\
&+\beta_{2,(13)(16)}+\beta_{2,(14)(17)}+\beta_{2,(15)(18)}\\
&+\beta_{2,(22)(25)}+\beta_{2,(23)(26)}+\beta_{2,(24)(27)})L_{KT_2}
\end{aligned}\right]}\\
\textcolor{green!60!black}{\eta\left[\begin{aligned}
(&\beta_{1,(1)(2)}+\beta_{1,(4)(5)}+\beta_{1,(7)(8)}\\
&+\beta_{1,(10)(11)}+\beta_{1,(13)(14)}+\beta_{1,(16)(17)}\\
&+\beta_{1,(19)(20)}+\beta_{1,(22)(23)}+\beta_{1,(25)(26)})L_{KT_1}
\end{aligned}\right]}&
\textcolor{green!60!black}{\eta\left[\begin{aligned}
(&\beta_{1,(2)(3)}+\beta_{1,(5)(6)}+\beta_{1,(8)(9)}\\
&+\beta_{1,(11)(12)}+\beta_{1,(14)(15)}+\beta_{1,(17)(18)}\\
&+\beta_{1,(20)(21)}+\beta_{1,(23)(24)}+\beta_{1,(26)(27)})L_{KT_1}
\end{aligned}\right]}&
\eta\left[0\right]
\end{pmatrix}$}
\end{gather}
Compare with Sec.~\ref{subsec:q=N partition function for key-ring}, we can divide $n^{(4-1)}$ terms into three kinds of colored regions and one uncolored single corner (see Fig.~\ref{fig:Quadripartite Keyring}). We obtain $3(n-1)^3$ independent indices in the upper-left-front $(n-1)\times(n-1)\times(n-1)$ 3D block (terms in the red region). Three $(n-1)^2$ blocks (blue regions) contain $2(n-1)^2$ independent indices each, and three lines (green regions) of length $(n-1)$ contain $(n-1)$ independent indices each. 

The number of remaining independent indices will be $3n^3-3(n-1)^3-6(n-1)^2-3(n-1)=3n^2$. We will get an extra $k^{3n^2}$ by summing over these indices. Summing over the indices in the expression will be just the product of the counting of each constraint. 
\begin{itemize}
    \item The number of solutions of $\beta$, $\gamma$ and $\mu$, which satisfies the $\eta$-constraints written in green in the expression \eqref{final contraints in 4 partite} are $(k,L_{KT_1})$, $(k,L_{KT_2})$ and $(k,L_{KT_3})$ respectively.
    \item The number of the pair of solutions $(\beta, \gamma)$, $(\gamma, \mu)$ and $(\beta, \mu)$, which satisfies the $\eta$- constraints written in blue in the expression \eqref{final contraints in 4 partite} are $k\cdot(k,L_{KT_1},L_{KT_2})$, $k\cdot(k,L_{KT_2},L_{KT_3})$ and $k\cdot(k,L_{KT_3},L_{KT_1})$ respectively.
    \item The number of the solutions of $(\beta, \gamma, \mu)$, which satisfies the $\eta$-constraints written in red in the expression \eqref{final contraints in 4 partite} is $k^2\cdot(k,L_{KT_1},L_{KT_2},L_{KT_3})$.
\end{itemize}
Therefore, the quadripartite partition function becomes
\begin{gather}
\mathcal{Z}^{(4)}_n(K:T_1:T_2:T_3)=\frac{k^{3n^2}}{k^{3n^3}}\notag\\\underbrace{(k,L_{KT_1})^{n-1}(k,L_{KT_2})^{n-1}(k,L_{KT_3})^{n-1}}_{\text{Green terms}}\notag\\\underbrace{\left(k\cdot(k,L_{KT_1},L_{KT_2})\right)^{(n-1)^2}\left(k\cdot(k,L_{KT_2},L_{KT_3})\right)^{(n-1)^2}\left(k\cdot(k,L_{KT_3},L_{KT_1})\right)^{(n-1)^2}}_{\text{Blue terms}}\notag\\
\underbrace{\left(k^2\cdot(k,L_{KT_1},L_{KT_2},L_{KT_3})\right)^{(n-1)^3}}_{\text{Red terms}}\notag\\
=\frac{1}{k^{n^3-1}}\left((k,L_{KT_1})(k,L_{KT_2})(k,L_{KT_3})\right)^{n-1}\notag\\\left((k,L_{KT_1},L_{KT_2})(k,L_{KT_2},L_{KT_3})(k,L_{KT_3},L_{KT_1})\right)^{(n-1)^2}
(k,L_{KT_1},L_{KT_2},L_{KT_3})^{(n-1)^3}
\label{eq:4 key partition}
\end{gather}

\section{Explicit evaluation of the effective linking number}
\label{Appendix_effective_linking_tripartite}
In this appendix, we explicitly illustrate how the $\eta$-constraints
appearing in the tripartite partition function for four-component link states are simplified by
repeated use of the identities in
\eqref{eta constraint properties}. This calculation justifies the
counting argument used in
Sec.~\ref{subsubsec:Adding key tori to a non-key-ring party}.

Consider the four-component key-ring state ($KT_1T_2T_3$) with the partition
\begin{align}
\mathcal{A}_1=K\,,\qquad\mathcal{A}_2=T_1\,,\qquad\mathcal{A}_3=T_2T_3.
\end{align}
From \eqref{eq:rho of key-ring}, the reduced
density matrix obtained by tracing out $\mathcal{A}_1=K$ is
\begin{align}
(\rho_{\mathcal{A}_2\mathcal{A}_3})^{\beta_1'\beta_2'\beta_3'}_{\beta_1\beta_2\beta_3}=(\rho_{T_1T_2T_3})^{\beta_1'\beta_2'\beta_3'}_{\beta_1\beta_2\beta_3}=\frac{1}{k^3}\eta\left[(\beta_1'-\beta_1)L_{KT_1}+(\beta_2'-\beta_2)L_{KT_2}+(\beta_3'-\beta_3)L_{KT_3}\right]
\end{align}
To make the contraction explicit, we consider $n=3$. The partition function $\mathcal Z_3^{(3)}(\mathcal{A}_1:\mathcal{A}_2:\mathcal{A}_3)$ contains nine copies of the reduced density matrix. For notational convenience, we display these
copies in a $3\times3$ array:
For $n=3$, the tripartite partition function becomes
\begin{align}
&\mathcal{Z}^{(3)}_3(\mathcal{A}_1:\mathcal{A}_2:\mathcal{A}_3)=\mathcal{Z}^{(3)}_3(K:T_1:T_2T_3)\notag\\
&=\begin{pmatrix}
(\rho_{T_1T_2T_3})^{\beta_{1,(1)}\beta_{2,(1)}\beta_{3,(1)}}_{\beta_{1,(2)}\beta_{2,(4)}\beta_{3,(4)}}&
(\rho_{T_1T_2T_3})^{\beta_{1,(2)}\beta_{2,(2)}\beta_{3,(2)}}_{\beta_{1,(3)}\beta_{2,(5)}\beta_{3,(5)}}&
(\rho_{T_1T_2T_3})^{\beta_{1,(3)}\beta_{2,(3)}\beta_{3,(3)}}_{\beta_{1,(1)}\beta_{2,(6)}\beta_{3,(6)}}\\
(\rho_{T_1T_2T_3})^{\beta_{1,(4)}\beta_{2,(4)}\beta_{3,(4)}}_{\beta_{1,(5)}\beta_{2,(7)}\beta_{3,(7)}}&
(\rho_{T_1T_2T_3})^{\beta_{1,(5)}\beta_{2,(5)}\beta_{3,(5)}}_{\beta_{1,(6)}\beta_{2,(8)}\beta_{3,(8)}}&
(\rho_{T_1T_2T_3})^{\beta_{1,(6)}\beta_{2,(6)}\beta_{3,(6)}}_{\beta_{1,(4)}\beta_{2,(9)}\beta_{3,(9)}}\\
(\rho_{T_1T_2T_3})^{\beta_{1,(7)}\beta_{2,(7)}\beta_{3,(7)}}_{\beta_{1,(8)}\beta_{2,(1)}\beta_{3,(1)}}&
(\rho_{T_1T_2T_3})^{\beta_{1,(8)}\beta_{2,(8)}\beta_{3,(8)}}_{\beta_{1,(9)}\beta_{2,(2)}\beta_{3,(2)}}&
(\rho_{T_1T_2T_3})^{\beta_{1,(9)}\beta_{2,(9)}\beta_{3,(9)}}_{\beta_{1,(7)}\beta_{2,(3)}\beta_{3,(3)}}
\end{pmatrix}
\label{eq:Z^3_3(K:T1:T2T3)}
\end{align}
Comparing with \eqref{eq:Z33-array-appendix}, we obtain that the $k$ factor we get from each reduced density operator becomes ($1/k^3$) instead of ($1/k^2$), and the $\eta$-constraint functions part becomes
\begin{gather}
\hspace{-8mm}\scalebox{0.71}{$ \left(
\begin{matrix}
\eta\left[\beta_{1,(1)(2)}L_{KT_1}+\beta_{2,(1)(4)}L_{KT_2}+\beta_{3,(1)(4)}L_{KT_3}\right]&
\eta\left[\beta_{1,(2)(3)}L_{KT_1}+\beta_{2,(2)(5)}L_{KT_2}+\beta_{3,(2)(5)}L_{KT_3}\right]&
\eta\left[\beta_{1,(3)(1)}L_{KT_1}+\beta_{2,(3)(6)}L_{KT_2}+\beta_{3,(3)(6)}L_{KT_3}\right]\\
\eta\left[\beta_{1,(4)(5)}L_{KT_1}+\beta_{2,(4)(7)}L_{KT_2}+\beta_{3,(4)(7)}L_{KT_3}\right]&
\eta\left[\beta_{1,(5)(6)}L_{KT_1}+\beta_{2,(5)(8)}L_{KT_2}+\beta_{3,(5)(8)}L_{KT_3}\right]&
\eta\left[\beta_{1,(6)(4)}L_{KT_1}+\beta_{2,(6)(9)}L_{KT_2}+\beta_{3,(6)(9)}L_{KT_3}\right]\\
\eta\left[\beta_{1,(7)(8)}L_{KT_1}+\beta_{2,(7)(1)}L_{KT_2}+\beta_{3,(7)(1)}L_{KT_3}\right]&
\eta\left[\beta_{1,(8)(9)}L_{KT_1}+\beta_{2,(8)(2)}L_{KT_2}+\beta_{3,(8)(2)}L_{KT_3}\right]&
\eta\left[\beta_{1,(9)(7)}L_{KT_1}+\beta_{2,(9)(3)}L_{KT_2}+\beta_{3,(9)(3)}L_{KT_3}\right]
\end{matrix} \right)$}
\end{gather}
By applying \eqref{eqC5}, we obtain
\begin{gather}
\hspace{-8mm}\scalebox{0.735}{$ \left(
\begin{matrix}
\eta\left[\beta_{1,(1)(2)}L_{KT_1}+\beta_{2,(1)(4)}L_{KT_2}+\beta_{3,(1)(4)}L_{KT_3}\right]&
\eta\left[\beta_{1,(2)(3)}L_{KT_1}+\beta_{2,(2)(5)}L_{KT_2}+\beta_{3,(2)(5)}L_{KT_3}\right]&
\eta\left[\begin{aligned}
(&\beta_{2,(1)(4)}+\beta_{2,(2)(5)}+\beta_{2,(3)(6)})L_{KT_2}\\&+(\beta_{3,(1)(4)}+\beta_{3,(2)(5)}+\beta_{3,(3)(6)})L_{KT_3}
\end{aligned}\right]\\
\eta\left[\beta_{1,(4)(5)}L_{KT_1}+\beta_{2,(4)(7)}L_{KT_2}+\beta_{3,(4)(7)}L_{KT_3}\right]&
\eta\left[\beta_{1,(5)(6)}L_{KT_1}+\beta_{2,(5)(8)}L_{KT_2}+\beta_{3,(5)(8)}L_{KT_3}\right]&
\eta\left[\begin{aligned}
(&\beta_{2,(4)(7)}+\beta_{2,(5)(8)}+\beta_{2,(6)(9)})L_{KT_2}\\&+(\beta_{3,(4)(7)}+\beta_{3,(5)(8)}+\beta_{3,(6)(9)})L_{KT_3}
\end{aligned}\right]\\
\eta\left[(\beta_{1,(1)(2)}+\beta_{1,(4)(5)}+\beta_{1,(7)(8)})L_{KT_1})\right]&
\eta\left[(\beta_{1,(2)(3)}+\beta_{1,(5)(6)}+\beta_{1,(8)(9)})L_{KT_1}\right]&
\eta\left[0\right]
\end{matrix} \right)$}
\end{gather}
Compared with \eqref{eq:eta-array-final}, the only new ingredients are the index of $T_3$ ($\beta_3$-index) and the linking number $L_{KT_3}$. However, they always appear together with the $\beta_2$-index of $T_2$ and $L_{KT_2}$. Namely, every term containing the $\beta_2$-index of $T_2$ and $L_{KT_2}$ in \eqref{eq:eta-array-final} now contains the $\beta_3$-index of $T_3$ and $L_{KT_3}$ as well. To count the corresponding index assignments, we use the following result derived in Appendix~\ref{app:one_eta_counting}: 
\begin{align}
\sum^{k-1}_{\alpha_{i}=0}\eta\left[\sum^{D}_{i=1}\alpha_iL_i\right]=k^{D-1}(k,L_1,L_2,\cdots,L_D) \,.
\end{align}
Adding one more linking number and summing over its corresponding index simply extends the greatest common divisor to include the additional linking number and contributes one extra factor of $k$. Moreover, similar to Sec.~\ref{subsec:tripartite partition function for key-ring}, we have $3n^2-3(n-1)^2-3(n-1)=3n$ remaining independent indices to be summed over, each contributing a factor of $k$. Therefore, for generic $n$, we obtain a result similar to \eqref{eq:Z^3_n our key} that is,
\begin{equation}
\scalebox{0.95}{$
\begin{aligned}
\mathcal{Z}^{(3)}_n(K:T_1:T_2T_3)=&\frac{k^{3n}}{k^{3n^2}}\underbrace{(k^2\cdot(k,L_{KT_1},L_{KT_2},L_{KT_3}))^{(n-1)^2}}_{\text{$(n-1)\times(n-1)$ block}}\underbrace{(k,L_{KT_1})^{n-1}}_{\text{last row}}\underbrace{(k\cdot(k,L_{KT_2},L_{KT_3}))^{n-1}}_{\text{last column}}\\
=&\frac{1}{k^{n^2-1}}(k,L_{KT_1},L_{KT_2},L_{KT_3})^{(n-1)^2}(k,L_{KT_1})^{n-1}(k,L_{KT_2},L_{KT_3})^{n-1} \,.
\end{aligned}$}
\end{equation}

\section{Detailed results of all partition functions, Genuine Rényi multi-entropy, and \texorpdfstring{$\partial GM^{(\mathtt{q})}_n$}{\partial GM^{q}_n} for key-ring states}
\label{app:G}

\subsection{Results for four-component key-ring state}
\label{app:G.1}
\begin{itemize}
\item Quadripartite partition function (1:1:1:1):\\
\begin{align}
\mathcal{Z}^{(4)}_n[1:1:1:1]=&\mathcal{Z}^{(4)}_n(K:T_1:T_2:T_3)\notag\\=&\frac{1}{k^{n^3-1}}\left((k,L_{KT_1})(k,L_{KT_2})(k,L_{KT_3})\right)^{n-1}\notag\\&\left((k,L_{KT_1},L_{KT_2})(k,L_{KT_2},L_{KT_3})(k,L_{KT_3},L_{KT_1})\right)^{(n-1)^2}\notag\\
&(k,L_{KT_1},L_{KT_2},L_{KT_3})^{(n-1)^3}.
\end{align}
\item Tripartite partition functions (2:1:1):\\
Consider the key-ring as a single party:
\begin{align}
\mathcal{Z}^{(3)}_n(K:T_1:T_2T_3)=\frac{1}{k^{n^2-1}}(k,L_{KT_1})^{n-1}(k,L_{KT_2},L_{KT_3})^{n-1}(k,L_{KT_1},L_{KT_2},L_{KT_3})^{(n-1)^2}
\end{align}
\begin{align}
\mathcal{Z}^{(3)}_n(K:T_2:T_3T_1)=\frac{1}{k^{n^2-1}}(k,L_{KT_2})^{n-1}(k,L_{KT_3},L_{KT_1})^{n-1}(k,L_{KT_1},L_{KT_2},L_{KT_3})^{(n-1)^2}
\end{align}
\begin{align}
\mathcal{Z}^{(3)}_n(K:T_3:T_1T_2)=\frac{1}{k^{n^2-1}}(k,L_{KT_3})^{n-1}(k,L_{KT_1},L_{KT_2})^{n-1}(k,L_{KT_1},L_{KT_2},L_{KT_3})^{(n-1)^2}
\end{align}
Consider the key-ring belongs to a two-component party:
\begin{align}
\mathcal{Z}^{(3)}_n(KT_1:T_2:T_3)
=\frac{1}{k^{n^2-1}}(k,L_{KT_2})^{n-1}(k,L_{KT_3})^{n-1}(k,L_{KT_2},L_{KT_3})^{(n-1)^2}
\end{align}
\begin{align}
\mathcal{Z}^{(3)}_n(KT_2:T_3:T_1)
=\frac{1}{k^{n^2-1}}(k,L_{KT_3})^{n-1}(k,L_{KT_1})^{n-1}(k,L_{KT_3},L_{KT_1})^{(n-1)^2}
\end{align}
\begin{align}
\mathcal{Z}^{(3)}_n(KT_3:T_1:T_2)
=\frac{1}{k^{n^2-1}}(k,L_{KT_1})^{n-1}(k,L_{KT_2})^{n-1}(k,L_{KT_1},L_{KT_2})^{(n-1)^2}
\end{align}
Combining all six, we obtain
\begin{align}
\mathcal{Z}^{(3)}_n[2:1:1]
=&\mathcal{Z}^{(3)}_n(K:T_1:T_2T_3)\mathcal{Z}^{(3)}_n(K:T_2:T_1T_3)\mathcal{Z}^{(3)}_n(K:T_3:T_1T_2)\notag\\
&\mathcal{Z}^{(3)}_n(T_1:T_3:KT_2)\mathcal{Z}^{(3)}_n(T_1:T_2:KT_3)\mathcal{Z}^{(3)}_n(T_2:T_3:KT_1)\notag\\
=&\frac{1}{k^{6(n^2-1)}}\left((k,L_{KT_1})(k,L_{KT_2})(k,L_{KT_3})\right)^{3(n-1)}\notag\\
& \quad \times \left((k,L_{KT_1},L_{KT_2})(k,L_{KT_2},L_{KT_3})(k,L_{KT_3},L_{KT_1})\right)^{(n-1)+(n-1)^2} \notag\\
& \quad  \times (k,L_{KT_1},L_{KT_2},L_{KT_3})^{3(n-1)^2}  
\end{align}
\item Bipartite partition functions (3:1):\\
Consider the key-ring as a single party:
\begin{gather}
\mathcal{Z}^{(2)}_n(K:T_1T_2T_3)
=\left(\frac{(k,L_{KT_1},L_{KT_2},L_{KT_3})}{k}\right)^{n-1}
\end{gather}
Consider the key-ring belongs to a three-component party:
\begin{gather}
\mathcal{Z}^{(2)}_n(KT_1T_2:T_3)
=\left(\frac{(k,L_{KT_3})}{k}\right)^{n-1}\\
\mathcal{Z}^{(2)}_n(KT_3T_1:T_2)
=\left(\frac{(k,L_{KT_2})}{k}\right)^{n-1}\\
\mathcal{Z}^{(2)}_n(KT_2T_3:T_1)
=\left(\frac{(k,L_{KT_1})}{k}\right)^{n-1}
\end{gather}
Combining all four, we obtain
\begin{align}
\mathcal{Z}^{(2)}_n[3:1]=&\mathcal{Z}^{(2)}_n(K:T_1T_2T_3)\mathcal{Z}^{(2)}_n(KT_1T_2:T_3)\mathcal{Z}^{(2)}_n(KT_3T_1:T_2)\mathcal{Z}^{(2)}_n(KT_2T_3:T_1)\notag\\
=&\left(\frac{(k,L_{KT_1},L_{KT_2},L_{KT_3})}{k}\frac{(k,L_{KT_1})}{k}\frac{(k,L_{KT_2})}{k}\frac{(k,L_{KT_3})}{k}\right)^{(n-1)}
\end{align}

\item Bipartite partition function (2:2):
\begin{gather}
\mathcal{Z}^{(2)}_n(KT_1:T_2T_3)=\left(\frac{(k,L_{KT_2},L_{KT_3})}{k}\right)^{n-1}\\
\mathcal{Z}^{(2)}_n(KT_2:T_3T_1)=\left(\frac{(k,L_{KT_3},L_{KT_1})}{k}\right)^{n-1}\\
\mathcal{Z}^{(2)}_n(KT_3:T_1T_2)=\left(\frac{(k,L_{KT_1},L_{KT_2})}{k}\right)^{n-1}
\end{gather}
Combining all three, we obtain
\begin{align}
\mathcal{Z}^{(2)}_n[2:2]=&\mathcal{Z}^{(2)}_n(KT_1:T_2T_3)\mathcal{Z}^{(2)}_n(KT_2:T_3T_1)\mathcal{Z}^{(2)}_n(KT_3:T_1T_2)\notag\\
=&\left(\frac{(k,L_{KT_2},L_{KT_3})}{k}\frac{(k,L_{KT_1},L_{KT_3})}{k}\frac{(k,L_{KT_1},L_{KT_2})}{k}\right)^{n-1}\label{eq:Z^2_n(2:2) product key}
\end{align}
\item  Genuine Rényi multi-entropy: 
\begin{align}
\left.\GM^{(4)}_n\right|_{a=0}=&-\frac{1}{n^2} \log \left(\left(\frac{1}{k}\right)^{n^2+n+1+2n(n-1)-4n^2+\frac{4n^2}{3}}\left((k,L_{KT_1})(k,L_{KT_2})(k,L_{KT_3})\right)^{1-n+\frac{n^2}{3}}\right.\notag\\
&\left. \hspace{-15mm} \left((k,L_{KT_1},L_{KT_2})(k,L_{KT_2},L_{KT_3})(k,L_{KT_3},L_{KT_1})\right)^{n-1-\frac{n^2}{3}}(k,L_{KT_1},L_{KT_2},L_{KT_3})^{1-n+\frac{n^2}{3}}\right)
\end{align}
\end{itemize}

\subsection{Results for five-component key-ring state}\label{app:G.2}

\begin{itemize}
    \item Partition functions:\\
    \\
Following eq.~\eqref{eq367} and by interchanging the labels $T_1, T_2, T_3, T_4$ among each other, we obtain the other similar type of five quadripartite partition functions $\mathcal{Z}^{(4)}_n(K:T_1T_2:T_3:T_4), \mathcal{Z}^{(4)}_n(K:T_1T_3:T_2:T_4), 
\mathcal{Z}^{(4)}_n(K:T_1 T_4:T_2:T_3),\mathcal{Z}^{(4)}_n(K:T_1:T_2T_3:T_4)$ and $\mathcal{Z}^{(4)}_n(K:T_1:T_2T_4:T_3)$.\\
Thus, combining all of them, we can obtain part of the quadripartite partition function
\begin{gather}
    \mathcal{Z}^{(4)}_{n \, {\rm part 1}}[2:1:1:1]= \frac{1}{k^{6({n^3-1})}}
    \left((k,L_{KT_3})(k,L_{KT_4})(k,L_{KT_1})(k,L_{KT_2})\right)^{3(n-1)}\notag\\
    \bigl((k,L_{KT_3},L_{KT_4})(k,L_{KT_2},L_{KT_1})(k,L_{KT_4},L_{KT_1})(k,L_{KT_2},L_{KT_3})\notag\\
    (k,L_{KT_3},L_{KT_1})(k,L_{KT_2},L_{KT_4})\bigr)^{(n-1)^2+(n-1)}\notag\\ \left((k,L_{KT_1},L_{KT_2},L_{KT_3})(k,L_{KT_1},L_{KT_2},L_{KT_4})(k,L_{KT_3},L_{KT_2},L_{KT_4})(k,L_{KT_1},L_{KT_3},L_{KT_4})\right)^{3(n-1)^2}\notag\\
    (k,L_{KT_1},L_{KT_2},L_{KT_3},L_{KT_4})^{6(n-1)^3}
    \label{eq:Z^4_n(1:2:1:1) key-ring summary}
\end{gather}

Similarly, following eq.~\eqref{eq368} and by interchanging the labels $T_1, T_2, T_3, T_4$ among each other, we obtain the other similar type of three quadripartite partition functions $\mathcal{Z}^{(4)}_n(KT_2:T_1:T_3:T_4),\mathcal{Z}^{(4)}_n(KT_3:T_1:T_2:T_4),
\mathcal{Z}^{(4)}_n(KT_4:T_1:T_2:T_3)$.

Thus, combining all of them, we can obtain another part of the quadripartite partition function
\begin{gather}
    \mathcal{Z}^{(4)}_{n \, {\rm part 2}}[2:1:1:1]=\frac{1}{k^{4({n^3-1})}}\left((k,L_{KT_1})(k,L_{KT_2})(k,L_{KT_3})(k,L_{KT_4})\right)^{3(n-1)}\notag\\
    \left((k,L_{KT_3},L_{KT_4})(k,L_{KT_2},L_{KT_1})(k,L_{KT_4},L_{KT_1})(k,L_{KT_2},L_{KT_3})(k,L_{KT_3},L_{KT_1})(k,L_{KT_2},L_{KT_4})\right)^{2(n-1)^2}\notag\\ \left((k,L_{KT_1},L_{KT_2},L_{KT_3})(k,L_{KT_1},L_{KT_2},L_{KT_4})(k,L_{KT_3},L_{KT_2},L_{KT_4})(k,L_{KT_1},L_{KT_3},L_{KT_4})\right)^{(n-1)^3}
    \label{eq:Z^4_n(2:1:1:1) key-ring summary}
\end{gather}
Combining these two, we obtain
\begin{equation}
    \mathcal{Z}^{(4)}_n[2:1:1:1]= \mathcal{Z}^{(4)}_{n \, {\rm part 1}}[2:1:1:1]\mathcal{Z}^{(4)}_{n \, {\rm part 2}}[2:1:1:1]
\end{equation}
which yields eq.~\eqref{eq369}.

Similarly, following eq.~\eqref{eq370} and by interchanging the labels $T_1, T_2, T_3, T_4$ among each other, we obtain the other similar type of two tripartite partition functions $\mathcal{Z}^{(3)}_n(K:T_1T_3:T_2T_4)$ and $\mathcal{Z}^{(3)}_n(K:T_1T_4:T_2T_3)$.\\
So, combining these, we can obtain part of the tripartite partition function
\begin{align}
\mathcal{Z}^{(3)}_{n \, {\rm part 1}}[2:2:1]& =\frac{1}{k^{3(n^2-1)}} 
    \left((k,L_{KT_1},L_{KT_2})(k,L_{KT_3},L_{KT_4})(k,L_{KT_1},L_{KT_3})(k,L_{KT_1},L_{KT_4}) \right. \notag \\
&\qquad \qquad \qquad  \left. (k,L_{KT_3},L_{KT_2})(k,L_{KT_4},L_{KT_2})\right)^{(n-1)}\notag\\
&   \qquad  \times (k,L_{KT_1},L_{KT_2},L_{KT_3},L_{KT_4})^{3(n-1)^2}
      \label{eq:Z^3_n(1:2:2) key-ring summary}
\end{align}
\\
Similarly, following eq.~\eqref{eq371} and by interchanging the labels $T_1, T_2, T_3, T_4$ among each other, we obtain the other similar type of eleven tripartite partition functions 
$\mathcal{Z}^{(3)}_n(KT_1:T_2T_4:T_3)$,
$\mathcal{Z}^{(3)}_n(KT_1:T_2:T_3T_4)$, $\mathcal{Z}^{(3)}_n(KT_2:T_1T_3:T_4)$, 
$\mathcal{Z}^{(3)}_n(KT_2:T_1T_4:T_3) $,
$\mathcal{Z}^{(3)}_n(KT_2:T_1:T_3T_4)$, 
$\mathcal{Z}^{(3)}_n(KT_3:T_1T_2:T_4)$, 
$\mathcal{Z}^{(3)}_n(KT_3:T_1T_4:T_2)$, 
$\mathcal{Z}^{(3)}_n(KT_3:T_1:T_2T_4)$,
$\mathcal{Z}^{(3)}_n(KT_4:T_1T_2:T_3)$, 
$\mathcal{Z}^{(3)}_n(KT_4:T_2:T_1T_3)$ and  $\mathcal{Z}^{(3)}_n(KT_4:T_1:T_2T_3)$.

So, combining all of these, we can obtain another part of the tripartite partition function
\begin{gather}
     \mathcal{Z}^{(3)}_{n \, {\rm part 2}}[2:2:1]=\frac{1}{k^{12(n^2-1)}}
     \left((k,L_{KT_1})(k,L_{KT_2})(k,L_{KT_3})(k,L_{KT_4})\right)^{3(n-1)}\notag\\
\hspace{-10mm} \left((k,L_{KT_1},L_{KT_2})(k,L_{KT_1},L_{KT_3})(k,L_{KT_1},L_{KT_4})(k,L_{KT_2},L_{KT_3})(k,L_{KT_2},L_{KT_4})(k,L_{KT_3},L_{KT_4})\right)^{2(n-1)}\notag\\
 \hspace{-1mm}    \left(
    (k,L_{KT_1},L_{KT_2},L_{KT_3})(k,L_{KT_2},L_{KT_3},L_{KT_4})(k,L_{KT_1},L_{KT_3},L_{KT_4})(k,L_{KT_1},L_{KT_2},L_{KT_4})\right)^{3(n-1)^2}
     \label{eq:Z^3_n(2:2:1) key-ring summary}
\end{gather}
Combining these two, we obtain
\begin{equation}
    \mathcal{Z}^{(3)}_n[2:2:1]= \mathcal{Z}^{(3)}_{n \, {\rm part 1}}[2:2:1]\mathcal{Z}^{(3)}_{n \, {\rm part 2}}[2:2:1]
\end{equation}
which yields eq.~\eqref{eq372}.

The tripartite partition function, which treats the key-ring $K$ as one party, $T_1$ as one party and $(T_2T_3T_4)$ as one party, (from eq.~\eqref{eq:q<N key partition})
\begin{align}
    & \hspace{-10mm} \mathcal{Z}^{(3)}_n(K:T_1:T_2T_3T_4)=\frac{1}{k^{n^2-1}}(k,L_{KT_1})^{(n-1)}
     \notag\\
     &(k,L_{KT_2},L_{KT_3},L_{KT_4})^{(n-1)}
     (k,L_{KT_1},L_{KT_2},L_{KT_3},L_{KT_4})^{(n-1)^2}
     \label{eq:Z^3_n(A:B:CDE) key-ring summary}
\end{align}
By interchanging the labels $T_1, T_2, T_3, T_4$ among each other, we obtain the other similar type of three tripartite partition functions $\mathcal{Z}^{(3)}_n(K:T_2:T_1T_3T_4), \mathcal{Z}^{(3)}_n(K:T_3:T_1T_2T_4),$ and $\mathcal{Z}^{(3)}_n(K:T_4:T_1T_2T_3)$ from eq.~\eqref{eq:Z^3_n(A:B:CDE) key-ring summary}.\\
So, combining all of them, we can obtain part of the tripartite partition function
\begin{gather}
    \mathcal{Z}^{(3)}_{n \, {\rm part 1} }[3:1:1]=\frac{1}{k^{4(n^2-1)}}
    \left((k,L_{KT_1})(k,L_{KT_2})(k,L_{KT_3})(k,L_{KT_4})\right)^{(n-1)}\notag\\
   \left( (k,L_{KT_1},L_{KT_2},L_{KT_3})(k,L_{KT_1},L_{KT_3},L_{KT_4})(k,L_{KT_2},L_{KT_3},L_{KT_4})(k,L_{KT_1},L_{KT_2},L_{KT_4})\right)^{(n-1)}\notag\\
   (k,L_{KT_1},L_{KT_2},L_{KT_3},L_{KT_4})^{4(n-1)^2}
    \label{eq:Z^3_n(1:1:3) key-ring summary}
\end{gather}
The tripartite partition function, which treats the $(KT_1T_2)$ as one party, $T_3$ as one party and $T_4$ as one party, (from eq.~\eqref{eq:q<N key partition})
\begin{gather}
     \mathcal{Z}^{(3)}_n(KT_1T_2:T_3:T_4)=\frac{1}{k^{n^2-1}} \left((k,L_{KT_3})(k,L_{KT_4})\right)^{(n-1)} (k,L_{KT_3},L_{KT_4})^{(n-1)^2}
    \label{eq:Z^3_n(ABC:D:E) key-ring summary}
\end{gather}
By interchanging the labels $T_1, T_2, T_3, T_4$ among each other, we obtain the other similar type of five tripartite partition functions 
$\mathcal{Z}^{(3)}_n(KT_1T_3:T_2:T_4)$,
 $\mathcal{Z}^{(3)}_n(KT_1T_4:T_2:T_3)$, 
$\mathcal{Z}^{(3)}_n(KT_2T_3:T_1:T_3T_4)$, 
$\mathcal{Z}^{(3)}_n(KT_2T_4:T_1:T_3)$
 and $\mathcal{Z}^{(3)}_n(KT_3T_4:T_1:T_2)$ from eq.~\eqref{eq:Z^3_n(ABC:D:E) key-ring summary}.
 
So, combining all of them, we can obtain another part of the tripartite partition function
\begin{align}
    \mathcal{Z}^{(3)}_{n \, {\rm part 2}}[3:1:1] &=\frac{1}{k^{6(n^2-1)}}
    \left((k,L_{KT_1})(k,L_{KT_2})(k,L_{KT_3})(k,L_{KT_4})\right)^{3(n-1)} \notag\\
  &  \left((k,L_{KT_1},L_{KT_2})(k,L_{KT_1},L_{KT_3})(k,L_{KT_1},L_{KT_4}) \right. \notag \\
& \left. \qquad (k,L_{KT_2},L_{KT_3})(k,L_{KT_2},L_{KT_4})(k,L_{KT_3},L_{KT_4})\right)^{(n-1)^2}
    \label{eq:Z^3_n(3:1:1) key-ring summary}
\end{align}
Now combining eq.~\eqref{eq:Z^3_n(1:1:3) key-ring summary} and \eqref{eq:Z^3_n(3:1:1) key-ring summary}, we obtain the tripartite partition function in eq.~\eqref{rest of the z}
\begin{align}\mathcal{Z}^{(3)}_n[3:1:1]& = \mathcal{Z}^{(3)}_{n \, {\rm part 1}}[3:1:1] \, \mathcal{Z}^{(3)}_{n \, {\rm part 2}}[3:1:1]   \notag \\
&=\frac{1}{k^{10(n^2-1)}}
    \left((k,L_{KT_1})(k,L_{KT_2})(k,L_{KT_3})(k,L_{KT_4})\right)^{4(n-1)}\notag\\
&     \left((k,L_{KT_1},L_{KT_2})(k,L_{KT_1},L_{KT_3})(k,L_{KT_1},L_{KT_4})(k,L_{KT_2},L_{KT_3}) \right. \notag \\
&\qquad \qquad \left. (k,L_{KT_2},L_{KT_4})(k,L_{KT_3},L_{KT_4})\right)^{(n-1)^2}\notag\\
  &  \left( (k,L_{KT_1},L_{KT_2},L_{KT_3})(k,L_{KT_1},L_{KT_3},L_{KT_4})(k,L_{KT_2},L_{KT_3},L_{KT_4}) \right. \notag \\
&\qquad \qquad \left. (k,L_{KT_1},L_{KT_2},L_{KT_4})\right)^{(n-1)}\notag\\
&  (k,L_{KT_1},L_{KT_2},L_{KT_3},L_{KT_4})^{4(n-1)^2} \,.
    \label{eq:Z^3_n(1:1:3) key-ring summary total}
\end{align}

The bipartite partition function, which treats $(KT_1)$ alone as one party and $(T_2T_3T_4)$ as one subsystem, (from eq.~\eqref{eq:q<N key partition})
\begin{gather}
    \mathcal{Z}^{(2)}_n(KT_1:T_2T_3T_4)=\frac{1}{k^{n-1}}(k,L_{KT_2},L_{KT_3},L_{KT_4})^{(n-1)}
    \label{eq:Z^2_n(AB:CDE) key-ring summary}
\end{gather}
By interchanging the labels $T_1, T_2, T_3, T_4$ among each other, we obtain the other similar type of three bipartite partition functions $\mathcal{Z}^{(2)}_n(KT_2:T_1T_3T_4), \mathcal{Z}^{(2)}_n(KT_3:T_1T_2T_4)$, and $\mathcal{Z}^{(2)}_n(KT_4:T_1T_2T_3)$ from eq.~\ref{eq:Z^2_n(AB:CDE) key-ring summary}. 

So, combining these, we can obtain part of the bipartite partition function
\begin{align}
\mathcal{Z}^{(2)}_{n \, {\rm part 1}}[3:2]& =\frac{1}{k^{4(n-1)}}    \left((k,L_{KT_1},L_{KT_2},L_{KT_3})(k,L_{KT_2},L_{KT_3},L_{KT_4}) \right. \notag \\
&\qquad \qquad \qquad \left. (k,L_{KT_1},L_{KT_3},L_{KT_4})(k,L_{KT_1},L_{KT_2},L_{KT_4})\right)^{(n-1)}
    \label{eq:Z^2_n(2:3) key-ring summary}
\end{align}
The bipartite partition function, which treats $(KT_1T_2)$ alone as one party and $(T_3T_4)$ as one subsystem, (from eq.~\eqref{eq:q<N key partition})
\begin{gather}
    \mathcal{Z}^{(2)}_n(KT_1T_2:T_3T_4)=\frac{1}{k^{n-1}}(k,L_{KT_3},L_{KT_4})^{(n-1)}
    \label{eq:Z^2_n(ABC:DE) key-ring summary}
\end{gather}

By interchanging the labels $T_1, T_2, T_3, T_4$ among each other, we obtain the other similar type of five bipartite partition functions 
$\mathcal{Z}^{(2)}_n(KT_1T_3:T_2T_4)$, 
$\mathcal{Z}^{(2)}_n(KT_1T_4:T_2T_3)$, 
$\mathcal{Z}^{(2)}_n(KT_2T_3:T_1T_4)$,
$\mathcal{Z}^{(2)}_n(KT_2T_4:T_1T_3)$ and 
$\mathcal{Z}^{(2)}_n(KT_3T_4:T_1T_2)$  from eq.~\eqref{eq:Z^2_n(ABC:DE) key-ring summary}.\\
So, combining all of them, we can obtain the bipartite partition function
\begin{align}
\mathcal{Z}^{(2)}_{n \, {\rm part 2}}[3:2]& =\frac{1}{k^{6(n-1)}}
    \left((k,L_{KT_1},L_{KT_2})(k,L_{KT_1},L_{KT_3})(k,L_{KT_1},L_{KT_4}) \right. \notag \\
&\qquad \qquad \qquad \left. (k,L_{KT_2},L_{KT_3})(k,L_{KT_2},L_{KT_4})(k,L_{KT_3},L_{KT_4})\right)^{(n-1)}
    \label{eq:Z^2_n(3:2) key-ring summary}
\end{align}
Now combining eq.~\eqref{eq:Z^2_n(2:3) key-ring summary} and \eqref{eq:Z^2_n(3:2) key-ring summary}, we obtain bipartite partition function in eq.~\eqref{rest of the z}
\begin{align}
\mathcal{Z}^{(2)}_n[3:2] & = \mathcal{Z}^{(2)}_{n \, {\rm part 1} }[3:2]\, 
\mathcal{Z}^{(2)}_{n \, {\rm part 2} }[3:2] \notag \\ 
&=\frac{1}{k^{10(n-1)}}   \left((k,L_{KT_1},L_{KT_2})(k,L_{KT_1},L_{KT_3})(k,L_{KT_1},L_{KT_4}) \right. \notag \\
&\qquad \qquad \left. (k,L_{KT_2},L_{KT_3})(k,L_{KT_2},L_{KT_4})(k,L_{KT_3},L_{KT_4})\right)^{(n-1)}\notag\\
  & \quad \left((k,L_{KT_1},L_{KT_2},L_{KT_3})(k,L_{KT_2},L_{KT_3},L_{KT_4}) \right. \notag \\
&\qquad \qquad \left. (k,L_{KT_1},L_{KT_3},L_{KT_4})(k,L_{KT_1},L_{KT_2},L_{KT_4})\right)^{(n-1)}
    \label{eq:Z^2_n(3:2) key-ring summary total}
\end{align}

Finally, another  bipartite partition function, which treats the key-ring $K$ alone as one party and $(T_1T_2T_3T_4)$ as one subsystem, (from eq.~\eqref{eq:q<N key partition})
\begin{gather}
    \mathcal{Z}^{(2)}_n(K:T_1T_2T_3T_4)=\frac{1}{k^{n-1}}(k,L_{KT_1},L_{KT_2},L_{KT_3},L_{KT_4})^{(n-1)}
    \label{eq:Z^2_n(A:BCDE) key-ring summary}
\end{gather}

The bipartite partition function, which treats $(KT_1T_2T_3)$ alone as one party and $T_4$ as one subsystem, (from eq.~\eqref{eq:q<N key partition})
\begin{gather}
    \mathcal{Z}^{(2)}_n(KT_1T_2T_3:T_4)=\frac{1}{k^{n-1}}(k,L_{KT_4})^{(n-1)}
    \label{eq:Z^2_n(ABCD:E) key-ring summary}
\end{gather}
By interchanging the labels $T_1, T_2, T_3, T_4$ among each other, we obtain the other similar type of three bipartite partition functions $\mathcal{Z}^{(2)}_n(KT_1T_2T_4:T_3), \mathcal{Z}^{(2)}_n(KT_1T_3T_4:T_2)$ and $\mathcal{Z}^{(2)}_n(KT_2T_3T_4:T_1)$ from eq.~\eqref{eq:Z^2_n(ABCD:E) key-ring summary}.
\begin{gather}
    \mathcal{Z}^{(2)}_{n \, {\rm part}}[4:1]=\frac{1}{k^{4(n-1)}}\left((k,L_{KT_1})(k,L_{KT_2})(k,L_{KT_3})(k,L_{KT_4})\right)^{(n-1)}
    \label{eq:Z^2_n(4:1) key-ring summary}
\end{gather}

Now combining eq.~\eqref{eq:Z^2_n(A:BCDE) key-ring summary} and \eqref{eq:Z^2_n(4:1) key-ring summary}, we obtain bipartite partition function in eq.~\eqref{rest of the z} as 
\begin{align}    
\mathcal{Z}^{(2)}_n[4:1]
&=\mathcal{Z}^{(2)}_n(K:T_1T_2T_3T_4) \, \mathcal{Z}^{(2)}_{n \, {\rm part}}[4:1]    
   \notag \\
&=\frac{1}{k^{5(n-1)}}\left((k,L_{KT_1})(k,L_{KT_2})(k,L_{KT_3})(k,L_{KT_4})\right)^{(n-1)}\notag\\    & \qquad (k,L_{KT_1},L_{KT_2},L_{KT_3},L_{KT_4})^{(n-1)}
    \label{eq:Z^2_n(4:1) key-ring summary total}
\end{align}
\item Genuine Rényi multi-entropy:
\begin{align}
&\left.GM^{(5)}_n\right|_{b=0}
={}-\frac{1}{n^3}\log \Bigg[
\left(\frac{1}{k}\right)^{(n^3+n^2+n+1)-\frac{5n}{2}(n^2+n+1)+\frac{3n^2}{2}(n+1)+\frac{n^2}{2}(n+1)-\frac{n^3}{2}}
\notag\\
&\left((k,L_{KT_1})(k,L_{KT_2})(k,L_{KT_3})(k,L_{KT_4})\right)^{1-\frac{3n}{4}-\frac{3n}{4}+\frac{3n^2}{10}+\frac{3n^2}{20}+\frac{n^2}{20}}
\notag\\
&\left((k,L_{KT_1},L_{KT_2})(k,L_{KT_1},L_{KT_3})(k,L_{KT_1},L_{KT_4})
(k,L_{KT_2},L_{KT_3}) \right. \notag \\
&\left. \qquad \qquad  (k,L_{KT_2},L_{KT_4})(k,L_{KT_3},L_{KT_4})\right)^{(n-1)-\frac{n^2}{4}-\frac{n}{2}(n-1)+\frac{n^2}{10}+\frac{n^2}{5}+\frac{n^2}{20}(n-1)-\frac{n^3}{20}}
\notag\\
&\left((k,L_{KT_1},L_{KT_2},L_{KT_3})(k,L_{KT_4},L_{KT_1},L_{KT_2})
(k,L_{KT_3},L_{KT_4},L_{KT_1}) \right. \notag \\ & \qquad \qquad \left. (k,L_{KT_2},L_{KT_3},L_{KT_4})\right)^{(n-1)^2-\frac{3n}{4}(n-1)-\frac{n}{4}(n-1)^2+\frac{3n^2}{10}(n-1)+\frac{n^2}{20}-\frac{n^3}{20}}
\notag\\
& (k,L_{KT_1},L_{KT_2},L_{KT_3},L_{KT_4})^{(n-1)^3-\frac{3n}{2}(n-1)^2+\frac{3n^2}{10}(n-1)+\frac{n^2}{5}(n-1)}
\Bigg] \,.
\label{GM^5 details}
\end{align}

\item \texorpdfstring{$\partial GM^{5}_n$}{partial GM5}:
\begin{align}
\partial_bGM^{(5)}_n
=&-\frac{1}{n}\log \Bigg[ 
\left(\frac{1}{k}\right)^{6(n+1)-8(n+1)-2n+5n} 
\left((k,L_{KT_1})(k,L_{KT_2})(k,L_{KT_3})(k,L_{KT_4})\right)^{\frac{6}{5}-\frac{16}{5}+n} 
\notag\\
&\left((k,L_{KT_1},L_{KT_2})(k,L_{KT_1},L_{KT_3})(k,L_{KT_1},L_{KT_4})
(k,L_{KT_2},L_{KT_3}) \right. \notag \\
&\qquad \qquad \left. (k,L_{KT_2},L_{KT_4})(k,L_{KT_3},L_{KT_4})\right)^{\frac{6}{5}-\frac{4}{5}(n-1)-\frac{n}{5}}
\notag\\
&\left((k,L_{KT_1},L_{KT_2},L_{KT_3})(k,L_{KT_4},L_{KT_1},L_{KT_2})
(k,L_{KT_3},L_{KT_4},L_{KT_1}) \right. \notag \\
&\qquad \qquad \left. (k,L_{KT_2},L_{KT_3},L_{KT_4})\right)^{\frac{6}{5}(n-1)-\frac{4}{5}-\frac{n}{5}}
\notag\\
&(k,L_{KT_1},L_{KT_2},L_{KT_3},L_{KT_4})^{\frac{6(n-1)}{5}-\frac{16(n-1)}{5}+n}
\Bigg] \,.
\label{partial GM^5 details}
\end{align}

\end{itemize}

\section{Equivalence relations for linking-number configurations}
\label{app:linking-equivalence}
In this appendix, we explain the equivalence relation used to classify the linking-number configurations in the numerical analysis. We identify configurations related by transformations that leave the multi-entropies, and hence the genuine multi-entropy, invariant. In this way, linking-number configurations related by these transformations are grouped into the same equivalence class, since they give the same values of the multi-entropies and the genuine multi-entropy.

To make this equivalence relation explicit, we consider its action on the linking-number data. The link state \eqref{eq:U1-link-state} is labelled by the pairwise linking numbers
\begin{equation}
L_{ij}=L_{ji},\qquad L_{ii}=0,\qquad L_{ij}\in\mathbb{Z}_k,
\end{equation}
where $i,j\in\{K,T_1,T_2,\ldots,T_{N-1}\}$. Thus, there are $k^{\binom{N}{2}}$ labelled linking-number configurations in the $N$-link case.

\paragraph{$S_N$ subsystem permutation}
First, a permutation $\pi\in S_N$ of the $N$ subsystems acts on the linking numbers as
\begin{equation}
L_{ij}\longrightarrow L_{\pi(i)\pi(j)}.
\end{equation}
This transformation simply relabels the subsystems and therefore does not change the underlying entanglement structure, apart from the corresponding relabeling of subsystem labels.

\paragraph{Local unit scaling}
Second, we consider independent rescalings of the basis label in each subsystem. For each subsystem $i$, we choose
\begin{equation}
u_i\in\mathbb{Z}_k^\times,\qquad \mathbb{Z}_k^\times\equiv\left\{u\in\mathbb{Z}_k\,\middle|\,\gcd(u,k)=1\right\}.
\end{equation}
Equivalently, $u\in\mathbb{Z}_k^\times$ if and only if there exists an inverse $u^{-1}\in\mathbb{Z}_k$ satisfying
\begin{equation}
u u^{-1}\equiv 1\pmod{k}.
\end{equation}
Here, an element of $\mathbb{Z}_k^\times$ is called a unit, or an invertible element. For example, $\mathbb{Z}_4^\times=\{1,3\}$ and $\mathbb{Z}_6^\times=\{1,5\}$, while for prime $k$, every nonzero element is a unit, so that $\mathbb{Z}_k^\times=\{1,2,\ldots,k-1\}$.

For each $u_i\in\mathbb{Z}_k^\times$, we define a local basis transformation by
\begin{equation}
U_i(u_i)\ket{\alpha_i}=\ket{u_i^{-1}\alpha_i},
\end{equation}
where all labels are understood modulo $k$. Since $u_i$ is invertible modulo $k$, the map $\alpha_i\mapsto u_i^{-1}\alpha_i$ is a bijection of $\mathbb{Z}_k$. It therefore only permutes the $k$ basis states of subsystem $i$ and defines a unitary transformation. Acting independently on the $N$ subsystems with $\bigotimes_i U_i(u_i)$ maps the linking numbers as
\begin{equation}
L_{ij}\longrightarrow u_i u_j L_{ij}\pmod{k}.
\end{equation}
Indeed, after the change of variables $\beta_i=u_i^{-1}\alpha_i$, one has $\alpha_i=u_i\beta_i$, so that the phase $L_{ij}\alpha_i\alpha_j$ is mapped to $u_i u_j L_{ij}\beta_i\beta_j$. Thus, linking-number configurations related by such local unit scalings describe states related by local unitary transformations and have the same multipartite entanglement structure.

Note that the restriction to $u_i\in\mathbb{Z}_k^\times$ is essential. If $u_i$ is not a unit, multiplication by $u_i$ modulo $k$ is not one-to-one and therefore does not define a permutation of the local basis states. Such a transformation is not an invertible local basis transformation and is not included in the equivalence relation.

\paragraph{Global complex conjugation}
Let $\mathsf{K}$ denote complex conjugation in the computational basis. Writing the link state \eqref{eq:U1-link-state} explicitly as $\ket{\mathcal{L}_L}$ to indicate its dependence on the linking matrix $L$, we have
\begin{equation}
\mathsf{K}\ket{\mathcal{L}_L}=\frac{1}{k^{N/2}}\sum_{\alpha_1,\ldots,\alpha_N=0}^{k-1}\exp\left[-\frac{2\pi i}{k}\sum_{i<j}L_{ij}\alpha_i\alpha_j\right]\ket{\alpha_1,\ldots,\alpha_N}=\ket{\mathcal{L}_{-L}}.
\end{equation}
Thus, global complex conjugation acts on the linking numbers as
\begin{equation}
L_{ij}\longrightarrow -L_{ij}\pmod{k}.
\end{equation}

To see why the multi-entropies are invariant under this transformation, consider the replica partition function
\begin{align}
\mathcal{Z}_n^{(\mathtt{q})}(L)=\bra{\mathcal{L}_L}^{\otimes n^{\mathtt{q}-1}}\Sigma_1(g_1)\Sigma_2(g_2)\cdots\Sigma_{\mathtt{q}}(g_{\mathtt{q}})\ket{\mathcal{L}_L}^{\otimes n^{\mathtt{q}-1}}.
\end{align}
Since the permutation operators have real matrix elements in the computational basis, namely $\Sigma_i(g_i)^*=\Sigma_i(g_i)$, global complex conjugation of the link state gives\footnote{For a real computational basis, $(\bra{\psi}A\ket{\phi})^*=\bra{\psi^*}A^*\ket{\phi^*}=\bra{\phi}A^\dagger\ket{\psi}$. Hence, if $A^*=A$, complex conjugating the scalar is equivalent to complex conjugating both states while leaving $A$ unchanged.}
\begin{equation}
\mathcal{Z}_n^{(\mathtt{q})}(-L)=\left[\mathcal{Z}_n^{(\mathtt{q})}(L)\right]^*.
\end{equation}

We next show that the replica partition function itself is real. Taking the complex conjugate of the replica partition function gives
\begin{align}
\left[\mathcal{Z}_n^{(\mathtt{q})}(L)\right]^*=\bra{\mathcal{L}_L}^{\otimes n^{\mathtt{q}-1}}\left[\Sigma_1(g_1)\Sigma_2(g_2)\cdots\Sigma_{\mathtt{q}}(g_{\mathtt{q}})\right]^\dagger\ket{\mathcal{L}_L}^{\otimes n^{\mathtt{q}-1}}.
\end{align}
Since each $\Sigma_{\mathtt{k}}(g_{\mathtt{k}})$ is a unitary permutation operator,
\begin{equation}
\Sigma_{\mathtt{k}}(g_{\mathtt{k}})^\dagger=\Sigma_{\mathtt{k}}(g_{\mathtt{k}}^{-1}).
\end{equation}
Moreover, the twist operators associated with different subsystems act on different subsystem factors and therefore commute. Hence,
\begin{align}
\left[\Sigma_1(g_1)\Sigma_2(g_2)\cdots\Sigma_{\mathtt{q}}(g_{\mathtt{q}})\right]^\dagger=\Sigma_1(g_1^{-1})\Sigma_2(g_2^{-1})\cdots\Sigma_{\mathtt{q}}(g_{\mathtt{q}}^{-1}).
\end{align}

We now use the explicit definition of the replica permutations in \eqref{gkdefinition}. Recall that the replicas are labelled by
\begin{equation}
(x_1,\ldots,x_{\mathtt{q}-1}),\qquad x_{\mathtt{k}}=1,\ldots,n,
\end{equation}
with periodic identification, and that $g_{\mathtt{k}}$ acts as a unit translation in the $x_{\mathtt{k}}$ direction for $1\leq\mathtt{k}\leq\mathtt{q}-1$. Consider the simultaneous reversal of all replica coordinates,
\begin{equation}
r:(x_1,\ldots,x_{\mathtt{q}-1})\longmapsto(n+1-x_1,\ldots,n+1-x_{\mathtt{q}-1}).
\end{equation}
This is an involution, $r^{-1}=r$. Since $g_{\mathtt{k}}$ shifts $x_{\mathtt{k}}$ by one unit, conjugation by $r$ reverses the direction of the shift,
\begin{equation}
r g_{\mathtt{k}} r^{-1}=g_{\mathtt{k}}^{-1},\qquad 1\leq\mathtt{k}\leq\mathtt{q}-1.
\end{equation}
For the last subsystem, $g_{\mathtt{q}}=e$, and therefore
\begin{equation}
r g_{\mathtt{q}} r^{-1}=e=g_{\mathtt{q}}^{-1}.
\end{equation}
Thus, the same replica relabeling $r$ simultaneously maps all the permutations $g_{\mathtt{k}}$ to their inverses.

Let $\mathcal{R}$ denote the permutation operator that implements this simultaneous relabeling of the replicas. It follows that
\begin{equation}
\mathcal{R}\Sigma_{\mathtt{k}}(g_{\mathtt{k}})\mathcal{R}^\dagger=\Sigma_{\mathtt{k}}(g_{\mathtt{k}}^{-1})
\end{equation}
for every $\mathtt{k}=1,\ldots,\mathtt{q}$. Consequently,
\begin{align}
\left[\Sigma_1(g_1)\Sigma_2(g_2)\cdots\Sigma_{\mathtt{q}}(g_{\mathtt{q}})\right]^\dagger=\mathcal{R}\Sigma_1(g_1)\Sigma_2(g_2)\cdots\Sigma_{\mathtt{q}}(g_{\mathtt{q}})\mathcal{R}^\dagger.
\end{align}

Since the replicated state consists of $n^{\mathtt{q}-1}$ identical copies of the same link state, a simultaneous permutation of the replica labels leaves it invariant,\footnote{More explicitly, $\ket{\mathcal{L}_L}^{\otimes n^{\mathtt{q}-1}}=\bigotimes_{x_1=1}^{n}\cdots\bigotimes_{x_{\mathtt{q}-1}=1}^{n}\ket{\mathcal{L}_L}_{(x_1,\ldots,x_{\mathtt{q}-1})}$. Since $r$ is a bijection of the replica lattice, the simultaneous relabeling $(x_1,\ldots,x_{\mathtt{q}-1})\mapsto(n+1-x_1,\ldots,n+1-x_{\mathtt{q}-1})$ leaves this tensor product unchanged.}
\begin{equation}
\mathcal{R}\ket{\mathcal{L}_L}^{\otimes n^{\mathtt{q}-1}}=\ket{\mathcal{L}_L}^{\otimes n^{\mathtt{q}-1}}.
\end{equation}
Therefore,
\begin{equation}
\begin{aligned}
\left[\mathcal{Z}_n^{(\mathtt{q})}(L)\right]^*&=\bra{\mathcal{L}_L}^{\otimes n^{\mathtt{q}-1}}\mathcal{R}\Sigma_1(g_1)\Sigma_2(g_2)\cdots\Sigma_{\mathtt{q}}(g_{\mathtt{q}})\mathcal{R}^\dagger\ket{\mathcal{L}_L}^{\otimes n^{\mathtt{q}-1}} \\
&=\mathcal{Z}_n^{(\mathtt{q})}(L).
\end{aligned}
\end{equation}
Combining this result with the transformation under global complex conjugation, we obtain
\begin{equation}
\mathcal{Z}_n^{(\mathtt{q})}(-L)=\mathcal{Z}_n^{(\mathtt{q})}(L).
\end{equation}
Hence, the multi-entropies are invariant under global complex conjugation. Since the genuine multi-entropy is constructed as a real linear combination of these multi-entropies, it is invariant as well.

\paragraph{Combining the three transformations}
Combining the transformations above, we regard two linking-number sets $L=(L_{ij})$ and $L'=(L'_{ij})$ as equivalent, $L\sim L'$, if there exist a permutation $\pi\in S_N$, units $u_i\in\mathbb{Z}_k^\times$, and $\epsilon=\pm1$ such that
\begin{equation}
L'_{ij}\equiv\epsilon\,u_i u_j L_{\pi(i)\pi(j)}\pmod{k}.
\label{eq:linking-equivalence}
\end{equation}
These transformations correspond to subsystem relabelings, local basis transformations, and global complex conjugation, respectively, and do not change the multi-entropies, apart from the corresponding relabeling of subsystem labels.

For each equivalence class, we choose one representative in the numerical tables. In the four-link case, the multiplicity assigned to a representative is the number of labelled linking-number configurations in its equivalence class. Accordingly, the sum of the multiplicities over all inequivalent representatives equals the total number $k^6$ of labelled configurations.

For $k=2$, the local unit scalings and global complex conjugation act trivially, since
\begin{equation}
\mathbb{Z}_2^\times=\{1\},\qquad -1\equiv1\pmod{2}.
\end{equation}
Hence, in this case the nontrivial equivalence relation reduces to the $S_N$ permutations of the subsystems, or to $S_4$ in the four-link case considered in the numerics.

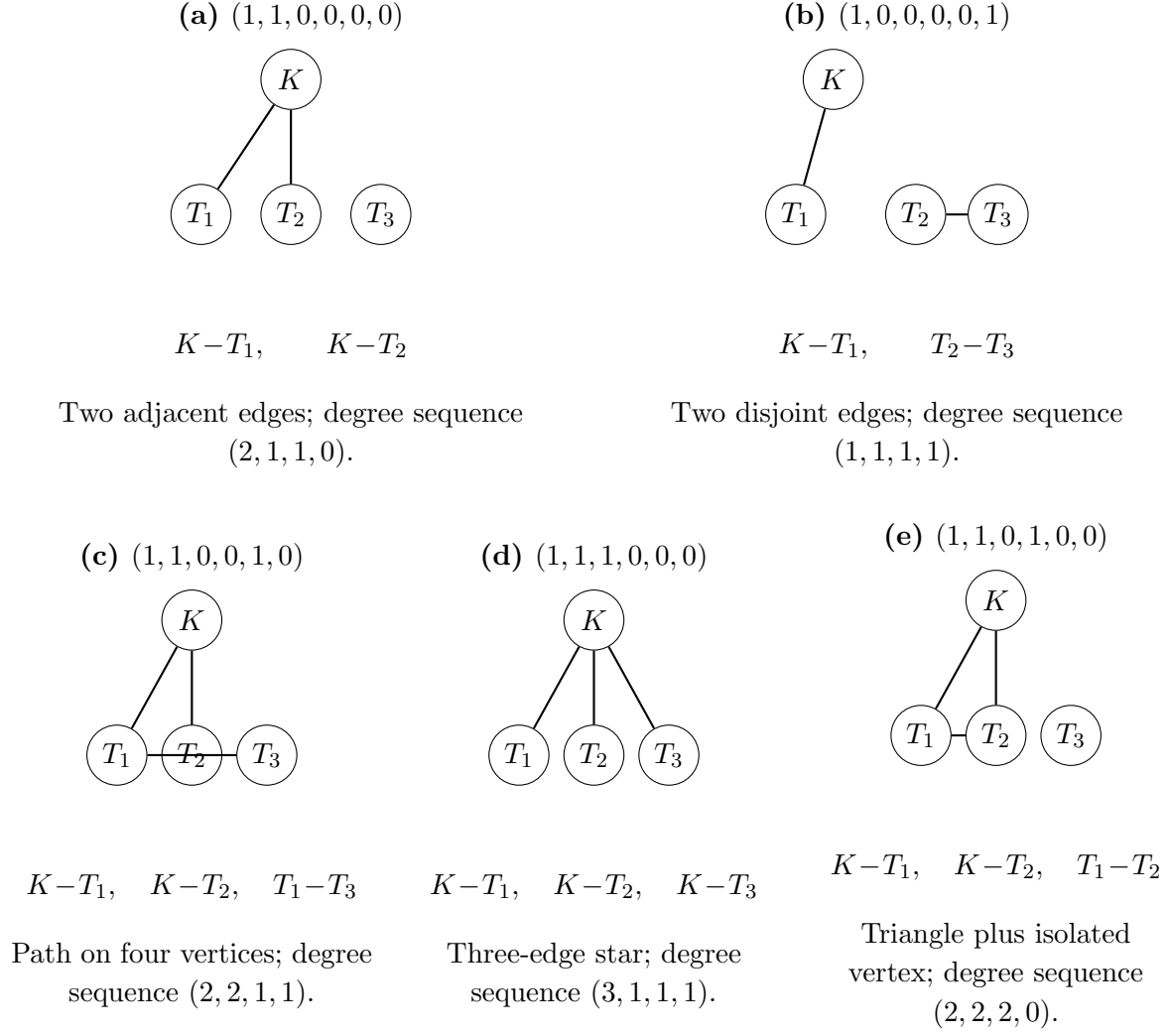
\begin{figure}[t]
\centering

\begin{minipage}{0.48\textwidth}
\centering
\textbf{(a) $(1,1,0,0,0,0)$}

\vspace{2mm}

\begin{tikzpicture}[
    vertex/.style={circle,draw,minimum size=8mm,inner sep=0pt},
    edge/.style={thick}
]
\node[vertex] (K) at (0,1.8) {$K$};
\node[vertex] (T1) at (-1.2,0) {$T_1$};
\node[vertex] (T2) at (0,0) {$T_2$};
\node[vertex] (T3) at (1.2,0) {$T_3$};

\draw[edge] (K)--(T1);
\draw[edge] (K)--(T2);
\end{tikzpicture}

\[
K\!-\!T_1,\qquad K\!-\!T_2
\]

Two adjacent edges; degree sequence $(2,1,1,0)$.
\end{minipage}
\hfill
\begin{minipage}{0.48\textwidth}
\centering
\textbf{(b) $(1,0,0,0,0,1)$}

\vspace{2mm}

\begin{tikzpicture}[
    vertex/.style={circle,draw,minimum size=8mm,inner sep=0pt},
    edge/.style={thick}
]
\node[vertex] (K) at (-0.7,1.8) {$K$};
\node[vertex] (T1) at (-1.2,0) {$T_1$};
\node[vertex] (T2) at (0.4,0) {$T_2$};
\node[vertex] (T3) at (1.5,0) {$T_3$};

\draw[edge] (K)--(T1);
\draw[edge] (T2)--(T3);
\end{tikzpicture}

\[
K\!-\!T_1,\qquad T_2\!-\!T_3
\]

Two disjoint edges; degree sequence $(1,1,1,1)$.
\end{minipage}

\vspace{7mm}

\begin{minipage}{0.31\textwidth}
\centering
\textbf{(c) $(1,1,0,0,1,0)$}

\vspace{2mm}

\begin{tikzpicture}[
    vertex/.style={circle,draw,minimum size=8mm,inner sep=0pt},
    edge/.style={thick}
]
\node[vertex] (K) at (0,1.8) {$K$};
\node[vertex] (T1) at (-1.0,0) {$T_1$};
\node[vertex] (T2) at (0,0) {$T_2$};
\node[vertex] (T3) at (1.0,0) {$T_3$};

\draw[edge] (K)--(T1);
\draw[edge] (K)--(T2);
\draw[edge] (T1)--(T3);
\end{tikzpicture}

\[
K\!-\!T_1,\quad K\!-\!T_2,\quad T_1\!-\!T_3
\]

Path on four vertices; degree sequence $(2,2,1,1)$.
\end{minipage}
\hfill
\begin{minipage}{0.31\textwidth}
\centering
\textbf{(d) $(1,1,1,0,0,0)$}

\vspace{2mm}

\begin{tikzpicture}[
    vertex/.style={circle,draw,minimum size=8mm,inner sep=0pt},
    edge/.style={thick}
]
\node[vertex] (K) at (0,1.8) {$K$};
\node[vertex] (T1) at (-1.0,0) {$T_1$};
\node[vertex] (T2) at (0,0) {$T_2$};
\node[vertex] (T3) at (1.0,0) {$T_3$};

\draw[edge] (K)--(T1);
\draw[edge] (K)--(T2);
\draw[edge] (K)--(T3);
\end{tikzpicture}

\[
K\!-\!T_1,\quad K\!-\!T_2,\quad K\!-\!T_3
\]

Three-edge star; degree sequence $(3,1,1,1)$.
\end{minipage}
\hfill
\begin{minipage}{0.31\textwidth}
\centering
\textbf{(e) $(1,1,0,1,0,0)$}

\vspace{2mm}

\begin{tikzpicture}[
    vertex/.style={circle,draw,minimum size=8mm,inner sep=0pt},
    edge/.style={thick}
]
\node[vertex] (K) at (0,1.8) {$K$};
\node[vertex] (T1) at (-1.0,0) {$T_1$};
\node[vertex] (T2) at (0,0) {$T_2$};
\node[vertex] (T3) at (1.0,0) {$T_3$};

\draw[edge] (K)--(T1);
\draw[edge] (K)--(T2);
\draw[edge] (T1)--(T2);
\end{tikzpicture}

\[
K\!-\!T_1,\quad K\!-\!T_2,\quad T_1\!-\!T_2
\]

Triangle plus isolated vertex; degree sequence $(2,2,2,0)$.
\end{minipage}
\caption{
Examples of inequivalent linking-number configurations, with entries ordered as
$(L_{KT_1},L_{KT_2},L_{KT_3},L_{T_1T_2},L_{T_1T_3},L_{T_2T_3})$.
Panels (a),(b) show the two inequivalent graph structures with two nonzero
links, while panels (c)--(e) show the three inequivalent structures with
three nonzero links. Their distinct incidence structures prevent them from
being related by subsystem permutations.
}
\label{fig:linking-equivalence-examples}
\end{figure}

It is useful to illustrate this equivalence relation explicitly in the
four-link case. Configurations with the same number of nonzero linking
numbers need not be equivalent, since subsystem permutations preserve
the incidence structure of the corresponding graph.
Figure~\ref{fig:linking-equivalence-examples} shows representative
two- and three-link examples, illustrating how distinct incidence structures among the linking numbers prevent configurations from being related by subsystem permutations.


\section{Numerical result for $k=4$}
\label{app:Numerical-table-k=4}
In this appendix, we present detailed numerical data for the inequivalent linking-number representatives.

\begin{longtable}{c c c c}
\toprule
$\left(L_{KT_1},L_{KT_2},L_{KT_3},L_{T_1T_2},L_{T_1T_3},L_{T_2T_3}\right)$  & multiplicity & $n=4$ & $n=5$ \\
\midrule
\endfirsthead
\toprule
$\left(L_{KT_1},L_{KT_2},L_{KT_3},L_{T_1T_2},L_{T_1T_3},L_{T_2T_3}\right)$  & multiplicity & $n=4$ & $n=5$ \\
\midrule
\endhead
$(0,0,0,0,0,0)$ & $1$ & $0$ & $0$ \\
$(1,0,0,0,0,0)$ & $12$ & $0$ & $0$ \\
$(0,0,0,0,0,2)$ & $6$ & $0$ & $0$ \\
$(1,1,0,0,0,0)$ & $48$ & $0$ & $0$ \\
$(1,0,0,0,2,0)$ & $48$ & $0$ & $0$ \\
$(0,0,0,0,2,2)$ & $12$ & $0$ & $0$ \\
$(1,0,0,0,0,1)$ & $12$ & $0$ & $0$ \\
$(1,0,0,0,0,2)$ & $12$ & $0$ & $0$ \\
$(1,1,0,0,1,0)$ & $96$ & $\frac{1}{24}$ & $\frac{2}{25}$ \\
$(1,1,0,0,0,2)$ & $96$ & $\frac{1}{48}$ & $\frac{1}{25}$ \\
$(1,0,0,0,2,1)$ & $48$ & $\frac{1}{48}$ & $\frac{1}{25}$ \\
$(1,0,0,0,2,2)$ & $48$ & $\frac{1}{48}$ & $\frac{1}{25}$ \\
$(0,0,2,2,0,0)$ & $3$ & $0$ & $0$ \\
$(1,0,2,2,0,0)$ & $24$ & $0$ & $0$ \\
$(0,0,2,2,0,2)$ & $12$ & $\frac{1}{48}$ & $\frac{1}{25}$ \\
$(1,1,1,0,0,0)$ & $32$ & $0$ & $0$ \\
$(1,1,2,0,0,0)$ & $48$ & $0$ & $0$ \\
$(1,0,0,2,2,0)$ & $24$ & $0$ & $0$ \\
$(1,1,0,1,0,0)$ & $32$ & $0$ & $0$ \\
$(1,1,1,1,0,0)$ & $192$ & $\frac{1}{24}$ & $\frac{2}{25}$ \\
$(1,1,0,1,0,2)$ & $96$ & $\frac{1}{48}$ & $\frac{1}{25}$ \\
$(1,1,0,2,0,0)$ & $48$ & $0$ & $0$ \\
$(1,1,0,2,1,0)$ & $192$ & $\frac{1}{24}$ & $\frac{2}{25}$ \\
$(1,1,0,2,0,2)$ & $96$ & $\frac{1}{48}$ & $\frac{1}{25}$ \\
$(1,1,1,0,0,2)$ & $96$ & $\frac{1}{48}$ & $\frac{1}{25}$ \\
$(1,1,2,2,0,0)$ & $48$ & $0$ & $0$ \\
$(1,0,2,0,2,0)$ & $24$ & $0$ & $0$ \\
$(1,1,2,0,0,2)$ & $96$ & $\frac{1}{48}$ & $\frac{1}{25}$ \\
$(1,0,2,2,2,0)$ & $48$ & $0$ & $0$ \\
$(0,0,2,0,2,2)$ & $4$ & $0$ & $0$ \\
$(1,0,0,2,2,1)$ & $48$ & $\frac{1}{48}$ & $\frac{1}{25}$ \\
$(1,0,2,0,2,2)$ & $24$ & $\frac{1}{48}$ & $\frac{1}{25}$ \\
$(0,0,0,2,2,2)$ & $4$ & $0$ & $0$ \\
$(1,0,0,2,2,2)$ & $24$ & $\frac{1}{48}$ & $\frac{1}{25}$ \\
$(0,0,2,2,2,2)$ & $12$ & $\frac{1}{48}$ & $\frac{1}{25}$ \\
$(1,1,0,0,1,1)$ & $24$ & $\frac{1}{24}$ & $\frac{2}{25}$ \\
$(1,1,0,0,1,2)$ & $96$ & $-\frac{3}{32}$ & $\frac{1}{25}$ \\
$(1,1,0,0,1,3)$ & $24$ & $-\frac{3}{32}$ & $\frac{1}{25}$ \\
$(1,1,0,0,2,2)$ & $48$ & $\frac{1}{48}$ & $\frac{1}{25}$ \\
$(1,0,2,2,0,1)$ & $24$ & $\frac{1}{24}$ & $\frac{2}{25}$ \\
$(1,0,2,2,0,2)$ & $24$ & $-\frac{1}{24}$ & $\frac{1}{25}$ \\
$(1,1,1,1,1,0)$ & $96$ & $\frac{1}{24}$ & $\frac{2}{25}$ \\
$(1,1,1,1,0,2)$ & $384$ & $-\frac{3}{32}$ & $\frac{1}{25}$ \\
$(1,1,1,1,0,3)$ & $96$ & $-\frac{3}{32}$ & $\frac{1}{25}$ \\
$(1,1,1,0,2,2)$ & $96$ & $\frac{1}{48}$ & $\frac{1}{25}$ \\
$(1,1,2,2,1,0)$ & $96$ & $\frac{1}{24}$ & $\frac{2}{25}$ \\
$(1,1,0,1,2,2)$ & $96$ & $\frac{1}{48}$ & $\frac{1}{25}$ \\
$(1,1,2,2,0,2)$ & $96$ & $\frac{1}{48}$ & $\frac{1}{25}$ \\
$(1,1,0,2,1,1)$ & $48$ & $\frac{1}{24}$ & $\frac{2}{25}$ \\
$(1,1,0,2,1,2)$ & $192$ & $-\frac{3}{32}$ & $\frac{1}{25}$ \\
$(1,1,0,2,1,3)$ & $48$ & $-\frac{3}{32}$ & $\frac{1}{25}$ \\
$(1,1,2,0,2,2)$ & $48$ & $\frac{1}{48}$ & $\frac{1}{25}$ \\
$(1,0,2,2,2,1)$ & $48$ & $\frac{1}{24}$ & $\frac{2}{25}$ \\
$(1,1,0,2,2,2)$ & $48$ & $\frac{1}{48}$ & $\frac{1}{25}$ \\
$(1,0,2,2,2,2)$ & $48$ & $-\frac{1}{24}$ & $\frac{1}{25}$ \\
$(0,2,2,2,2,0)$ & $3$ & $\frac{1}{48}$ & $\frac{1}{25}$ \\
$(1,2,2,2,2,0)$ & $12$ & $0$ & $0$ \\
$(0,2,2,2,2,2)$ & $6$ & $\frac{1}{48}$ & $\frac{1}{25}$ \\
$(1,1,1,1,1,1)$ & $16$ & $0$ & $0$ \\
$(1,1,1,1,1,2)$ & $96$ & $\frac{1}{24}$ & $\frac{2}{25}$ \\
$(1,1,1,1,1,3)$ & $48$ & $\frac{1}{48}$ & $\frac{1}{25}$ \\
$(1,1,1,1,2,2)$ & $192$ & $\frac{1}{24}$ & $\frac{2}{25}$ \\
$(1,1,1,1,2,3)$ & $96$ & $-\frac{3}{32}$ & $\frac{1}{25}$ \\
$(1,1,2,2,1,1)$ & $24$ & $\frac{1}{24}$ & $\frac{2}{25}$ \\
$(1,1,2,2,1,2)$ & $96$ & $-\frac{3}{32}$ & $\frac{1}{25}$ \\
$(1,1,2,2,1,3)$ & $24$ & $-\frac{3}{32}$ & $\frac{1}{25}$ \\
$(1,1,2,1,2,2)$ & $32$ & $0$ & $0$ \\
$(1,1,1,2,2,2)$ & $32$ & $0$ & $0$ \\
$(1,1,2,2,2,2)$ & $48$ & $\frac{1}{48}$ & $\frac{1}{25}$ \\
$(1,2,2,2,2,1)$ & $12$ & $\frac{1}{48}$ & $\frac{1}{25}$ \\
$(1,2,2,2,2,2)$ & $12$ & $\frac{1}{48}$ & $\frac{1}{25}$ \\
$(2,2,2,2,2,2)$ & $1$ & $0$ & $0$ \\
\midrule
\multicolumn{2}{c}{Violation rate} & $85.77\%$ & $85.77\%$ \\
\bottomrule
\caption{
Numerical results for all inequivalent linking-number representatives at $k=4$.
The entries show $\Delta_n/\log 4$. For $n=2$ and $n=3$, $\Delta_n=0$ for all representatives and these columns are therefore omitted. For each equivalence class, the representative is chosen with entries equal to $1$ placed as far to the left as possible. The multiplicities sum to $4^6=4096$.
}
\label{tab:k4-all-linkings-full}
\end{longtable}


\bibliographystyle{JHEP}

\bibliography{reference}

@article{Schlingemann:2001zyo,
    author = "Schlingemann, D.",
    title = "{Stabilizer codes can be realized as graph codes}",
    eprint = "quant-ph/0111080",
    archivePrefix = "arXiv",
    month = "11",
    year = "2001"
}

@article{Bahramgiri:2006yab,
    author = "Bahramgiri, Mohsen and Beigi, Salman",
    title = "{Graph States Under the Action of Local Clifford Group in Non-Binary Case}",
    eprint = "quant-ph/0610267",
    archivePrefix = "arXiv",
    month = "10",
    year = "2006"
}

@article{Nest:2004khg,
    author = "Nest, Maarten Van den and Dehaene, Jeroen and Moor, Bart De",
    title = "{Graphical description of the action of local Clifford transformations on graph states}",
    eprint = "quant-ph/0308151",
    archivePrefix = "arXiv",
    doi = "10.1103/PhysRevA.69.022316",
    journal = "Phys. Rev. A",
    volume = "69",
    number = "2",
    pages = "022316",
    year = "2004"
}

@article{Akella:2026xza,
    author = "Akella, Sriram and Gadde, Abhijit and Pandey, Jay",
    title = "{Multi-invariants in stabilizer states}",
    eprint = "2601.16258",
    archivePrefix = "arXiv",
    primaryClass = "quant-ph",
    reportNumber = "TIFR/TH/26-4",
    month = "1",
    year = "2026"
}

@article{Akella:2025owv,
    author = "Akella, Sriram",
    title = "{Tripartite entanglement in the HaPPY code is not holographic}",
    eprint = "2510.08520",
    archivePrefix = "arXiv",
    primaryClass = "hep-th",
    month = "10",
    year = "2025"
}

@article{Iizuka:2025caq,
    author = "Iizuka, Norihiro and Lin, Simon and Nishida, Mitsuhiro",
    title = "{More on genuine multientropy and holography}",
    eprint = "2504.16589",
    archivePrefix = "arXiv",
    primaryClass = "hep-th",
    doi = "10.1103/x76v-mr6n",
    journal = "Phys. Rev. D",
    volume = "112",
    number = "6",
    pages = "066014",
    year = "2025"
}

@article{Balasubramanian:2025kaf,
    author = "Balasubramanian, Vijay and Cummings, Charlie",
    title = "{Multipartite entanglement structure of fibered link states}",
    eprint = "2502.19466",
    archivePrefix = "arXiv",
    primaryClass = "hep-th",
    doi = "10.1103/PhysRevD.111.105020",
    journal = "Phys. Rev. D",
    volume = "111",
    number = "10",
    pages = "105020",
    year = "2025"
}

@article{Iizuka:2025ioc,
    author = "Iizuka, Norihiro and Nishida, Mitsuhiro",
    title = "{Genuine multientropy and holography}",
    eprint = "2502.07995",
    archivePrefix = "arXiv",
    primaryClass = "hep-th",
    doi = "10.1103/714c-byxq",
    journal = "Phys. Rev. D",
    volume = "112",
    number = "2",
    pages = "026011",
    year = "2025"
}

@article{Harper:2024ker,
    author = "Harper, Jonathan and Takayanagi, Tadashi and Tsuda, Takashi",
    title = "{Multi-entropy at low Renyi index in 2d CFTs}",
    eprint = "2401.04236",
    archivePrefix = "arXiv",
    primaryClass = "hep-th",
    reportNumber = "YITP-24-02",
    doi = "10.21468/SciPostPhys.16.5.125",
    journal = "SciPost Phys.",
    volume = "16",
    number = "5",
    pages = "125",
    year = "2024"
}

@article{Gadde:2022cqi,
    author = "Gadde, Abhijit and Krishna, Vineeth and Sharma, Trakshu",
    title = "{New multipartite entanglement measure and its holographic dual}",
    eprint = "2206.09723",
    archivePrefix = "arXiv",
    primaryClass = "hep-th",
    reportNumber = "TIFR/TH/22-34",
    doi = "10.1103/PhysRevD.106.126001",
    journal = "Phys. Rev. D",
    volume = "106",
    number = "12",
    pages = "126001",
    year = "2022"
}

@article{Dwivedi:2020jyx,
    author = "Dwivedi, Siddharth and Addazi, Andrea and Zhou, Yang and Sharma, Puneet",
    title = "{Multi-boundary entanglement in Chern-Simons theory with finite gauge groups}",
    eprint = "2003.01404",
    archivePrefix = "arXiv",
    primaryClass = "hep-th",
    reportNumber = "CTP-SCU/2020004",
    doi = "10.1007/JHEP04(2020)158",
    journal = "JHEP",
    volume = "04",
    pages = "158",
    year = "2020"
}

@article{Balasubramanian:2018por,
    author = "Balasubramanian, Vijay and DeCross, Matthew and Fliss, Jackson and Kar, Arjun and Leigh, Robert G. and Parrikar, Onkar",
    title = "{Entanglement Entropy and the Colored Jones Polynomial}",
    eprint = "1801.01131",
    archivePrefix = "arXiv",
    primaryClass = "hep-th",
    doi = "10.1007/JHEP05(2018)038",
    journal = "JHEP",
    volume = "05",
    pages = "038",
    year = "2018"
}

@article{Dwivedi:2017rnj,
    author = "Dwivedi, Siddharth and Singh, Vivek Kumar and Dhara, Saswati and Ramadevi, P. and Zhou, Yang and Joshi, Lata Kh",
    title = "{Entanglement on linked boundaries in Chern-Simons theory with generic gauge groups}",
    eprint = "1711.06474",
    archivePrefix = "arXiv",
    primaryClass = "hep-th",
    reportNumber = "CTP-SCU/2017036; TIFR/TH/17-41, CTP-SCU-2017036, TIFR-TH-17-41",
    doi = "10.1007/JHEP02(2018)163",
    journal = "JHEP",
    volume = "02",
    pages = "163",
    year = "2018"
}

@article{Balasubramanian:2016sro,
    author = "Balasubramanian, Vijay and Fliss, Jackson R. and Leigh, Robert G. and Parrikar, Onkar",
    title = "{Multi-Boundary Entanglement in Chern-Simons Theory and Link Invariants}",
    eprint = "1611.05460",
    archivePrefix = "arXiv",
    primaryClass = "hep-th",
    doi = "10.1007/JHEP04(2017)061",
    journal = "JHEP",
    volume = "04",
    pages = "061",
    year = "2017"
}

@article{Salton:2016qpp,
    author = "Salton, Grant and Swingle, Brian and Walter, Michael",
    title = "{Entanglement from Topology in Chern-Simons Theory}",
    eprint = "1611.01516",
    archivePrefix = "arXiv",
    primaryClass = "quant-ph",
    doi = "10.1103/PhysRevD.95.105007",
    journal = "Phys. Rev. D",
    volume = "95",
    number = "10",
    pages = "105007",
    year = "2017"
}

@article{Hein:2004zjp,
    author = "Hein, M. and Eisert, J. and Briegel, H. J.",
    title = "{Multiparty entanglement in graph states}",
    eprint = "quant-ph/0307130",
    archivePrefix = "arXiv",
    doi = "10.1103/PhysRevA.69.062311",
    journal = "Phys. Rev. A",
    volume = "69",
    number = "6",
    pages = "062311",
    year = "2004"
}

@article{Witten:1988hf,
    author = "Witten, Edward",
    editor = "Mitra, Asoke N.",
    title = "{Quantum Field Theory and the Jones Polynomial}",
    reportNumber = "IASSNS-HEP-88-33",
    doi = "10.1007/BF01217730",
    journal = "Commun. Math. Phys.",
    volume = "121",
    pages = "351--399",
    year = "1989"
}

@article{Balasubramanian:2014hda,
    author = "Balasubramanian, Vijay and Hayden, Patrick and Maloney, Alexander and Marolf, Donald and Ross, Simon F.",
    title = "{Multiboundary Wormholes and Holographic Entanglement}",
    eprint = "1406.2663",
    archivePrefix = "arXiv",
    primaryClass = "hep-th",
    doi = "10.1088/0264-9381/31/18/185015",
    journal = "Class. Quant. Grav.",
    volume = "31",
    pages = "185015",
    year = "2014"
}

@article{Akella:2026rbe,
    author = "Akella, Sriram and Iizuka, Norihiro and Miyata, Akihiro",
    title = "{Genuine Multi-Entropy in the Toric Code}",
    eprint = "2607.06050",
    archivePrefix = "arXiv",
    primaryClass = "hep-th",
    month = "7",
    year = "2026"
}

@article{Yuan:2025dgx,
    author = "Yuan, Ma-Ke and Li, Mingyi and Zhou, Yang",
    title = "{Multi-entropy from Linking in Chern-Simons Theory}",
    eprint = "2510.18408",
    archivePrefix = "arXiv",
    primaryClass = "hep-th",
    month = "10",
    year = "2025"
}

@article{Iizuka:2026ahd,
    author = "Iizuka, Norihiro and Miyata, Akihiro",
    title = "{The junction law for multipartite entanglement in confining holographic backgrounds}",
    eprint = "2604.10583",
    archivePrefix = "arXiv",
    primaryClass = "hep-th",
    doi = "10.1007/JHEP08(2026)060",
    journal = "JHEP",
    volume = "08",
    pages = "060",
    year = "2026"
}

@article{Iizuka:2026qqg,
    author = "Iizuka, Norihiro and Miyata, Akihiro",
    title = "{Where Multipartite Entanglement Localizes: The Junction Law for Genuine Multi-Entropy}",
    eprint = "2602.16331",
    archivePrefix = "arXiv",
    primaryClass = "hep-th",
    month = "2",
    year = "2026"
}

@article{Iizuka:2025pqq,
    author = "Iizuka, Norihiro and Lin, Simon",
    title = "{Symmetry-resolved genuine multientropy: Random Haar and graph states}",
    eprint = "2511.00905",
    archivePrefix = "arXiv",
    primaryClass = "hep-th",
    doi = "10.1103/yts4-fldf",
    journal = "Phys. Rev. D",
    volume = "113",
    number = "2",
    pages = "026016",
    year = "2026"
}
\end{document}